\documentclass[twocolumn]{aastex631}

\usepackage{amssymb}
\usepackage{rotating}

\usepackage{booktabs}

\newcommand{\ron}{}

\shorttitle{WASP-166b}
\shortauthors{Mayo et al.}

\usepackage{hyperref}

\makeatletter
\newcommand\footnoteref[1]{\protected@xdef\@thefnmark{\ref{#1}}\@footnotemark}
\makeatother

\begin{document}

\title{Detection of H$_2$O and CO$_2$ in the Atmosphere of the Hot Super-Neptune WASP-166b with \textit{JWST}}

\author[0000-0002-7216-2135]{Andrew W. Mayo}
\affiliation{Department of Physics and Astronomy, San Francisco State University, San Francisco, CA 94132, USA}

\author[0000-0001-5286-639X]{Charles D. Fortenbach}
\affiliation{Department of Physics and Astronomy, San Francisco State University, San Francisco, CA 94132, USA}

\author[0000-0002-2457-272X]{Dana R. Louie}
\affiliation{Catholic University of America, Department of Physics, Washington, DC, 20064, USA}
\affiliation{Exoplanets and Stellar Astrophysics Laboratory (Code 667), NASA Goddard Space Flight Center, Greenbelt, MD 20771, USA}
\affiliation{Center for Research and Exploration in Space Science and Technology II, NASA/GSFC, Greenbelt, MD 20771, USA}

\author[0000-0001-8189-0233]{Courtney D. Dressing}
\affiliation{Department of Astronomy, University of California, Berkeley, Berkeley, CA 94720, USA}

\author[0000-0002-1845-2617]{Emma V. Turtelboom}
\affiliation{Department of Astronomy, University of California, Berkeley, Berkeley, CA 94720, USA}

\author[0000-0002-8965-3969]{Steven Giacalone}
\altaffiliation{NSF Astronomy and Astrophysics Postdoctoral Fellow}
\affiliation{Department of Astronomy, California Institute of Technology, Pasadena, CA 91125, USA}

\author[0000-0001-5737-1687]{Caleb K. Harada}
\altaffiliation{NSF Graduate Research Fellow}
\affiliation{Department of Astronomy, University of California, Berkeley, Berkeley, CA 94720, USA}

\begin{abstract}
We characterize the atmosphere of the hot super-Neptune WASP-166b ($P = 5.44$ d, $R_p = 6.9 \pm 0.3$ R$_\oplus$, $M_p = 32.1 \pm 1.6$ M$_\oplus$, $T_\mathrm{eq} = 1270 \pm 30$ K) orbiting an F9V star using \textit{JWST} transmission spectroscopy with NIRISS and NIRSpec ($0.85-5.17$ $\mu$m). {\ron With this broad wavelength range, NIRISS provides strong constraints on H$_2$O and clouds (where NIRSpec performs poorly) while NIRSpec captures CO$_2$ and NH$_3$ (where NIRISS performs poorly).} Our \texttt{POSEIDON} free chemistry retrievals confirm the detection of H$_2$O ($15.2\sigma$ significance) and detect CO$_2$ ($14.7\sigma$) for the first time. We also find a {\ron possible} hint of NH$_3$ ($2.3\sigma$) and an intermediate pressure cloud deck ($2.6\sigma$). Finally, {\ron we report inconclusive support for the presence of SO$_2$, CO, and Na, as well as non-detections of CH$_4$, C$_2$H$_2$, HCN, H$_2$S, and K}. We verify our results using a \texttt{TauREx} free chemistry retrieval. We also measure with \texttt{POSEIDON} {\ron equilibrium chemistry retrievals a superstellar} planetary atmospheric metallicity ($\log(Z) = 1.57^{+0.17}_{-0.18}$, $Z = 37^{+18}_{-13}$) and planetary C/O ratio ($C/O = 0.282^{+0.078}_{-0.053}$) {\ron consistent} with the stellar C/O ratio ($C/O_* = 0.41 \pm 0.08$). {\ron These results are compatible with various planetary formation pathways, especially those that include planetesimal accretion followed by core erosion or photoevaporation. WASP-166b also resides near the edge of the Hot Neptune Desert, a scarcity of intermediate-sized planets at high insolation fluxes; thus, these results and further atmospheric observations of Hot Neptunes will help determine the driving processes in the formation of the Hot Neptune Desert.}

\end{abstract}


\section{Introduction} \label{intro}

Although more than 5000 exoplanets have been detected to date, only a small fraction have had any constituents of their atmospheres measured. According to the NASA Exoplanet Archive (accessed {\ron 2025 Mar 24}), less than $5\%$ of known exoplanets have an observed planetary spectrum of 5 or more points. And yet, exoplanet atmospheres are quickly becoming one of the most fruitful areas of study for understanding planet formation and evolution.

Conducting transmission spectroscopy with \textit{JWST} has opened the door to placing firm atmospheric constraints on the smallest {\ron{observed}} planets to date, including a rare class of planets called Hot Neptunes. Hot Neptunes, as their name would suggest, are planets of intermediate radii ($\sim 1.5 - 7.5$ R$_\oplus$) found at high insolation fluxes ($\gtrsim 400$ S$_\oplus$). Hot Neptunes, even those with similar bulk properties, can exhibit a wide variety of atmospheric features and compositions. {\ron (We discuss other Hot Neptunes and planets similar to WASP-166b in Sections~\ref{ltt9779b} and~\ref{other_planets} and present a list of key planetary parameters and atmospheric properties in Table~\ref{tab:similar_planets}.)}

\begin{figure}
    \centering
    \includegraphics[width=0.5\textwidth]{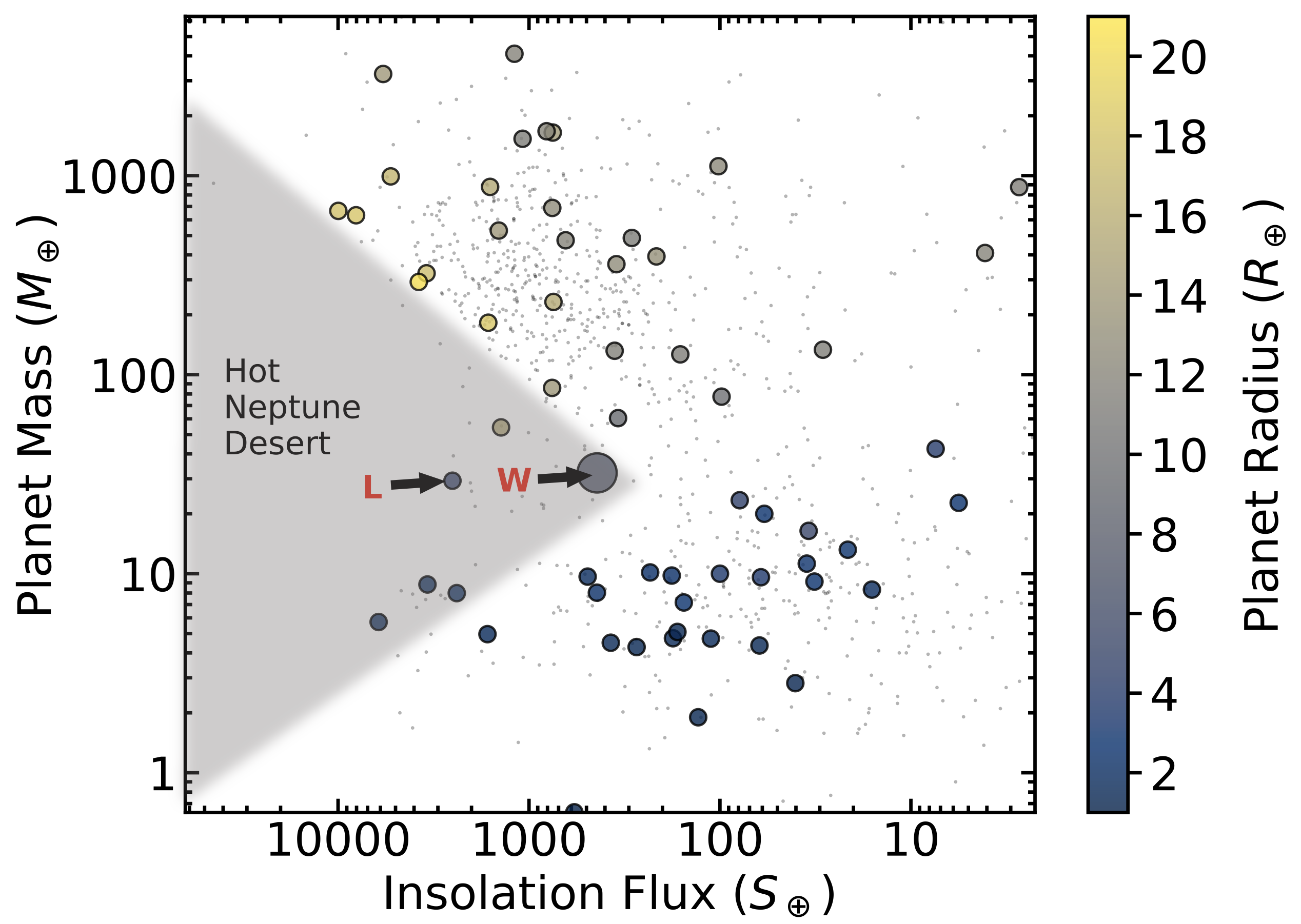}
    \caption{Insolation flux versus planet mass, with the Hot Neptune Desert \citep{mazehetal2016} labeled on the left. Small black dots are confirmed exoplanets; large dots colored by planet radius are confirmed planets with mass (or M$\sin(i)$) and radius uncertainties $< 10\%$ and bright host stars (J $< 9$). A rough outline of the Hot Neptune Desert is designated by the gray triangular region (to guide the eye only, not to serve as a sharp delineation). WASP-166b ({\ron large and labeled} ``W'') is on the edge of the desert. LTT 9779b (``L'') is deep within the desert and labeled for comparison; see Section~\ref{ltt9779b} for further discussion.} 
    \label{fig:HND_mass_v_insol}
\end{figure}

In this paper we present the results of an atmospheric analysis of WASP-166b, {\ron a.k.a. Catalineta\footnote{\url{https://wasp-planets.net/2023/06/25/the-iau-names-more-wasp-exoplanets/}\label{iau_names}}}, a hot, puffy super-Neptune ($P = 5.44$ d, $R_p = 6.9 \pm 0.3$ R$_\oplus$, $M_p = 32.1 \pm 1.6$ M$_\oplus$; \citealt{Hellieretal2019,Doyleetal2022}). {\ron This planet orbits the F9V star WASP-166 ($M_* = 1.19 \pm 0.06$ M$_\odot$, $R_* = 1.22 \pm 0.06$ R$_\odot$, $T_\mathrm{eff} = 6050 \pm 50$ K; \citealt{Hellieretal2019}), a.k.a. Filetdor\textsuperscript{\ref{iau_names}}.} WASP-166b is located at the edge of the Hot Neptune Desert \citep{mazehetal2016}, a region of parameter space at high insolation flux and intermediate planet radii that is very sparsely populated (see Figure~\ref{fig:HND_mass_v_insol}). 

WASP-166b was originally detected in transit photometry collected by the Wide Angle Search for Planets (WASP) survey over 2006-2012. It was then followed up with a radial velocity (RV) mass measurement and confirmation from CORALIE and HARPS over 2014-2018 (\citealt{Hellieretal2019}; see Table~\ref{tab:system_params}). \citet{bryantetal2020} analyzed additional transits of WASP-166b collected in February 2019 with the Next Generation Transit Survey (NGTS) and the Transiting Exoplanet Survey Satellite (TESS) in order to further refine the transit ephemeris and planet radius. 

\begin{table*}[bt]
\begin{center}
\caption{System parameters used in this paper. 
(1)	\citet{Hellieretal2019}, (2) \citet{Doyleetal2022}.}

\begin{tabular}{lcc}
\hline \hline
Parameter & Value & Reference \\
\hline
WASP-166 \\
\hspace{3mm} $M_*$ (M$_\odot$) & $1.19 \pm 0.06$ & (1) \\
\hspace{3mm} $R_*$ (R$_\odot$) & $1.22 \pm 0.06$ & (1) \\
\hspace{3mm} $T_\mathrm{eff}$ (K) & $6050 \pm 50$ & (1) \\
\hspace{3mm} $\log g_*$ (log$_\mathrm{10}$(cm/s$^\mathrm{2}$)) & $4.5 \pm 0.1$ & (1) \\
\hspace{3mm} [Fe/H] & $0.19 \pm 0.05$ & (1) \\
\hspace{3mm} {\ron Distance (pc)} & {\ron $113 \pm 1$} & {\ron (1)} \\
\tableline
WASP-166b \\
\hspace{3mm} $P$ (days) & $5.44354215^{+0.00000307}_{-0.00000297}$ & (2) \\
\hspace{3mm} $t_0$ (BJD-2457000) & $1524.40869201^{+0.00030021}_{-0.00029559}$ & (2) \\
\hspace{3mm} $a/R_*$ & $11.83^{+0.29}_{-0.68}$ & (2) \\
\hspace{3mm} $i$ (deg) & $88.85^{+0.74}_{-0.94}$ & (2) \\
\hspace{3mm} $R_p$ (R$_\mathrm{Jup}$) & $0.6155^{+0.0306}_{-0.0307}$ & (2) \\
\hspace{3mm} $M_p$ (M$_\mathrm{Jup}$) & $0.101 \pm 0.005$ & (1) \\
\hspace{3mm} $T_\mathrm{eq}$ (K)\footnote{\ron{Equilibrium temperature ($T_\mathrm{eq}$) calculated assuming zero albedo and efficient heat redistribution}} & $1270 \pm 30$ & (1) \\
\hspace{3mm} $e$ & $0$ & (1) \\
\hspace{3mm} $\omega$ ($^\circ$) & $90$ & (1) \\
\hline

\label{tab:system_params}
\end{tabular}
\end{center}
\end{table*}

The first report of an atmospheric constituent for WASP-166b came from \citet{seideletal2020}, who reported a tentative ($3.4\sigma$) detection of neutral sodium through ground-based HARPS transmission spectroscopy. This result was subsequently confirmed ($7.3\sigma$) with additional ground-based transmission spectroscopy from ESPRESSO \citep{seideletal2022}, helping demonstrate the planet's suitability for additional atmospheric characterization. {\ron Both studies focused on the Na I doublet at $588.9950$ and $589.5924$ nm. {\ron The prominence of this} doublet is extremely sensitive to even small quantities of sodium, and thus all prior sodium detections have been accomplished through {\ron observations of} the Na I doublet \citep{charbonneauetal2002,redfieldetal2008,nikolovetal2016,singetal2016,casasayas-barrisetal2017,casasayas-barrisetal2019,wyttenbachetal2017,chenetal2018,chenetal2020,jensenetal2018,deibertetal2019,hoeijmakersetal2019,seideletal2019,cabotetal2020}.} Those same ESPRESSO observations were used by \citet{Doyleetal2022} to employ the reloaded Rossiter McLaughlin technique \citep{ceglaetal2016} to measure the sky-projected star-planet obliquity; they found the planet orbit to be well aligned with the stellar spin axis ($\lambda = -15.52^{+2.85}_{-2.76}$ degrees). Further analysis of the same ESPRESSO observations led to a tentative detection of water vapor, with \citet{lafargaetal2023} finding that models with intermediate water vapor abundances and cloud deck pressures were preferred at $4\sigma$ significance or better.

We present here the first observations of WASP-166b with \textit{JWST}, a near infrared (NIR) transmission spectrum that we reduce and analyze with the goal of constraining molecular abundances, detecting atmospheric features, and estimating the atmospheric metallicity and C/O ratio. This is the first publication providing a thorough description of the NIRSpec G395M transmission spectroscopy data reduction process, and demonstrating the combination of NIRSpec G395M with NIRISS SOSS. Our analysis of these data allows us to investigate the formation pathway of WASP-166b and explore the origins of planets inside or near the boundary of the Hot Neptune Desert.

{\ron In Section~\ref{obs}, we present our observations of WASP-166b. In Section~\ref{analysis}, we summarize our data reduction procedure, reserving full descriptions for Appendices~\ref{appA_g395m_reduction} (NIRSpec) and~\ref{appB_soss_reduction} (NIRISS). We then discuss our atmospheric modeling and retrievals in Section~\ref{retrievals}. We verify our results and explore various aspects of our analysis with additional retrievals, which we present and discuss in Appendix~\ref{ap:free_retrievals} for free chemistry and Appendix~\ref{ap:eq_retrievals} for equilibrium chemistry. Then, we discuss how our results compare to prior observations of WASP-166b and other Hot Neptunes in Section~\ref{results}. Finally, we present our overall summary and conclusions in Section ~\ref{conclusion}.}


\section{Observations} \label{obs}

We observed two transits of WASP-166b as part of \textit{JWST} Cycle 1 General Observer (GO) program 2062 (PI: Mayo, Co-PI: Dressing). The first observation was obtained on 31 Dec 2023 at 03:59 - 12:54 UTC with the Near Infrared Imager and Slitless Spectrograph (NIRISS; \citealt{albertetal2023,doyonetal2023}). It was collected in Single Object Slitless Spectroscopy (SOSS) mode covering a wavelength range of $0.85$ to $2.81\ \mu$m across Order 1 at a native Spectral Resolving Power ($R$) of $\sim 650$ at $1.25\ \mu$m. 

The NIRISS SOSS science observation employed the GR700XD grism combined with the clear filter, making use of the SUBSTRIP96 subarray (2048 columns by 96 rows). Use of the larger SUBSTRIP256 subarray was not feasible for our observations because the brightness of the host star, WASP-166, would saturate the detector in this configuration. Instead, we used the SUBSTRIP96 subarray in the NISRAPID read mode. This subarray does not adequately capture the shorter wavelength, Order 2 spectral trace, so our observations were limited to the Order 1 wavelength range. The transit observation comprised 4836 integrations, with 2 groups per integration. Our effective exposure time during the observational window was $21413.808$ s, yielding an effective integration time of $4.428$ s ($21413.808 / 4836$). We added a recommended GR700XD/F277W exposure following our science exposure,\footnote{See NIRISS SOSS recommended strategies at \href{https://jwst-docs.stsci.edu/jwst-near-infrared-imager-and-slitless-spectrograph/niriss-observing-strategies/niriss-soss-recommended-strategies}{https://jwst-docs.stsci.edu/}} which included a total of 20 integrations, with 2 groups per integration.

The second observation was obtained on 16 Jan 2024 at 12:06 - 21:01 UTC with the Near Infrared Spectrograph (NIRSpec; \citealt{jakobsenetal2022, BirkmanetalIV2022, Bokeretal2023, Espinozaetal2023}). It was collected in the Bright Object Time Series (BOTS) mode using the G395M grating and F290LP filter, covering a wavelength range from $2.80$ to $5.17\ \mu$m with $R \sim 1000$ at $3.95\ \mu$m. 

The NIRSpec data were taken using the 1.6'' square aperture (S1600A1) with the SUB2048 subarray (2048 pixel columns by 32 rows on the NRS1 detector), and NRSRAPID readout pattern. The spectra were dispersed across approximately 1320-pixel columns of the subarray, with moderate curvature (central row of the raw spectral image shifts along the columns) of the trace. The observation comprised a total of 8880 integrations taken in a single exposure, roughly centered around the $3.6$ h transit. Our effective exposure time during the observational window was $24029.28$ s, yielding an effective integration time of $2.706$ s ($24029.28/ 8880$). The observation was set up for maximum efficiency while remaining below an 80\% full-well threshold to avoid detector non-linearity. Due to the brightness of WASP-166 we used 3 groups per integration to avoid saturation. 

Our choice of the medium-resolution NIRSpec G395M mode rather than the high resolution G395H mode was driven by several considerations. First, the nominal $R \sim 1000$ resolving power of the M mode is sufficient for our science goals. In general, we are not trying to discriminate between very closely packed or overlapping features. Secondly, the M mode has slightly higher throughput, and does not have a complicating detector gap at $3.72$ to $3.82\ \mu$m. Third, an information content analysis by \citet{guzmanmesaetal2020} revealed that improving the spectral resolution from the intermediate value of $R \sim 1000$ covered by G395M to the higher value of $R \sim 2700$ accorded by G395H results in negligible improvement in abundance constraints.

\section{{\ron Observational Data} Analysis} \label{analysis}
For both the NIRSpec G395M and NIRISS SOSS observations, we generated a planetary transmission spectrum by following a specific sequence of processing steps. 

In Section~\ref{NIRSpec_Data_Reduction}, we describe our application of the \textit{JWST} \texttt{Science Calibration Pipeline} \citep{bushouse2023}, and portions of the \texttt{Eureka!\hspace{0.2em}}pipeline \citep{belletal2022} to the NIRSpec G395M observational data; likewise in Section~\ref{NIRISS_SOSS_Data_Reduction}, we describe our application of the tested \texttt{Ahsoka} pipeline \citep[][Macdonald et al., in prep]{Louie_2025,Gressier_2025}, and again portions of the \texttt{Eureka!\hspace{0.2em}}pipeline to the NIRISS SOSS observational data.

Throughout the data reduction process, our goal was to properly calibrate the data and minimize correlated or systematic noise without degrading or introducing bias into the ``true'' signal.

The final output transmission spectra for both instruments have been reduced (pre-binned) in three forms: (1) a constant ($\sim 0.01797 \ \mu$m) bin width, (2) R = 100 binning, and (3) 2.2-pixel bins for NIRSpec G395M and 1-pixel bins for NIRISS SOSS. {\ron A more detailed explanation of the binning rationale for each instrument is provided in Sections~\ref{subsubsec:NIRSpec_stage4} and~\ref{subsubsec:NIRISS_Stages456} for NIRSpec and NIRISS, respectively.}

All data reduction control files to reproduce these results and the reduced data are available on Zenodo.\footnote{\url{https://doi.org/10.5281/zenodo.14503925}}

\subsection{NIRSpec Data Reduction} \label{NIRSpec_Data_Reduction}
{\ron Our NIRSpec G395M data reduction uses the default \textit{JWST} \texttt{Science Calibration Pipeline} for Stage 1, which starts with 2-D uncalibrated data image frames (\textit{uncal.fits} data), calibrates the raw data and produces \textit{rateints.fits} files, and then makes use of \texttt{Eureka!\hspace{0.1em}}(v0.10), for Stages 2 through 6. The \texttt{Eureka!\hspace{0.2em}}pipeline has been used on many \textit{JWST} atmospheric analyses and has produced reliable results \citep{JWSTERSTEAM2023,LustigYaegeretal2023,Ahreretal2023,Aldersonetal2023,Moranetal2023,Rustamkulovetal2023}.}

\begin{figure*}
    \includegraphics[width=\textwidth]{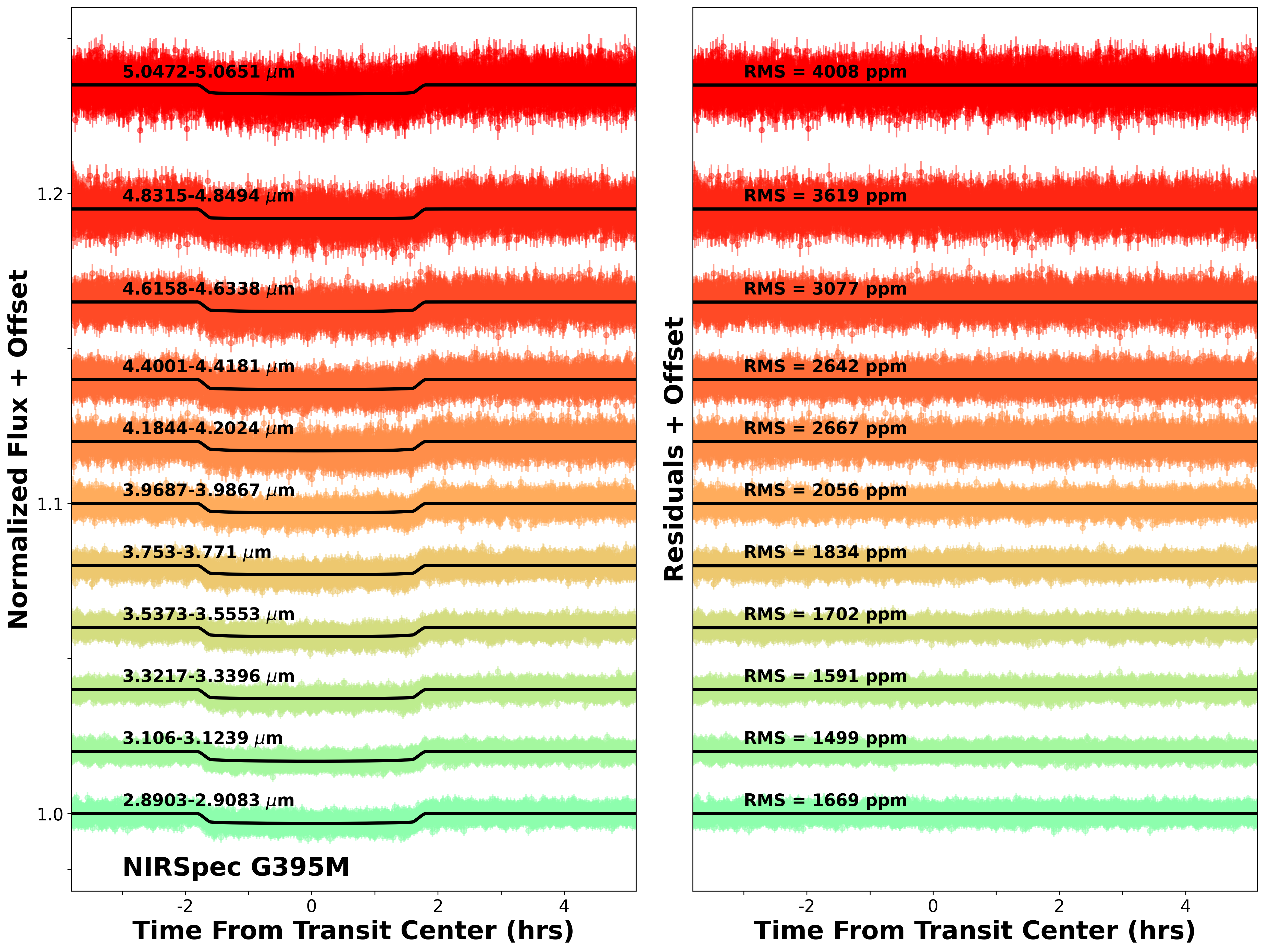}
        \caption{Examples of the \texttt{Eureka!\hspace{0.2em}}\citep{belletal2022} Stage 5 spectroscopic light curves and \texttt{dynesty} \citep{Speagle2020} fits for the NIRSpec G395M constant ($\sim 0.01797 \ \mu$m) bin-width data reduction. \textbf{\underline{Left:}} Data corrected with polynomial systematics models (colored points), and overplotted with the best fit transit models (black curves). Wavelengths for each spectral channel are shown above the corresponding transit light curve. \textbf{\underline{Right:}} Residuals for the corresponding data, with RMS scatter in black text. Data and best fit transit curves are offset for each spectral channel for clarity.}
    \label{fig:EurekaNIRSpecFitSpectroscopicLCs}
\end{figure*}

{\ron Appendix \ref{appA_g395m_reduction} presents in detail our application of the \textit{JWST} \texttt{Science Calibration Pipeline} and the \texttt{Eureka!\hspace{0.1em}}pipeline to our WASP-166b NIRSpec G395M data. In Section \ref{subsubsec:NIRSpec_stage1} we detail our use of the \textit{JWST} \texttt{Science Calibration Pipeline} to do the initial Stage 1 detector processing and calibration. We then move on to Stage 2 with \texttt{Eureka!\hspace{0.2em}}for further calibrations, wavelength mapping, etc., described in \ref{subsubsec:NIRSpec_stage2}. In \ref{subsubsec:NIRSpec_stage3} we describe the \texttt{Eureka!\hspace{0.2em}}Stage 3 background subtraction, optimal spectral extraction, and generation of a time series of 1D spectra. We then use \texttt{Eureka!\hspace{0.2em}}to generate spectroscopic light curves, fit light curves, and produce the transmission spectrum (Stages 4, 5, and 6). The processing details of these stages are discussed in \ref{subsubsec:NIRSpec_stage4} and \ref{subsubsec:NIRSpec_stage56}. A sample of the spectroscopic light curve fits from \texttt{Eureka!\hspace{0.2em}}Stage 5 are shown for NIRSpec in Figure~\ref{fig:EurekaNIRSpecFitSpectroscopicLCs}.}

\subsection{NIRISS SOSS Data Reduction} \label{NIRISS_SOSS_Data_Reduction}
We used the \texttt{Ahsoka} pipeline \citep[][Macdonald et al., in prep]{Louie_2025,Gressier_2025} in our NIRISS SOSS analysis. The \texttt{Ahsoka} pipeline is comprised of six separate stages, and combines software modules from the \textit{JWST} Science Calibration Pipeline, {\ron \texttt{exoTEDRF} \citep[formerly known as \texttt{supreme-SPOON};][]{feinsteinetal2023,radicaetal2023,Radica2024JOSS}}, \texttt{nirHiss} \citep{feinsteinetal2023}, and \texttt{Eureka!\hspace{0.2em}}pipelines.

\begin{figure*} 
    \includegraphics[width=\textwidth]{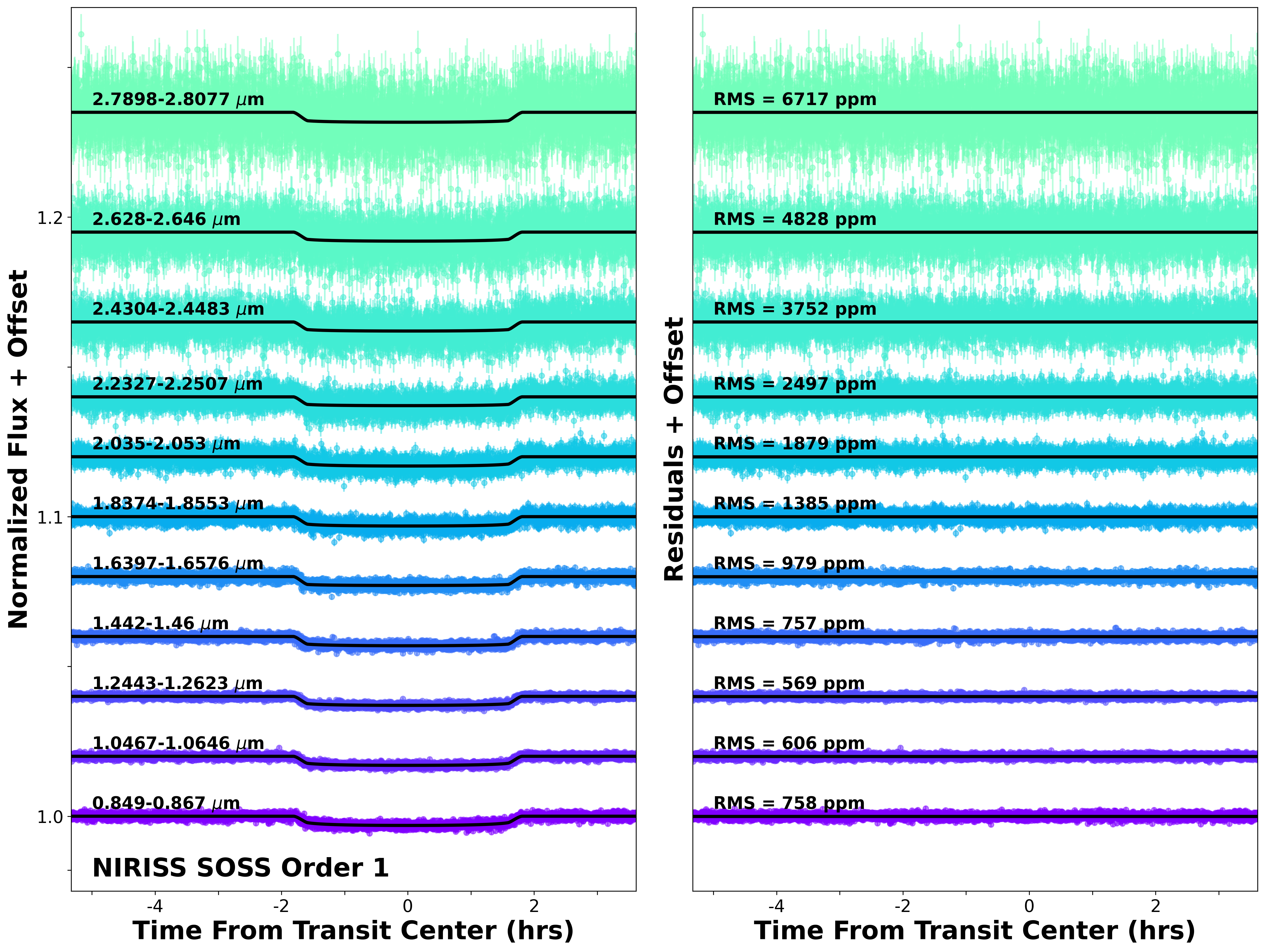}
        \caption{Examples of our \texttt{Eureka!\hspace{0.2em}}\citep{belletal2022} Stage 5 spectroscopic light curve \texttt{emcee} \citep{Foreman-Mackey2013} fits to the NIRISS SOSS \texttt{Ahsoka} data reduction {\ron for constant ($\sim 0.01797 \ \mu$m) bin-widths}. \textbf{\underline{Left:}} Data corrected with polynomial systematics models (colored points), and overplotted with the best fit transit models (black curves). Wavelengths for each spectral channel are shown above the corresponding transit light curve. \textbf{\underline{Right:}} Residuals for the corresponding data, with RMS scatter in black text. Data and best fit transit curves are offset for each spectral channel for clarity.}
    \label{fig:EurekaFitSpectroscopicLCs}
\end{figure*}

{\ron Appendix \ref{appB_soss_reduction} describes in detail our application of \texttt{Ahsoka} to our WASP-166b NIRISS SOSS data.} In Sections \ref{subsubsec:NIRISS_Stage1}, \ref{subsubsec:NIRISS_Stage2}, and \ref{subsubsec:NIRISS_Stage3} we describe our application of \texttt{Ahsoka} detector-level processing, spectroscopic processing, and spectral extraction, respectively, to the WASP-166b NIRISS SOSS data. The output product from \texttt{Ahsoka} Stage 3 is a time series of 1D (flux versus wavelength) stellar spectra. \texttt{Ahsoka} exclusively employs \texttt{Eureka!\hspace{0.2em}}to generate spectroscopic light curves, fit light curves, and produce the transmission spectrum (\texttt{Eureka!\hspace{0.2em}}Stages 4, 5, and 6). We describe our application of \texttt{Eureka!\hspace{0.2em}}to NIRISS SOSS stellar spectra in Section \ref{subsubsec:NIRISS_Stages456}. {\ron In Figure~\ref{fig:EurekaFitSpectroscopicLCs}, we show a sample of our \texttt{Eureka!\hspace{0.2em}}Stage 5 spectroscopic light curves spaced across the NIRISS SOSS bandpass, overplotted with the corresponding fits.}

\section{Atmospheric Modeling and Retrievals} \label{retrievals}
{\ron After reducing the NIRISS and NIRSpec data as described in Section~\ref{analysis}, we used the resulting spectra to explore the atmospheric composition of WASP-166b with forward modeling and retrievals. To ensure that our results were robust, we used a variety of modeling assumptions and three different analysis packages: \texttt{POSEIDON} v1.2.1 \citep{macdonaldandmadhusudhan2017, macdonald2023}, \texttt{PLATON} v5.3 \citep{zhangetal2019,zhangetal2020}, and \texttt{TauREx} v3 \citep{alrefaieetal2021}. These packages, of course, have differences. One of the more significant ones is that they draw on slightly different absorption (opacity) line lists, which could potentially induce minor changes in the spectra generated for the same atmospheric conditions. Table~\ref{tab:opacity_data} summarizes the line lists used by each of the modeling packages.}

{\ron We initially used \texttt{POSEIDON} with the combined (NIRISS and NIRSpec) instrument data using the $\sim 0.01797 \ \mu$m bin width data reduction. As discussed in Section~\ref{subsubsec:NIRSpec_stage4}, the motivation for constant bin widths was to provide a consistent set of data from both instruments, and balance the need for high enough resolution to capture the important spectral features, with reasonable computation time. We also retrieved the systematic transit depth offset ($\delta_\mathrm{rel}$) between the two instruments as part of our analysis for \texttt{POSEIDON} (as well as for \texttt{PLATON}, as discussed in Appendix~\ref{ap:eq_retrievals}).

Our final reported parameters for the atmospheric composition of WASP-166b come from \texttt{POSEIDON} modeling and retrievals (namely B1, E1, and the R1 sequence); we used the other two packages to verify our results. See Table~\ref{tab:retrieval_cases} to associate the reference ID assigned to each retrieval case with the defining parameters (e.g., Instrument, Code, Chemistry, Binning, etc.).} 

{\ron In Section~\ref{poseidon}, we first describe the assumptions and procedures used for our retrievals with \texttt{POSEIDON}. We then present our use of \texttt{POSEIDON} (including settings and results) to investigate the atmospheric composition of WASP-166b using both free chemistry (Section~\ref{free_retrievals}) and equilibrium chemistry (Section~\ref{equilibrium_retrievals}). 

To confirm the results of our retrievals, we conducted additional retrievals of the atmosphere of WASP-166b using different assumptions and modeling packages. We present these modeling approaches in Appendix~\ref{ap:free_retrievals} for free chemistry (with \texttt{POSEIDON} and \texttt{TauREx}) and Appendix~\ref{ap:eq_retrievals} for equilibrium chemistry (with \texttt{POSEIDON} and \texttt{PLATON}). In both appendices, we explore retrievals using the combined NIRSpec and NIRISS data, the NIRSpec data alone, and the NIRISS data alone, with both fixed ($0.01797 \ \mu$m) bins as well as at constant {\ron ($R = 100$)} spectral resolution. } All retrieval inputs and results are available on Zenodo.\footnote{\url{https://doi.org/10.5281/zenodo.14503925}}

\begin{table*}
    \centering
    \caption{Opacity Data for Retrievals}
    \begin{tabular}{ll|ll|ll|l}
        \multicolumn{2}{c}{\texttt{POSEIDON} (v1.2.1)} & \multicolumn{2}{c}{\texttt{PLATON} (v5.3)} & \multicolumn{2}{c}{\texttt{TauREx} (v3)} & \\
        Molecule\footnote{The \texttt{POSEIDON} and \texttt{PLATON} molecular constituents are those used during our equilibrium chemistry retrievals. Only molecules shown in bold are used in free chemistry retrievals and forward models.} & Opacity Reference & Molecule & Opacity Reference & Molecule & Opacity Reference & Note\footnote{(1) \texttt{PLATON} and \texttt{POSEIDON} equilibrium chemistry sources are different; (2) \texttt{PLATON} and \texttt{POSEIDON} equilibrium chemistry sources are the same; and (a) \texttt{POSEIDON} and \texttt{TauREx} free chemistry sources are different.} \\
        \hline
        C$_2$ & \citet{yurchenkoetal2018c} &  &  &  &  & \\
        \textbf{C}$_\mathbf{2}$\textbf{H}$_\mathbf{2}$ & \citet{chubb_et_al2020} & C$_2$H$_2$ & \citet{gordonetal2017} &  &  & 1 \\
        C$_2$H$_4$ & \citet{gordonetal2022} & C$_2$H$_4$ & \citet{mantetal2018} &  &  & 1 \\
         &  & C$_2$H$_6$ & \citet{gordonetal2017} &  &  & \\
        \textbf{CH}$_\mathbf{4}$ & \citet{yurchenko_et_al2024} & CH$_4$ & \citet{reyetal2017} &  &  & 1 \\
        \textbf{CO} & \citet{li_et_al2015} & CO & \citet{li_et_al2015} &  &  & 2 \\
        \textbf{CO}$_\mathbf{2}$ & \citet{yurchenko_et_al2020} & CO$_2$ & \citet{tashkun+perevalov2011} & \textbf{CO$_\mathbf{2}$} & \citet{rothmanetal2010}  & 1a \\
        CaH & \citet{Owensetal2022} &  &  &  &  & \\
        CrH & \citet{Bernath2020} &  &  &  &  & \\
        FeH & \citet{Bernath2020} &  &  &  &  & \\
        \textbf{H$_\mathbf{2}$O} & \citet{polyansky_et_al2018} & H$_2$O & \citet{polyansky_et_al2018} & \textbf{H$_\mathbf{2}$O} & \citet{polyansky_et_al2018},  & 2a \\
          &  &  &  &  &  \citet{bartonetal2017}  & \\
        \textbf{H$_\mathbf{2}$S} & \citet{azzam_et_al2016} & H$_2$S & \citet{azzam_et_al2016} &  &  & 2 \\
         &  & H$_2$CO & \citet{al-refaieetal2015} &  &  & \\
         &  & HCl & \citet{lietal2013} &  &  & \\
        \textbf{HCN} & \citet{barber_et_al2014} & HCN & \citet{barber_et_al2014} &  &  & 2 \\
         &  & HF & \citet{gordonetal2017} &  &  & \\
        \textbf{K} & \citet{ryabchikova_2015} & K & \citet{kramida_et_al2013} &  &  & 1 \\
        MgH & \citet{Owensetal2022} & MgH & \citet{gharibnezhadetal2013} &  &  & 1 \\
         &  & N$_2$ & \citet{gordonetal2017} &  &  & \\
        \textbf{NH$_\mathbf{3}$} & \citet{coles_et_al2019} & NH$_3$ & \citet{coles_et_al2019} & \textbf{NH$_\mathbf{3}$} & \citet{yurchenkoetal2011}  & 2a \\
        NO & \citet{Quetal2021} & NO & \citet{wongetal2017} &  &  & 1 \\
        NO$_2$ & \citet{Hargreavesetal2019} & NO$_2$ & \citet{gordonetal2017} &  &  & 1 \\
        \textbf{Na} & \citet{ryabchikova_2015} & Na & \citet{kramida_et_al2013} &  &  & 1 \\
        O$_2$ & \citet{gordonetal2022} & O$_2$ & \citet{gordonetal2017} &  &  & 1 \\
        O$_3$ & \citet{gordonetal2022}, & O$_3$ & \citet{gordonetal2017} &  &  & 1 \\
        & \citet{Serdyuchenkoetal2014} &  &  &  &  & \\
        OCS & \citet{Owensetal2024} & OCS & \citet{gordonetal2017} &  &  & 1 \\
        OH & \citet{Bernath2020} & OH & \citet{brookeetal2016} &  &  & 1 \\
        PH$_3$ & \citet{Sousasilvaetal2015} & PH$_3$ & \citet{sousa-silvaetal2014} &  &  & 2 \\
        SH & \citet{Gormanetal2019} & SH & \citet{yurchenkoetal2018a} &  &  & 1 \\
        SO & \citet{Bradyetal2024} &  &  &  &  & \\
        SiH & \citet{Yurchenkoetal2018b} & SiH & \citet{Yurchenkoetal2018b} &  &  & 2 \\
        SiO & \citet{Yurchenkoetal2022} & SiO & \citet{bartonetal2013} &  &  & 1 \\
        \textbf{SO}$_\mathbf{2}$ & \citet{underwood_et_al2016} & SO$_2$ & \citet{underwood_et_al2016} &  &  & 2 \\
        TiH & \citet{Bernath2020} &  &  &  &  & \\
        TiO & \citet{Mckemmishetal2019} & TiO & \citet{Mckemmishetal2019} &  &  & 2 \\
        VO & \citet{mckemmishetal2016} & VO & \citet{mckemmishetal2016} &  &  &  2\\
    \end{tabular}
    \label{tab:opacity_data}
\end{table*}

\begin{table*}
\begin{center}
\caption{Retrieval Cases}

\begin{tabular}{llllllll}
\hline \hline
ID   & Instrument & Binning          & Code  & Chem   & P-T model   & Clouds       & Constituents\footnote{For \texttt{POSEIDON} and \texttt{TauREx} retrievals, the bulk gases, H$_2$ and He, are present at an assumed primordial solar ratio of $X_\mathrm{He}/X_\mathrm{H_2}$ = 0.17. The remainder of the atmosphere not made up of trace gases is filled by the bulk gases.} \\
\hline
R1   & Combined   & $0.01797 \ \mu$m & POS   & Free   & 'isotherm'  & 'deck'       & Ref1\footnote{Free parameters: log\_vmr of H$_2$O, CO$_2$, NH$_3$, SO$_2$, CO, C$_2$H$_2$, H$_2$S, CH$_4$, HCN, and log\_cloud top pressure}       \\
R1a  & Combined   & $0.01797 \ \mu$m & POS   & Free   & 'isotherm'  & 'deck'       & Ref1 - H$_2$O \\
R1b  & Combined   & $0.01797 \ \mu$m & POS   & Free   & 'isotherm'  & 'deck'       & Ref1 - CO$_2$ \\
R1c  & Combined   & $0.01797 \ \mu$m & POS   & Free   & 'isotherm'  & 'deck'       & Ref1 - NH$_3$ \\
R1d  & Combined   & $0.01797 \ \mu$m & POS   & Free   & 'isotherm'  & 'deck'       & Ref1 - SO$_2$ \\
R1e  & Combined   & $0.01797 \ \mu$m & POS   & Free   & 'isotherm'  & 'deck'       & Ref1 - CO \\
R1f  & Combined   & $0.01797 \ \mu$m & POS   & Free   & 'isotherm'  & 'deck'       & Ref1 - C$_2$H$_2$ \\
R1g  & Combined   & $0.01797 \ \mu$m & POS   & Free   & 'isotherm'  & 'deck'       & Ref1 - H$_2$S \\
R1h  & Combined   & $0.01797 \ \mu$m & POS   & Free   & 'isotherm'  & 'deck'       & Ref1 - CH$_4$ \\
R1i  & Combined   & $0.01797 \ \mu$m & POS   & Free   & 'isotherm'  & 'deck'       & Ref1 - HCN \\
R1j  & Combined   & $0.01797 \ \mu$m & POS   & Free   & 'isotherm'  & 'cloud-free' & Ref1 - Clouds \\
& & & & & & &  \\
B1   & Combined   & $0.01797 \ \mu$m & POS   & Free   & 'isotherm'  & 'deck'       & Baseline\footnote{Free parameters: log\_vmr of H$_2$O, CO$_2$, NH$_3$, and log\_cloud top pressure. This is the same as the R3 sequence case with CO + SO$_2$ removed.} \\
B2   & Combined   & $0.01797 \ \mu$m & POS   & Free   & 'gradient'  & 'deck'       & Baseline \\
B3   & NIRISS     & $0.01797 \ \mu$m & POS   & Free   & 'isotherm'  & 'deck'       & Baseline \\
B4   & NIRSpec    & $0.01797 \ \mu$m & POS   & Free   & 'isotherm'  & 'deck'       & Baseline \\
B5   & Combined   & $R=100$          & POS   & Free   & 'isotherm'  & 'deck'       & Baseline \\
B6   & Combined   & $0.01797 \ \mu$m & TRx   & Free   & 'isotherm'  & 'deck'       & Baseline \\
B7   & Combined   & $0.01797 \ \mu$m & POS   & Free   & 'isotherm'  & 'deck-haze'  & Baseline+haze\footnote{Free parameters: log\_vmr of H$_2$O, CO$_2$, NH$_3$, log\_a, gamma, and log\_cloud top pressure} \\
& & & & & & &  \\
R2   & Combined   & $0.01797 \ \mu$m & POS   & Free   & 'isotherm'  & 'deck'       & Ref2\footnote{Free parameters: log\_vmr of H$_2$O, CO$_2$, NH$_3$, Na, K, and log\_cloud top pressure} \\
R2a  & Combined   & $0.01797 \ \mu$m & POS   & Free   & 'isotherm'  & 'deck'       & Ref2 - Na \\
R2b  & Combined   & $0.01797 \ \mu$m & POS   & Free   & 'isotherm'  & 'deck'       & Ref2 - K \\
& & & & & & &  \\
R3   & Combined   & $0.01797 \ \mu$m & POS   & Free   & 'isotherm'  & 'deck'       & Ref3\footnote{Free parameters: log\_vmr of H$_2$O, CO$_2$, NH$_3$, SO$_2$, CO, and log\_cloud top pressure} \\
& & & & & & &  \\
E1   & Combined   & $0.01797 \ \mu$m & POS   & Eq     & 'isotherm'  & 'deck'       & POS Eq Chem \\
E2   & Combined   & $0.01797 \ \mu$m & PLA   & Eq     & 'isotherm'  & 'deck'       & PLA Eq Chem \\
E3   & NIRISS     & $0.01797 \ \mu$m & PLA   & Eq     & 'isotherm'  & 'deck'       & PLA Eq Chem \\
E4   & NIRSpec    & $0.01797 \ \mu$m & PLA   & Eq     & 'isotherm'  & 'deck'       & PLA Eq Chem \\
E5   & Combined   & $R=100$          & PLA   & Eq     & 'isotherm'  & 'deck'       & PLA Eq Chem \\
E6   & Combined   & $0.01797 \ \mu$m & PLA   & Eq     & 'isotherm'  & 'deck'       & PLA Eq Chem\footnote{{\ron Same as E2 except metallicity ($\log(Z)$) is fixed to the B1 retrieval value.}} \\

\hline
\label{tab:retrieval_cases}
\end{tabular}
\end{center}
\end{table*}

\subsection{Settings and Assumptions for Atmospheric Modeling and Retrievals}
\label{poseidon}
{\ron For our retrievals, we configured \texttt{POSEIDON} to model clear and cloudy 1-D atmospheres (i.e., variation in only the radial direction). While \texttt{POSEIDON} has the ability to model 2-D and 3-D atmospheres, we restricted ourselves to the 1-D case.} \texttt{POSEIDON} models an atmosphere using a grid of pressure and temperature. We used a $100$-layer grid, spaced uniformly in $\log_{10}$ pressure space, with a maximum pressure of $100$ bar and a minimum pressure of $10^{-9}$ bar for free chemistry {\ron and $10^{-7}$ bar for equilibrium chemistry.} We used a relatively low minimum pressure to avoid clipping strong line cores and distorting our results.\footnote{See \texttt{POSEIDON} High-Resolution Spectroscopy at \href{https://poseidon-retrievals.readthedocs.io/en/latest/content/notebooks/high_res.html}{https://poseidon-retrievals.readthedocs.io}}

{\ron \texttt{POSEIDON} (like \texttt{TauREx} and \texttt{PLATON}, which we present in Appendices \ref{ap:free_retrievals} and \ref{ap:eq_retrievals}, respectively)} incorporates Rayleigh scattering and absorption from pair processes, collision-induced absorption (CIA), and free-free absorption consistent with the fill gases (i.e., the dominant atmospheric constituents) present and other molecules specified in each model atmosphere considered (including for the complex equilibrium chemistry situation). 

{\ron For cloudy \texttt{POSEIDON} retrievals,} we used the simple ``MacMad17'' 1-D cloud model \citep{macdonaldandmadhusudhan2017} that is characterized by a fixed cloud `type' (e.g., opaque cloud deck) and a free cloud top pressure parameter (log\_P\_cloud). For the comparative free chemistry retrievals with \texttt{TauREx} and equilibrium chemistry retrievals with \texttt{PLATON}, we used similar cloud models as discussed in Appendices~\ref{ap:free_retrievals} and \ref{ap:eq_retrievals}, respectively. {\ron We tested the effects of our cloud assumptions by also exploring a} deck haze cloud model with \texttt{POSEIDON}, which we present in Section~\ref{deck_haze_retrieval}. 

{\ron Past work} \citep[e.g.,]{greeneetal2016,guzmanmesaetal2020,howeetal2017} {\ron has shown that} an isothermal profile is a reasonable assumption when analyzing transmission spectroscopy. {\ron Accordingly, our retrievals assumed an} isothermal pressure-temperature (P-T) profile. However, in some cases even small shifts in temperature in the pressure range probed by transmission spectroscopy can have a significant effect on retrieved abundances \citep{macdonaldandmadhusudhan2017}. {\ron In order to verify that our retrieved abundances are robust to small variations in temperature, we break our assumption of an isothermal atmosphere and explore a two-parameter ``gradient'' P-T profile in Section~\ref{Non-isothermal}.}

In the {\ron transmission spectrum} fitting process, \texttt{POSEIDON} uses \texttt{MultiNest} \citep{ferozetal2009,ferozetal2013} to sample the Bayesian evidence, and determine posterior distributions. \texttt{MultiNest} is used through its Python implementation \texttt{PyMultiNest} \citep{buchner2016}. As suggested by the \texttt{POSEIDON} documentation, we used $400$ live points (which determines how finely the parameter space will be sampled) for our exploratory retrievals, and for our final published work we used $2000$ live points.

\texttt{POSEIDON} generates output parameters to assess the quality of the model fits, including $\chi^2$, reduced $\chi^2$, and the natural log of the Bayesian evidence (ln ($Z$)). We will discuss how the Bayesian evidence can be used in Bayesian model comparisons in Section~\ref{Bayesian_Model_Comparison}.

\begin{table*}
\begin{center}
\caption{Model Parameters and Prior Bounds Used in the Atmospheric Retrievals of WASP-166b}

\begin{tabular}{lcc}
\hline \hline
Parameter & Symbol\footnote{Parameter names used in corresponding model packages.} & Prior \\
\hline
\textit{Free Chemistry (\texttt{POSEIDON)}} \\
\hspace{3mm} Planet radius\footnote{The underlying package expresses this parameter in meters.\label{m_note}} at $1$ bar reference pressure (R$_\mathrm{J}$) & R\_p\_ref & \textit{U}($0.50$, $0.70$) \\
\hspace{3mm} Planet mass\footnote{The underlying package expresses this parameter in kilograms.\label{kg_note}} (M$_\mathrm{J}$) & M\_p & \textit{N}($0.101$, $0.005$) \\
\hspace{3mm} Temp (K); ``isotherm'' & T\_iso & \textit{U}($400$, $1800$) \\
\hspace{3mm} Temp, min pressure (K); ``gradient'' & T\_high & \textit{U}($400$, $1400$) \\
\hspace{3mm} Temp, max pressure (K); ``gradient'' & T\_deep & \textit{U}($400$, $1400$) \\
\hspace{3mm} $\log_{10}(vmr_X)$\footnote{$vmr_x$ is the volume mixing ratio of the constituent species, x.} & log\_X & \textit{U}($-10$, $-0.5$) \\
\hspace{3mm} $\log_{10}(P_\mathrm{cloud})$ ($\log_{10}$(bar)) & log\_P\_cloud & \textit{U}($-9$, $2$) \\
\hspace{3mm} Dataset offset (ppm) & delta\_rel & \textit{U}($-250$, $250$) \\
\tableline

\textit{Free Chemistry (\texttt{TauREx)}} \\
\hspace{3mm} Planet radius at $1$ bar reference pressure (R$_\mathrm{J}$) & R\_p\_ref & \textit{U}($0.50$, $0.70$) \\
\hspace{3mm} Planet mass (M$_\mathrm{J}$) & M\_p & \textit{N}($0.101$, $0.005$) \\
\hspace{3mm} Temp (K); ``isotherm'' & T\_iso & \textit{U}($400$, $1800$) \\
\hspace{3mm} $\log_{10}(vmr_X)$ & log\_X & \textit{U}($-7$, $-1$) \\
\hspace{3mm} $\log_{10}(P_\mathrm{cloud})$\footnote{The underlying package expresses this parameter in Pascals.\label{pa_note}} ($\log_{10}$(bar)) & $\log_{10}(\mathrm{clouds\_pressure})$ & \textit{U}($-6$, $0$) \\
\tableline

\textit{Equilibrium Chemistry (\texttt{POSEIDON)}} \\
\hspace{3mm} Planet radius\footnoteref{m_note} at $1$ bar reference pressure (R$_\mathrm{J}$) & R\_p\_ref & \textit{U}($0.50$, $0.70$) \\
\hspace{3mm} Planet mass\footnoteref{kg_note} (M$_\mathrm{J}$) & M\_p & \textit{N}($0.101$, $0.005$) \\
\hspace{3mm} Temp (K); `isotherm' & T\_iso & \textit{U}($400$, $1800$) \\
\hspace{3mm} Carbon/Oxygen ratio & C\_to\_O & \textit{U}($0.2$, $1.2$) \\
\hspace{3mm} $\log_{10}(\mathrm{metallicity})$\footnote{Metallicity measured relative to Solar.} & log\_Met & \textit{U}($-1$, $3$) \\
\hspace{3mm} $\log_{10}(P_\mathrm{cloud})$ ($\log_{10}$(bar)) & log\_P\_cloud & \textit{U}($-7$, $2$) \\
\hspace{3mm} Dataset offset (ppm) & delta\_rel & \textit{U}($-250$, $250$) \\
\tableline

\textit{Equilibrium Chemistry (\texttt{PLATON)}} \\
\hspace{3mm} Stellar radius\footnoteref{m_note} (R$_*$) & Rs & \textit{N}($1.22$, $0.06$) \\
\hspace{3mm} Planet radius\footnoteref{m_note} at $1$ bar reference pressure (R$_\mathrm{J}$) & Rp & \textit{U}($0.50$, $0.70$) \\
\hspace{3mm} Planet mass\footnoteref{kg_note} (M$_\mathrm{J}$) & Mp & \textit{N}($0.101$, $0.005$) \\
\hspace{3mm} Temp (K) & T & \textit{U}($600$, $1600$) \\
\hspace{3mm} Carbon/Oxygen ratio & CO\_ratio & \textit{U}($0.05$, $2$) \\
\hspace{3mm} $\log_{10}(\mathrm{metallicity})$ & logZ & \textit{U}($0.5$, $3$) \\
\hspace{3mm} $\log_{10}(P_\mathrm{cloud})$\footnoteref{pa_note} ($\log_{10}$(bar)) & log\_cloudtop\_P & \textit{U}($-5$, $3$) \\
\hspace{3mm} $\log_{10}(\mathrm{scattering})$ & log\_scatt\_factor & \textit{U}($0$, $4$) \\
\hspace{3mm} Error multiple & error\_multiple & \textit{U}($0.5$, $4$) \\
\hspace{3mm} Dataset offset (ppm) & wfc3\_offset\_transit\footnote{To offset our NIRISS SOSS data correctly, the WFC3 wavelength limits were overwritten to the NIRISS SOSS wavelength range for this parameter.} & \textit{U}($-60$, $60$) \\
\tableline

\label{tab:retrieval_priors}
\end{tabular}
\end{center}
Note: \textit{U}(a,b) is the uniform distribution between values a and b (inclusive), and \textit{N}($\mu$, $\sigma$) is the normal distribution with mean $\mu$ and standard deviation $\sigma$. $X$ are the volume mixing ratios of the various molecular species being considered (e.g., H$_2$O). The \texttt{POSEIDON}, \texttt{PLATON}, and \texttt{TauREx} reference pressure is $1$ bar (i.e., $100$ kPa). $T_\mathrm{eq}$ is the planetary equilibrium temperature ($1270$ K){\ron{, assuming zero albedo and efficient heat redistribution}}. $R_p$ is the broadband planetary radius (0.6155 R$_\mathrm{J}$, from Table~\ref{tab:system_params}).
\end{table*}

\subsection{Free Chemistry}
\label{free_retrievals}
{\ron To determine an initial starting point for more rigorous analysis of the atmospheric composition of WASP-166b, we began by running a series of \texttt{POSEIDON} free chemistry retrievals testing various scenarios.}
For these {\ron exploratory} retrievals, we fixed system parameters ($R_*$, $\log g_*$, [$Fe$/$H$]$_*$, $T_\mathrm{eff}$, and $T_\mathrm{eq}$) to those shown in Table~\ref{tab:system_params} and fit the free atmospheric model parameters presented in Table~\ref{tab:retrieval_priors}. We also used forward modeling to examine the essential shape and character of spectral features resulting from various trace gases, and cloud models. By considering spectra for these gases individually, we could identify candidate molecules to include (or exclude) in further analysis. This was more of a qualitative exercise than a detailed analytical comparison of a grid of models to the observational data.

{\ron Based on these studies we constructed a ``Reference'' model assuming an isothermal atmosphere and considering a} broad set of trace molecular constituents (H$_2$O, CO$_2$, CO, NH$_3$, SO$_2$, CH$_4$, H$_2$S, HCN, and C$_2$H$_2$) as well as a 1-D, opaque, cloud deck (\texttt{POSEIDON}:`MacMad17'). The remainder of the atmosphere was composed of the fill gases He and H$_2$ in a primordial solar ratio ($X_\mathrm{He}/X_\mathrm{H_2}$ = 0.17), an appropriate assumption for this class of planet (i.e., planets ranging from Neptune to Jupiter in mass and radius, with equilibrium temperature less than $2000$ K; \citealt{DAngeloetal2010,Heng2017,DAngeloetal2018,macdonaldlewis2022}). {\ron Next, using this model, we ran our primary Reference retrieval case (denoted as ``R1;'' see Table~\ref{tab:retrieval_cases}) using the fixed $0.01797 \ \mu$m bin instrument data.}

\begin{figure*}[p] 
    \centering
    \includegraphics[width=\textwidth]{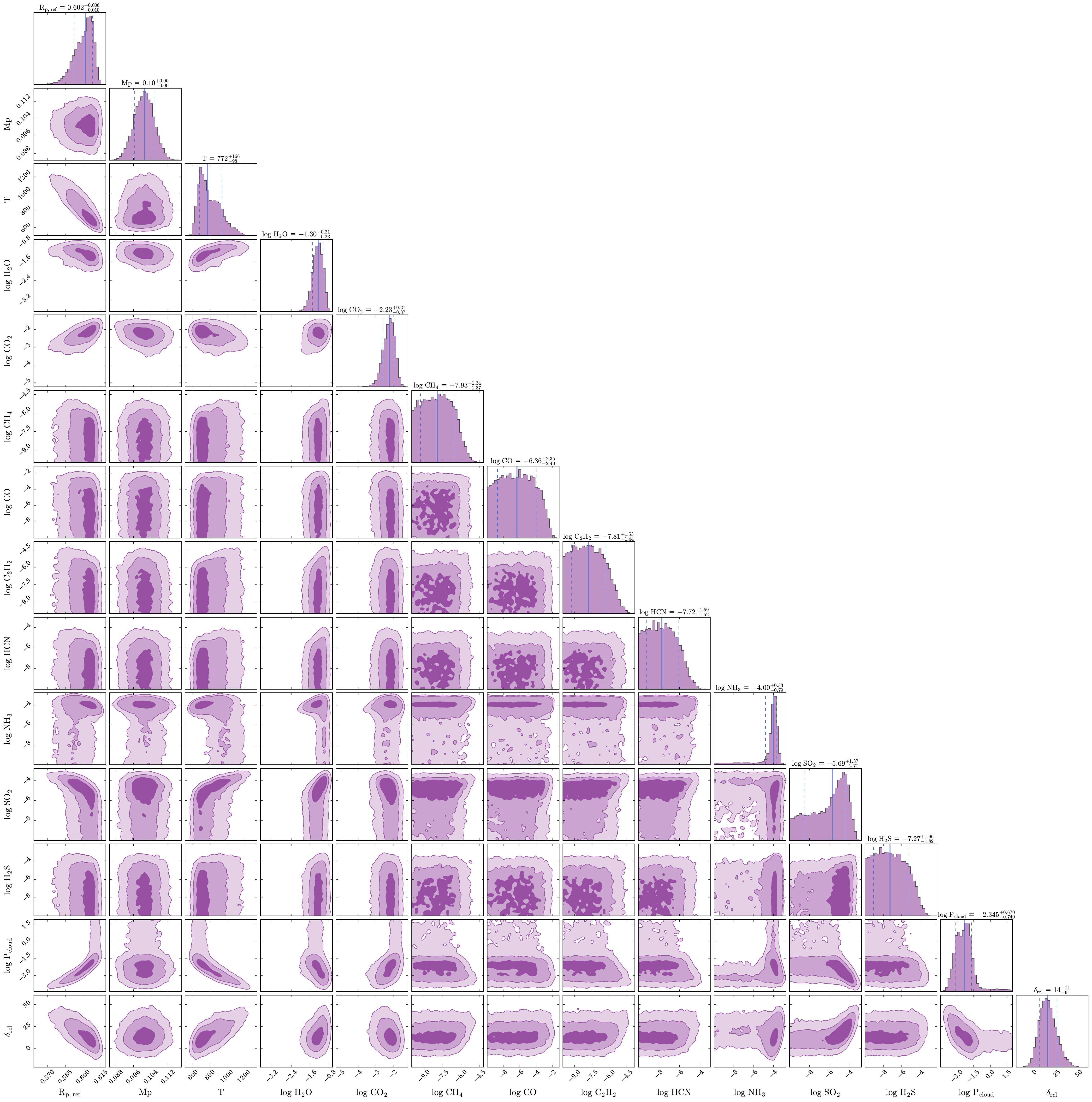}
    \caption{{\ron{Posterior distributions of free parameters (corner plot) for \texttt{POSEIDON} Reference retrieval case, R1.}}}
    \label{fig:POS_v62_R1_corner}
\end{figure*} 

\begin{table*}[tb]
\caption{Free Parameter Posteriors from \texttt{POSEIDON} Reference Retrieval {\ron{(R1)}}}
\begin{center}

\begin{tabular}{llcc}
\hline \hline
Parameter & \texttt{POSEIDON} Symbol & Units & Value \\
\hline
Planet reference radius\footnote{At reference pressure of $1$ bar.} & R\_p\_ref & R$_\mathrm{J}$ & $0.60^{+0.01}_{-0.01}$ \\
Planet mass & M\_p & M$_\mathrm{J}$ & $0.101^{+0.005}_{-0.005}$ \\
Planet isothermal temperature & T & K & $772^{+166}_{-98}$ \\
$X_{\mathrm{H_2O}}$\footnote{All abundance parameters ($X$) are $\log_{10}$ volume mixing ratios.} & log\_H2O & - & $-1.30^{+0.21}_{-0.23}$ \\
$X_{\mathrm{CO_2}}$ & log\_CO2 & - & $-2.23^{+0.31}_{-0.37}$ \\
$X_{\mathrm{CH_4}}$ & log\_CH4 & - & $-7.93^{+1.34}_{-1.37}$ \\
$X_{\mathrm{CO}}$ & log\_CO & - & $-6.36^{+2.35}_{-2.40}$ \\
$X_{\mathrm{C_2H_2}}$ & log\_C2H2 & - & $-7.81^{+1.53}_{-1.44}$ \\
$X_{\mathrm{HCN}}$ & log\_HCN & - & $-7.72^{+1.59}_{-1.52}$ \\
$X_{\mathrm{NH_3}}$ & log\_NH3 & - & $-4.00^{+0.33}_{-0.79}$ \\
$X_{\mathrm{SO_2}}$ & log\_SO2 & - & $-5.69^{+1.37}_{-2.77}$ \\
$X_{\mathrm{H_2S}}$ & log\_H2S & - & $-7.27^{+1.96}_{-1.82}$ \\
$\log_{10}(P_\mathrm{cloud})$ & log\_P\_{clouds} & $\log_{10}$(bar) & $-2.34^{+0.67}_{-0.74}$ \\
Instrumental offset & $\delta_\mathrm{rel}$ & ppm & $14.4^{+10.7}_{-8.9}$ \\
\hline
\label{tab:reference_retrieval_posteriors}
\end{tabular}
\end{center}
\end{table*}

{\ron The median abundances and $1\sigma$ bounds for the atmospheric constituents of the R1 retrieval are shown in Table~\ref{tab:reference_retrieval_posteriors}. It is evident that H$_2$O and CO$_2$ show very high abundance ($> 4500$ ppm) compared to the other constituents ($< 100$ ppm). The full posterior distributions of all free parameters are shown in the corner plot, Figure~\ref{fig:POS_v62_R1_corner}. This emphasizes the well constrained distributions for H$_2$O, CO$_2$, NH$_3$, and cloud top pressure compared to the very poorly constrained distributions for the other constituents.}

In Section~\ref{Spectral_Decomp}, we use spectral decomposition to assess and display the spectral feature strength of each {\ron constituent} as a function of wavelength. Guided by those results, in Section~\ref{Bayesian_Model_Comparison} we describe a Bayesian model comparison procedure to determine the detection significance of each atmospheric component. {\ron In Section~\ref{baseline_free_retrieval}, we then present our Baseline retrieval (B1) using only the atmospheric constituents for which Bayesian model comparison suggests at least weak support (see Table~\ref{tab:translation})}.

\subsubsection{\texttt{POSEIDON} Spectral Decomposition of Reference Case (R1)} \label{Spectral_Decomp}

\begin{figure*}
    \centering
    \includegraphics[width=\textwidth]{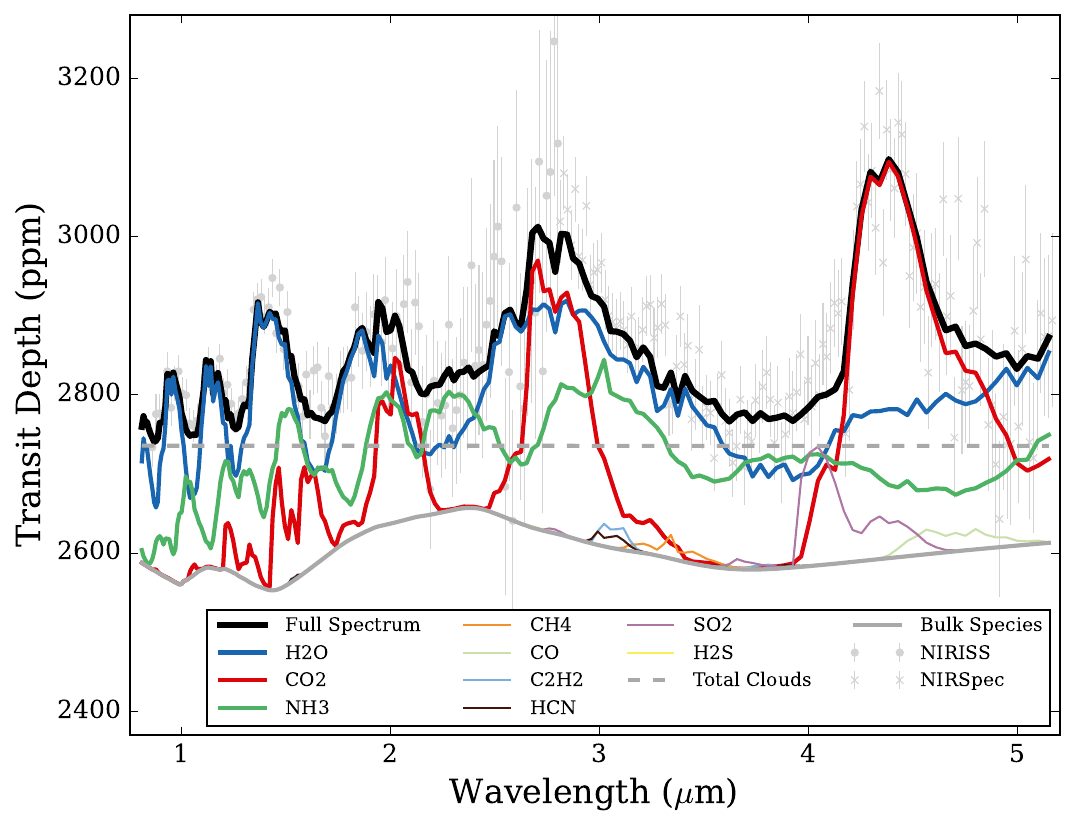} 
    \caption{The spectral decomposition of the Reference retrieval (R1) of WASP-166b with the absorption contributions of the key atmospheric constituents shown. {\ron Our reduced NIRISS and NIRSpec observational data (fixed $0.01797 \ \mu$m bins) are plotted as grey circles and ``x'' markers, respectively. Post-retrieval, we have subtracted the median transit depth offset value (14.4 ppm) for the plotted NIRISS SOSS data (the NIRSpec data is the relative zero anchor).} Spectra are plotted at $R = 100$ for clarity. The combined forward model, shown as a heavy black line, is the median Reference retrieval {\ron{(R1)}} result (see Table~\ref{tab:reference_retrieval_posteriors}). The colored lines correspond to the individual opacity/absorption contributions (including Rayleigh scattering and collision absorption) of the various constituent molecules and clouds at specific wavelengths. (H$_2$S is plotted but is not distinguishable from the bulk species line.) The transmission spectrum is dominated by contributions from H$_2$O ($15.2\sigma$), and CO$_2$ ($14.7\sigma$). The more subtle effects of NH$_3$ ($2.3\sigma$) may be apparent at $\sim 2.3 \ \mu$m, and possibly at $\sim 3.9 \ \mu$m.
    The effects of the cloud deck ({\ron{dashed grey}} line at a transit depth of {\ron $\sim 2740$ ppm}) are apparent in clipping the troughs of the water features from $0.85$ to $1.7 \ \mu$m.}
    \label{fig:POS_Spectral_Decomp_v42}
\end{figure*}

In order to gain insight into how the various absorption features combined to produce the observed spectrum, we took advantage of the Spectral Decomposition function within \texttt{POSEIDON} v1.2.1 {\ron \citep{mullensetal2024}}. Based on the median values of the retrieved posteriors for the Reference (R1) case (see Table~\ref{tab:reference_retrieval_posteriors}), we generated a plot (Figure~\ref{fig:POS_Spectral_Decomp_v42}) that shows the combined forward model as well as the forward models of the individual spectral contributions of the various absorbers (fill and trace gases, and the effects of clouds) included in the {\ron{primary}} Reference spectrum {\ron{(R1)}}.

It is clear from Figure~\ref{fig:POS_Spectral_Decomp_v42} that H$_2$O and CO$_2$ dominate the spectrum. The water features are the main drivers in the $0.85$ to $3.0 \ \mu$m range; however, the effects of the cloud deck are apparent in clipping the lower extent (transit depth) of the water features from $0.85$ to $1.8 \ \mu$m. CO$_2$ combined with H$_2$O drive the peak at $\sim 2.8 \ \mu$m, while CO$_2$ is responsible for the very strong feature at $\sim 4.4 \ \mu$m. The effects of NH$_3$ are more subtle, but it appears to have the most significant impact on the shape of the combined spectrum at $\sim 2.3 \ \mu$m, and possibly at $\sim 3.9 \ \mu$m.

{\ron We noticed hints of a small absorption feature around $4.1 \ \mu$m (see the primary Reference case (R1) fit to the data in Figure~\ref{fig:POS_Spectral_Decomp_v42}). We were aware that a more prominent (likely SO$_2$) feature had been observed for WASP-39b \citep{tsaietal2023}, and that we might be observing a similar, but more subtle effect. The presence of SO$_2$ could potentially be shaping our combined spectrum at this wavelength, although the spectrum is heavily dominated by CO$_2$ and H$_2$O in this region. The primary Reference case (R1) shows an abundance for SO$_2$ of $-5.69^{+1.96}_{-2.77}$ $\log_{10}$(vmr) (very small and poorly constrained; see Figure~\ref{fig:POS_v62_R1_corner} and Table~\ref{tab:reference_retrieval_posteriors}), with a detection significance of $1.5\sigma$ (inconclusive) as shown in Table~\ref{tab:BMC}. We explore this issue further in Section~\ref{CO_SO2_retrievals}.}

\begin{table*}
\begin{center}
\caption{Results of Bayesian Model Comparisons for Observations of WASP-166b}
\begin{tabular}{llccl}
\hline \hline
ID\footnote{See Table~\ref{tab:retrieval_cases} for retrieval case definitions. Results are based on \texttt{POSEIDON} retrievals using Free chemistry with a \texttt{Multinest} sampling parameter of 2000 live points. The bulk gases, H$_2$ and He, are present in all cases at an assumed primordial ratio of $X_{He}/X_{H_2} = 0.17$. The remainder of the atmosphere not made up of trace gases is filled by the bulk gases. Four additional free parameters: $R_\mathrm{p,ref}$, $M_p$, $T$, and the data offset between instruments ($\delta_\mathrm{rel}$), are included in all cases. We are using an isothermal P-T profile, and a simple, 1-D, opaque cloud-deck model (`MacMad17'). We are also using data reduced with constant ($0.01797 \ \mu$m) wavelength bins.} & Retrieval Model\footnote{Summary constituent configuration. {\ron Except for R1, R2, and R3, each model corresponds to an individual molecule or feature that was removed from the Reference space.}} & Evidence\footnote{Bayesian evidence with constituent removed ($Z_i$); Bayesian evidence of Ref. case ($Z_0$)} & Best fit & Bayes Factor\footnote{Bayes factor ($B_i$) and associated detection significance ($\sigma$). Detection is considered inconclusive for $B_i < 2.9$. For situations where $B_i < 1$, we follow the documentation for \texttt{POSEIDON} \citep{macdonald2023} and categorize it as a non-detection condition (ND).} \\
&  & ($\ln{Z_i}$) & ($\chi^2$) & ($B_i = Z_0/Z_i$) \\
\hline
R1   & Reference (1)\footnote{Free parameters: volume mixing ratios of H$_2$O, CO$_2$, NH$_3$, SO$_2$, CO, C$_2$H$_2$, H$_2$S, CH$_4$, HCN, and cloud top pressure.}            
                           & $2019.89$ & $224$ & Ref (1) \\
R1a  & H$_2$O removed                                      
                           & $1907.56$ & $454$ & \textbf{\textit{B}}$\mathbf{_{H_2O} = 6.09e48}$ ($\mathbf{15.2 \sigma}$) \\
R1b  & CO$_2$ removed      & $1915.36$ & $431$ & \textbf{\textit{B}}$\mathbf{_{CO_2} = 2.49e45}$ ($\mathbf{14.7\sigma}$) \\
R1c  & NH$_3$ removed      & $2018.43$ & $230$ & \textbf{\textit{B}}$\mathbf{_{NH_3} = 4.30}$ ($\mathbf{2.3\sigma}$)     \\
R1d  & SO$_2$ removed      & $2019.60$ & $226$ & $B_{\mathrm{SO_2}} = 1.34$ ($1.5\sigma$)                       \\
R1e  & CO removed          & $2019.85$ & $225$ & $B_{\mathrm{CO}} = 1.04$ ($1.1\sigma$)                        \\
R1f  & C$_2$H$_2$ removed  & $2020.30$ & $224$ & $B_{\mathrm{C_2H_2}} = 0.66$ (ND)                                \\
R1g  & H$_2$S removed      & $2020.33$ & $224$ & $B_{\mathrm{H_2S}} = 0.64$ (ND)                                 \\
R1h  & CH$_4$ removed      & $2020.42$ & $225$ & $B_{\mathrm{CH_4}} = 0.59$ (ND)                                 \\
R1i  & HCN removed         & $2020.46$ & $225$ & $B_{\mathrm{HCN}} = 0.57$ (ND)                                 \\
R1j  & Clouds removed      & $2017.67$ & $231$ & \textbf{\textit{B}}$\mathbf{_{Clouds} = 9.21}$ ($\mathbf{2.6\sigma}$)           \\
& & & & \\
\hline
R2   & Reference (2)\footnote{Free parameters: volume mixing ratios of H$_2$O, CO$_2$, NH$_3$, Na, K, and cloud top pressure.}                       
                           & $2021.83$ & $225$ & Ref (2)                                               \\
R2a  & Na removed          & $2021.30$ & $226$ & $B_{\mathrm{Na}} = 1.70$ ($1.7\sigma$)                        \\
R2b  & K removed           & $2021.93$ & $225$ & $B_{\mathrm{K}} = 0.90$ (ND)                                   \\
& & & & \\
\hline
R3   & Reference (3)\footnote{Free parameters: volume mixing ratios of H$_2$O, CO$_2$, NH$_3$, SO$_2$, CO, and cloud top pressure.}                      
                           & $2021.73$ & $224$ & Ref (3)                                               \\
B1   & CO + SO$_2$ removed    & $2021.39$ & $225$ & $B_{\mathrm{CO + SO_2}} = 1.40$ ($1.5\sigma$)                  \\
\tableline

\label{tab:BMC}
\end{tabular}
\end{center}
\end{table*}

\subsubsection{Bayesian Model Comparison}\label{Bayesian_Model_Comparison}

{\ron In order to determine the detection significance of possible atmospheric constituents of WASP-166b, we performed a Bayesian model comparison \citep{BennekeandSeager2013} using free chemistry with the combined instruments. 

We started with the primary Reference retrieval case (R1) that included an extensive set of potential constituent trace molecules (H$_2$O, CO$_2$, CO, NH$_3$, SO$_2$, CH$_4$, H$_2$S, HCN, and C$_2$H$_2$), and a 1-D, opaque, cloud-deck model (\texttt{POSEIDON}: ‘MacMad17’). We assumed an isothermal atmosphere for this retrieval. See Table~\ref{tab:retrieval_priors} for a listing of the free parameter priors for this case. The results of this retrieval included, among other things, the Bayesian evidence.

We then removed one component/molecule (e.g., H$_2$O) from the list of constituents and ran a new retrieval (i.e., R1a) with otherwise the same settings as the original R1 case. Now, given the Bayesian evidence for this case and for the R1 case, we could determine the detection significance of the removed constituent.

As previously mentioned, a by-product of the \texttt{POSEIDON}/\texttt{MultiNest}, nested sampling process is the natural log of the Bayesian evidence ($Z$). This parameter is similar to other parameters (e.g., chi-squared) that help gauge the goodness of fit between the observational data and our retrieval model. From the ln $Z_0$ of the Reference case (R1) and the ln $Z_i$ of the cases with one component removed (e.g., R1a), one can compute the Bayes factor ($B_i$ = exp(ln $Z_0$ - ln $Z_i$)). From the Bayes factor one can then compute the detection significance for the removed component, which can be put on a relative interpretation scale, in equivalent $\sigma$ (\citealp[see Table~\ref{tab:translation} adopted from][]{gordonandtrotta2007}). The \texttt{POSEIDON} package has the appropriate functions for making these transformations. 

This process was repeated for the entire sequence of retrievals (R1a through R1j), removing one constituent at a time from the full Reference (R1) constituent list. The detection significance of all of the atmospheric constituents (molecules and cloud model) are summarized in Table~\ref{tab:BMC}.}

\begin{table*}
\begin{center}
\caption{Translation from Bayes Factor to Detection Significance (sigma)}
\begin{tabular}{cccl}
\hline \hline
$B$     &  $ln \:B$   &  sigma ($\sigma$)  &  Category     \\
\hline
$2.5$ & $0.9$ & $2.0$ &                   \\
$2.9$ & $1.0$ & $2.1$ & 'weak' at best    \\
$8.0$ & $2.1$ & $2.6$ &                   \\
$12$ & $2.5$ & $2.7$ & 'moderate' at best \\
$21$ & $3.0$ & $3.0$ &                    \\
$53$ & $4.0$ & $3.3$ &                    \\
$150$ & $5.0$ & $3.6$ & 'strong' at best  \\
$43000$ & $11$ & $5.0$ &                  \\
\hline
\label{tab:translation}
\end{tabular}
\end{center}
Notes: Adopted from \citet{gordonandtrotta2007}. A Bayes factor ($B$) of 12, {\ron{for example}}, can be considered a “moderate" detection, corresponding to approximately $2.7\sigma$ significance. {\ron{A Bayes factor of $2.9$ (or $2.1\sigma$), is considered "weak" support at best.}} Detection is considered inconclusive for $ln \:B < 1$. For situations where $B < 1$, we follow the documentation for \texttt{POSEIDON} \citep{macdonald2023} and categorize it as a non-detection condition.

\end{table*}

We find that H$_2$O and CO$_2$ show robust detections ($15.2$ and $14.7\sigma$, respectively), while NH$_3$ shows weak support ($2.3\sigma$). We also show weak support for an opaque cloud deck ($2.6\sigma$) {\ron{at an intermediate pressure}}. All other constituents {\ron show detection as} inconclusive ($< 2.1\sigma$), or {\ron show no detection at all} ($B_i < 1$).

\subsubsection{Baseline Free Chemistry Retrieval}\label{baseline_free_retrieval}

{\ron {The atmospheric constituents that showed weak support ($2.1\sigma$) or better in the Bayesian model comparison (of R1 to R1a-R1j models) were included in our ``Baseline'' case (B1).}} From {\ron{Table~\ref{tab:BMC}}}, these constituents are H$_2$O, CO$_2$, NH$_3$, and an opaque cloud deck. {\ron Like the R1 series of retrievals, the B1 retrieval used \texttt{POSEIDON} to fit an isothermal atmosphere with free chemistry to the combined NIRISS and NIRSpec dataset with fixed ($0.01797 \ \mu$m) bins.} 

{\ron As described in Appendix~\ref{ap:free_retrievals}, we verified that the 
results from our B1 analysis were robust by running additional retrievals (denoted B2 - B7). With these retrievals, we explored the effects of using a P-T gradient rather than an isothermal atmosphere (B2; see Section~\ref{Non-isothermal}), fitting each instrument separately (B3 for NIRISS and B4 for NIRSpec; see Section~\ref{pos_single_instrument}), using fixed spectral resolution rather than fixed wavelength bins (B5; see Section~\ref{pos_r100}), using a different analysis framework (\texttt{TauREX}; B6; see Section~\ref{TauREx_Retrieval}), and incorporating hazes (B7; see Section~\ref{deck_haze_retrieval}). With the exception of B6, we used \texttt{POSEIDON} for all of these retrievals. For additional details about the B2-B7 retrievals and a thorough comparison of their results to the Baseline (B1) results, see Appendix~\ref{ap:free_retrievals}.}

\begin{figure*}
    \centering
    \includegraphics[width=\textwidth]{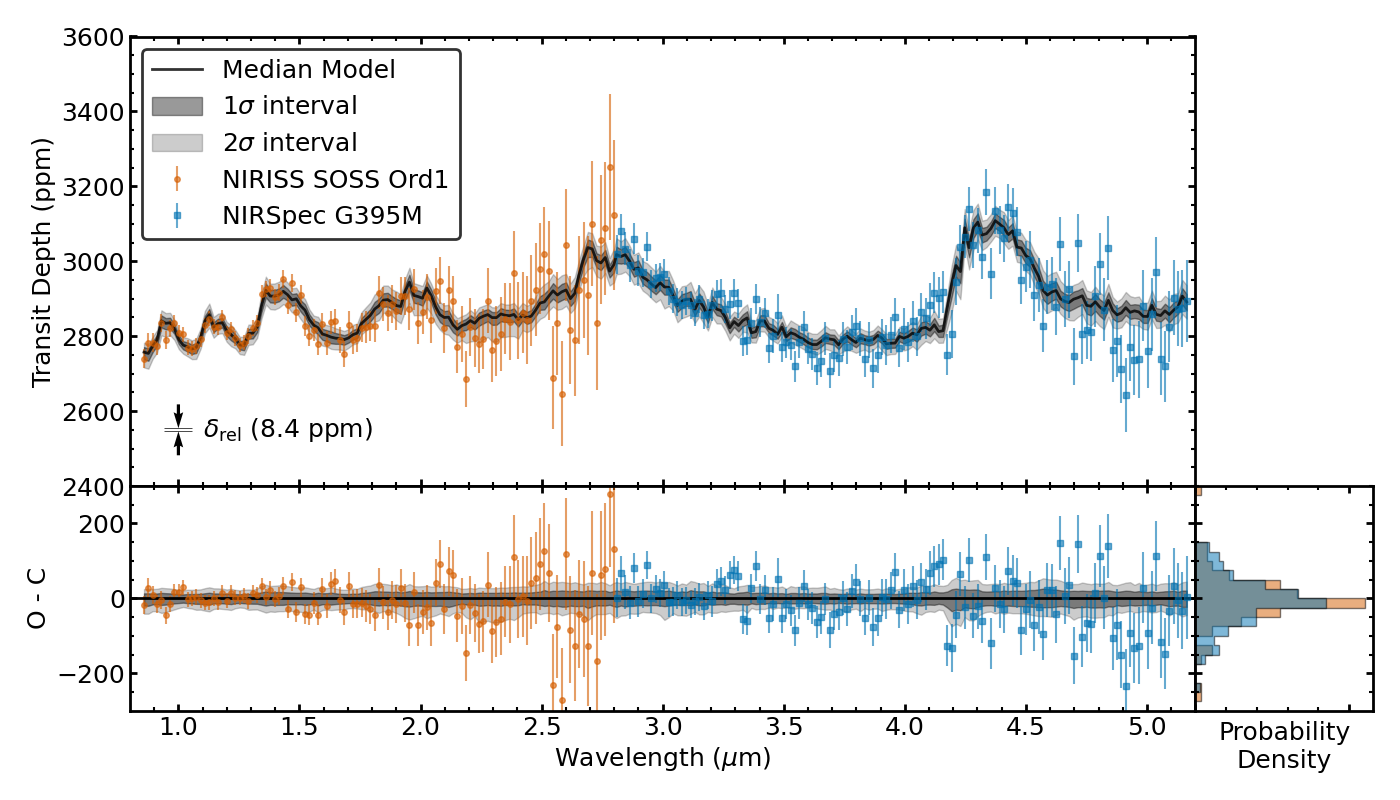}
    \caption{WASP-166b transmission spectrum from \texttt{POSEIDON} Baseline free chemistry retrieval {\ron (B1). \textbf{Top:} observational data shown for NIRISS SOSS (orange circles) and NIRSpec G395M (blue squares)}, for the case of fixed ($0.01797 \ \mu$m) bins. The retrieval is using an isothermal P-T profile with an opaque cloud deck. Fill gases (H$_2$ and He) are present at the primordial solar ratio. The trace gases H$_2$O, CO$_2$, and NH$_3$ are included in the model. \ron{The plotted NIRISS data have been adjusted down (post-retrieval) by a small retrieved instrument offset (see plotted $\delta_\mathrm{rel}$ value). The median retrieved spectrum for this case is shown as a fine black line with the $1\sigma$ and $2\sigma$ confidence intervals shown in dark and light grey shading respectively. \textbf{Bottom:} residuals between observational data and retrieval model from Top panel plot. \textbf{Bottom Right:} histogram of residuals for each instrument wavelength range, with orange for NIRISS data and blue for NIRSpec data.}}
    \label{fig:POS_Baseline_Free_Chem_Retrieval_v71}
\end{figure*}

\begin{figure*}
    \centering
    \includegraphics[width=\textwidth]{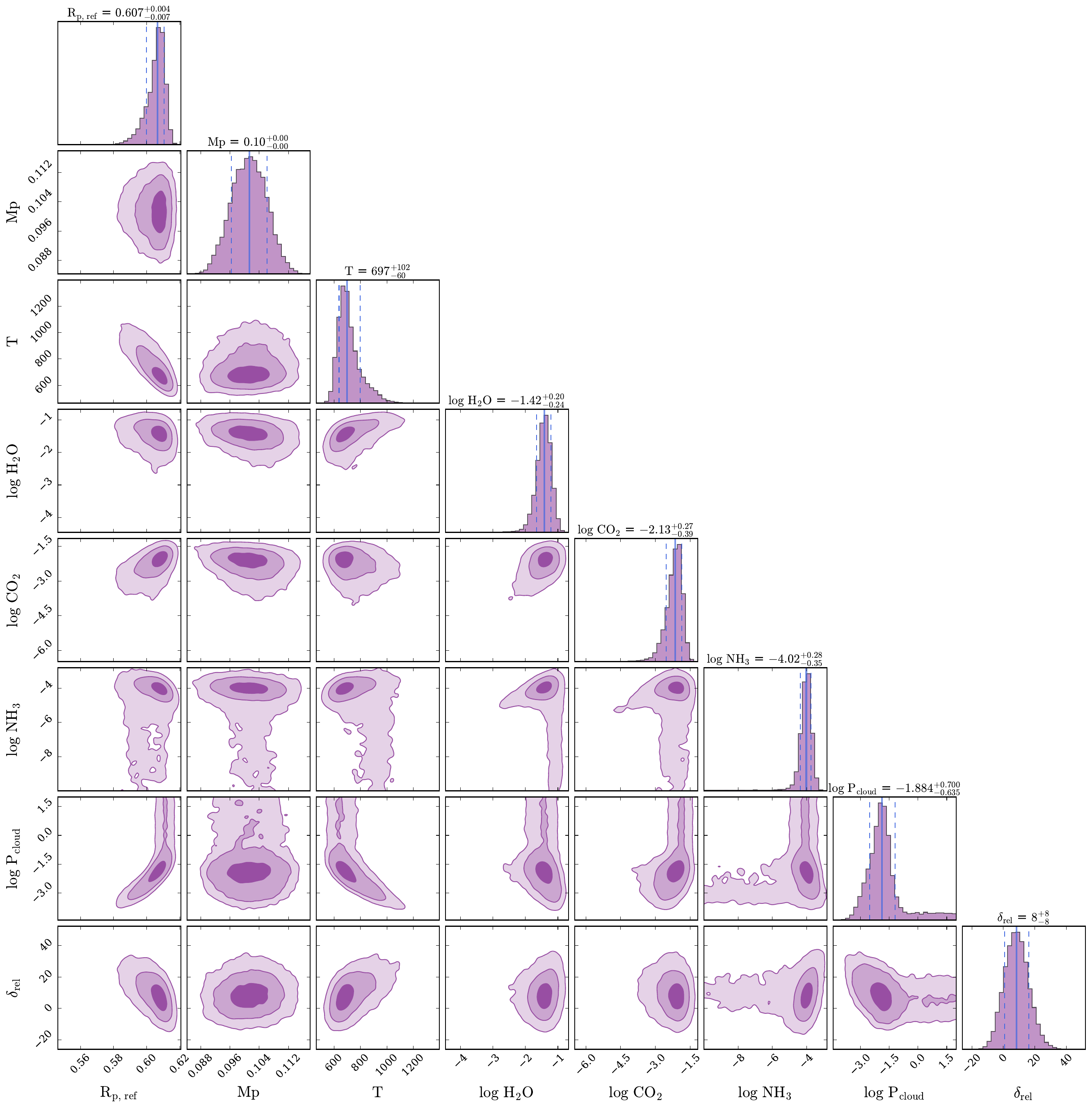}
    \caption{{\ron{Posterior distributions of free parameters (corner plot) for \texttt{POSEIDON} Baseline retrieval case, B1.}}}
    \label{fig:POS_v71_B1_corner}
\end{figure*}

{\ron For B1,} we present the retrieved spectrum {\ron{and reduced observational data}} in Figure~\ref{fig:POS_Baseline_Free_Chem_Retrieval_v71}. The free parameter posteriors for B1 are compiled in Table~\ref{tab:free_chem_posteriors} {\ron{and shown in corner plot form in Figure~\ref{fig:POS_v71_B1_corner}}}. {\ron Comparing the posteriors for the R1 (Figure~\ref{fig:POS_v62_R1_corner}) and B1 (Figure~\ref{fig:POS_Baseline_Free_Chem_Retrieval_v71}) retrievals, there} are small differences in the retrieved posterior {\ron abundances of} H$_2$O, CO$_2$, and NH$_3$ and the cloud deck pressure. {\ron In the R1 retrieval}, the very small, poorly constrained abundances shown for some of the constituents with $< 2.1\sigma$ detection significance have a non-zero (but small) effect on the overall model fit.

\begin{table*}
\footnotesize
\begin{center}
\caption{Free Parameter Posteriors from Free Chemistry Retrievals}

\begin{tabular}{lcccccccccc}
\hline \hline
ID\footnote{See Table~\ref{tab:retrieval_cases} for retrieval case definitions.} & $R_\mathrm{p,ref}$\footnote{We retrieve $R_\mathrm{p,ref}$, the planet radius at the reference pressure of $1$ bar, for \texttt{POSEIDON}, and \texttt{TauREx}.} & $M_p$ & $T_\mathrm{iso}$\footnote{We retrieve the isothermal temperature ($T_\mathrm{iso}$) in all cases but one (B2, the 'gradient' P-T case).} & & & $X_\mathrm{H_2O}$\footnote{This and all other abundance parameters ($X$) are $\log_{10}$ volume mixing ratios. The bulk gases, H$_2$ and He, are present in all cases at an assumed primordial solar ratio of $X_\mathrm{He}/X_\mathrm{H_2}$ = 0.17.} & $X_\mathrm{CO_2}$ & $X_\mathrm{NH_3}$ & $\log_{10}(P_\mathrm{cloud})$ & $\delta_\mathrm{rel}$\footnote{We retrieve the data offset ($\delta_\mathrm{rel}$) between instruments for the combined instrument cases. Since \texttt{TauREx} does not directly retrieve an instrument offset, the actual data has been adjusted in the TRx case, pre-retrieval (B6; all NIRISS transit depth values were reduced by $\sim 8.4$ ppm to account for the Baseline offset).}\\

& R$_\mathrm{J}$ & M$_\mathrm{J}$ & K & & & - & - & - & $\log_{10}$(bar) & ppm \\
\hline
\\[-3mm]
B1 & $0.6066^{+0.0039}_{-0.0067}$ & $0.1013^{+0.0049}_{-0.0049}$ & $697^{+101}_{-59}$ & - & - & $-1.42^{+0.20}_{-0.24}$ & $-2.13^{+0.27}_{-0.39}$ & $-4.02^{+0.28}_{-0.34}$ & $-1.88^{+0.71}_{-0.63}$ & $8.4^{+7.8}_{-7.6}$ \\[1.25mm]

& & & & & & & & & & \\
& & & & $T_\mathrm{high}$\footnote{For the `gradient’ P-T profile case (B2) we retrieve the temperatures $T_\mathrm{high}$ and $T_\mathrm{deep}$, associated with the pressures $10^{-5}$ bar and 10 bar, respectively.} & $T_\mathrm{deep}$ & & & & & \\
& & & & K & K & & & & & \\
\hline
\\[-3mm]
B2 & $0.603^{+0.010}_{-0.025}$ & $0.1012^{+0.0049}_{-0.0048}$ & - & $700^{+101}_{-91}$ & $780^{+820}_{-320}$ & $-1.48^{+0.29}_{-0.70}$ & $-2.28^{+0.39}_{-0.93}$ & $-4.09^{+0.36}_{-0.76}$ & $-2.09^{+1.74}_{-0.66}$ & $11.1^{+9.4}_{-8.8}$ \\[2mm]
B3 & $0.5980^{+0.0079}_{-0.0093}$ & $0.1021^{+0.0041}_{-0.0044}$ & $990^{+140}_{-130}$ & - & - & $-0.96^{+0.24}_{-0.43}$ & $-3.2^{+1.3}_{-4.6}$ & $-6.5^{+1.9}_{-2.1}$ & $-3.05^{+0.75}_{-0.53}$ & - \\[2mm]
B4 & $0.6034^{+0.0044}_{-0.0048}$ & $0.1015^{+0.0047}_{-0.0048}$ & $682^{+52}_{-50}$ & - & - & $-1.77^{+0.27}_{-0.45}$ & $-2.33^{+0.34}_{-0.46}$ & $-3.58^{+0.44}_{-0.49}$ & $-0.2^{+1.2}_{-1.3}$ & - \\[2mm]
B5 & $0.6097^{+0.0036}_{-0.0047}$ & $0.1012^{+0.0048}_{-0.0049}$ & $697^{+78}_{-56}$ & - & - & $-1.40^{+0.20}_{-0.22}$ & $-1.86^{+0.23}_{-0.28}$ & $-4.29^{+0.36}_{-0.74}$ & $-1.70^{+0.71}_{-0.50}$ & $2.8^{+7.6}_{-7.3}$ \\[2mm]
B6 & $0.6056^{+0.0035}_{-0.0049}$ & $0.1011^{+0.0047}_{-0.0048}$ & $708^{+76}_{-58}$ & - & - & $-1.37^{+0.17}_{-0.22}$ & $-2.09^{+0.27}_{-0.34}$ & $-4.11^{+0.26}_{-0.36}$ & $-1.99^{+0.50}_{-0.50}$ & - \\[1.25mm]

& & & & & & & & & & \\
& & & & $\log(a)$ & gamma & & & & & \\
\hline
\\[-3mm]
B7 & $0.6069^{+0.0036}_{-0.0052}$ & $0.1010^{+0.0047}_{-0.0045}$ & $693^{+85}_{-55}$ & $0.7^{+2.8}_{-2.9}$ & $-9.8^{+7.1}_{-7.0}$ & $-1.41^{+0.19}_{-0.22}$ & $-2.12^{+0.26}_{-0.33}$ & $-3.98^{+0.26}_{-0.32}$ & $-1.64^{+2.14}_{-0.70}$ & $7.2^{+8.2}_{-7.4}$ \\[1.25mm]

& & & & & & & & & & \\
& & & & $X_\mathrm{Na}$ & $X_\mathrm{K}$ & & & & & \\
\hline
\\[-3mm]
R2 & $0.6075^{+0.0032}_{-0.0058}$ & $0.1010^{+0.0048}_{-0.0048}$ & $694^{+84}_{-53}$ & $-5.3^{+3.2}_{-3.2}$ & $-7.0^{+1.9}_{-2.0}$ & $-1.44^{+0.20}_{-0.23}$ & $-2.13^{+0.27}_{-0.34}$ & $-4.00^{+0.26}_{-0.31}$ & $-1.62^{+2.20}_{-0.76}$ & $8.0^{+7.7}_{-7.2}$ \\[1.25mm]
R2a & $0.6068^{+0.0037}_{-0.0066}$ & $0.1011^{+0.0048}_{-0.0048}$ & $697^{+101}_{-57}$ & - & $-6.9^{+1.9}_{-2.1}$ & $-1.40^{+0.20}_{-0.22}$ & $-2.11^{+0.26}_{-0.38}$ & $-3.99^{+0.27}_{-0.34}$ & $-1.81^{+1.74}_{-0.68}$ & $8.4^{+8.0}_{-7.6}$ \\[1.25mm]
R2b & $0.6077^{+0.0032}_{-0.0061}$ & $0.1010^{+0.0048}_{-0.0050}$ & $691^{+87}_{-54}$ & $-5.0^{+3.1}_{-3.4}$ & - & $-1.47^{+0.22}_{-0.25}$ & $-2.15^{+0.28}_{-0.36}$ & $-4.05^{+0.28}_{-0.32}$ & $-1.68^{+2.02}_{-0.69}$ & $7.7^{+7.6}_{-7.4}$ \\[1.25mm]

& & & & & & & & & & \\
& & & & $X_\mathrm{CO}$ & $X_\mathrm{SO_2}$ & & & & & \\
\hline
\\[-3mm]
R3 & $0.6041^{+0.0052}_{-0.0092}$ & $0.1011^{+0.0048}_{-0.0048}$ & $732^{+158}_{-75}$ & $-6.5^{+2.4}_{-2.3}$ & $-6.1^{+1.5}_{-2.5}$ & $-1.36^{+0.21}_{-0.24}$ & $-2.21^{+0.31}_{-0.41}$ & $-3.99^{+0.30}_{-0.42}$ & $-2.12^{+0.65}_{-0.77}$ & $11.4^{+9.9}_{-8.3}$ \\[1.25mm]
\hline

\tableline

\label{tab:free_chem_posteriors}
\end{tabular}
\end{center}
\end{table*}

{\ron For the B1 case, our} results show a {\ron retrieved terminator} temperature of {\ron $697^{+101}_{-59}$} K (see Table~\ref{tab:free_chem_posteriors}), significantly below the simple equilibrium temperature benchmark of $1270$ K (from Table~\ref{tab:system_params}). This {\ron benchmark} equilibrium temperature value assumes zero albedo and efficient heat redistribution across the atmosphere. These may not be particularly good assumptions for WASP-166b. Given the lack of constraints on WASP-166b's albedo, we cannot evaluate the accuracy of the zero albedo assumption. However, the planet is likely tidally locked, which could potentially reduce the efficiency of heat redistribution \citep{wordsworth2015, Koll2022}. In addition, \citet{macdonaldetal2020} suggest that ``most retrieved temperatures are far colder than expected,'' and they also suggest that ``erroneously cold temperatures result when 1-D atmospheric models are applied to spectra of planets with differing morning–evening terminator compositions.'' Future studies may consider 2-D, or even 3-D atmospheric modeling, but are beyond the scope of this work.

\subsection{Equilibrium Chemistry} \label{equilibrium_retrievals}

The simplifying assumption of equilibrium chemistry can be useful for assessing the chemical composition of hot ($> 2000$K) and relatively massive exoplanets (\citealt{Mosesetal2011, Madhusudhanetal2016}). {\ron However,} equilibrium chemistry is a big assumption {\ron and} it is not at all clear that the atmosphere of WASP-166b is in chemical equilibrium, particularly at the pressures probed by transmission spectroscopy. Therefore, {\ron we primarily relied on free chemistry models when assessing the atmospheric composition of WASP-166b. However, we also fit models using equilibrium chemistry for completeness and to estimate the C/O ratio and metallicity of the atmosphere.}

{\ron In this section, we describe equilibrium chemistry modeling with \texttt{POSEIDON}. As we did for the free chemistry modeling in Section~\ref{free_retrievals}, we fit the combined NIRISS and NIRSpec data using fixed ($0.01797 \ \mu$m) bin widths assuming a 1-D atmosphere with an isothermal P-T profile and opaque ``MacMad17'' clouds. For compatibility with the various grids used in \texttt{POSEIDON} equilibrium retrievals,} we used a slightly higher minimum pressure for the atmosphere grid of $10^{-7}$ bar ({\ron versus the} $10^{-9}$ bar {\ron minimum pressure used for free chemistry retrievals}). To model the complex equilibrium atmospheric chemistry we used the \texttt{POSEIDON} implementation of the \texttt{FastChem} code \citep{Kitzmannetal2018} with all 31 constituent species included. As with free chemistry, we assumed that the bulk atmosphere was composed of the fill gases He and H$_2$ in a primordial solar ratio ($X_\mathrm{He}/X_\mathrm{H_2}$ = 0.17).

\begin{figure*}
    \centering
    \includegraphics[width=\textwidth]{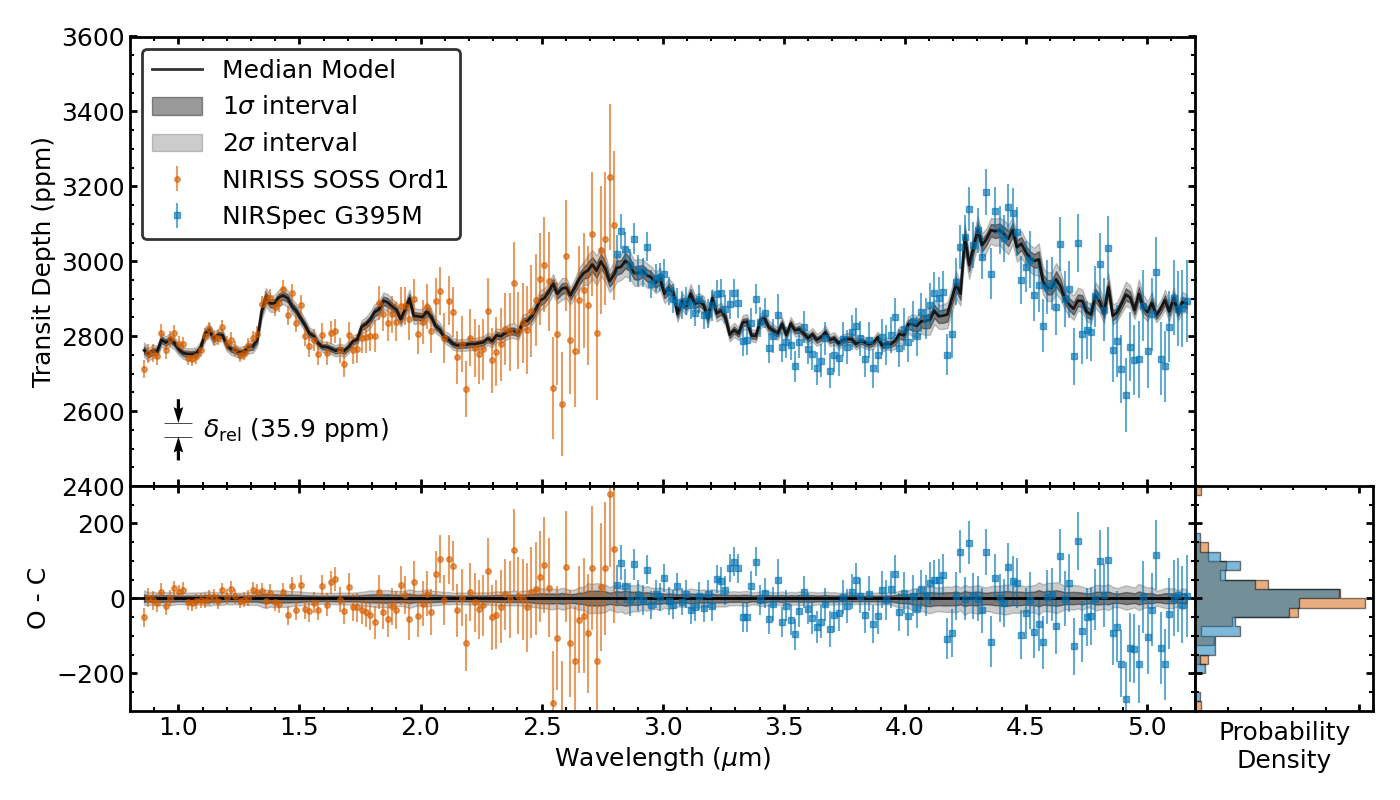}
    \caption{WASP-166b transmission spectrum from \texttt{POSEIDON} equilibrium chemistry retrieval {\ron (E1). \textbf{Top:} observational data shown for NIRISS SOSS (orange circles) and NIRSpec G395M (blue squares)}, for the case of fixed ($0.01797 \ \mu$m) bins. The retrieval is using an isothermal P-T profile with an opaque cloud deck. \ron{The 31 molecular species that are included in our \texttt{POSEIDON} equilibrium chemistry model are shown in Table~\ref{tab:opacity_data}. The plotted NIRISS data have been adjusted down (post-retrieval) by a small retrieved instrument offset (see plotted $\delta_\mathrm{rel}$ value). The median retrieved spectrum for this case is shown as a fine black line with the $1\sigma$ and $2\sigma$ confidence intervals shown in dark and light grey shading respectively. \textbf{Bottom:} residuals between observational data and retrieval model from Top panel plot. \textbf{Bottom Right:} histogram of residuals for each instrument wavelength range, with orange for NIRISS data and blue for NIRSpec data.}}
    \label{fig:POS_Eq_Chem_v76}
\end{figure*}

\begin{figure*}
    \centering
    \includegraphics[width=\textwidth]{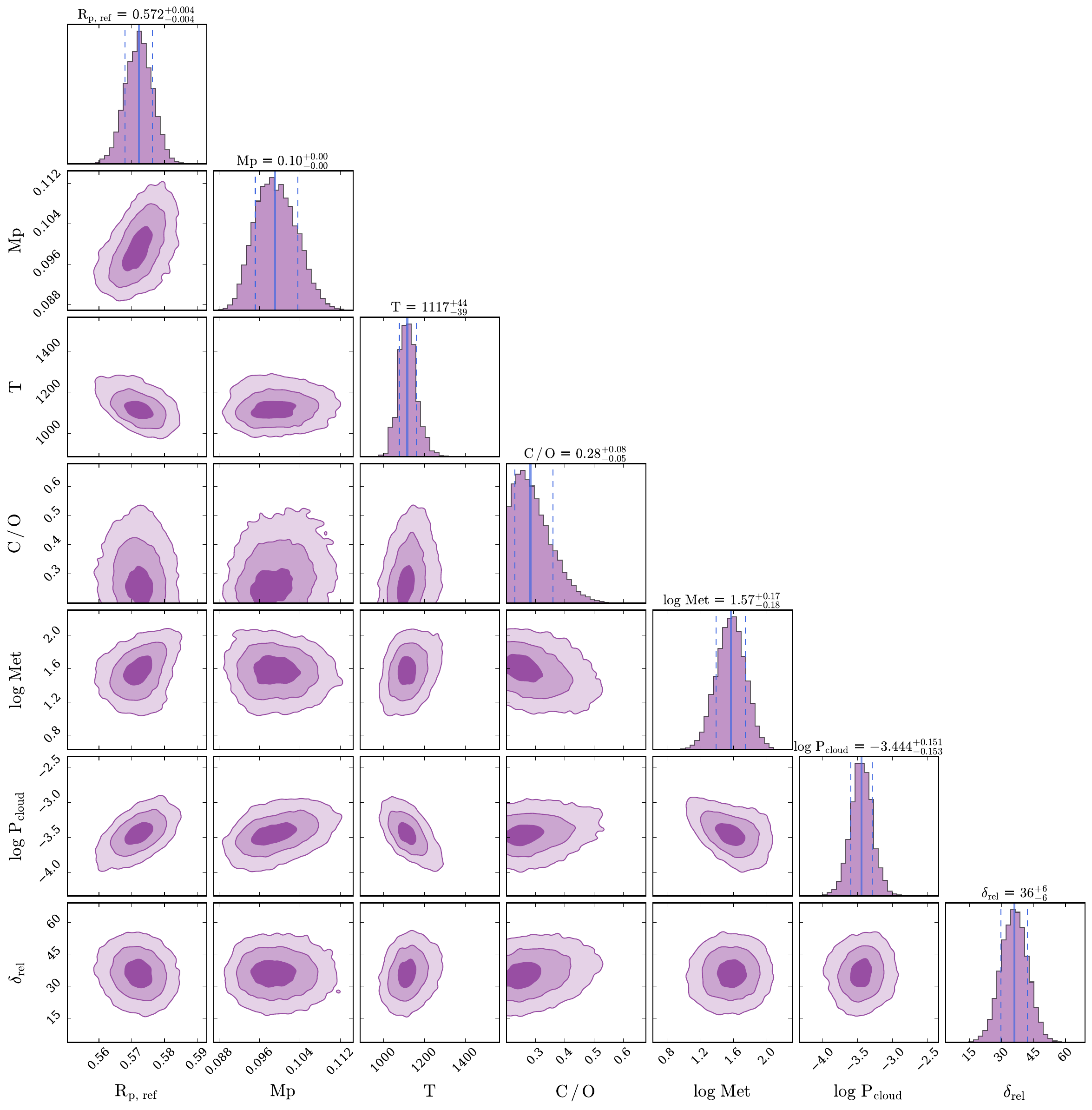}
    \caption{{\ron{Posterior distributions of free parameters (corner plot) for \texttt{POSEIDON} equilibrium chemistry case, E1.}}}
    \label{fig:POS_v76_E1_corner}
\end{figure*}

{\ron We denote our primary equilibrium chemistry retrieval as E1. We provide the priors used in this analysis in} Table~\ref{tab:retrieval_priors} and {\ron present} the resulting {\ron posteriors} in Table~\ref{tab:eq_chem_posteriors}. We also {\ron display} the retrieved spectrum for this case in Figure~\ref{fig:POS_Eq_Chem_v76}, and the posterior distributions of free parameters (corner plot) in Figure~\ref{fig:POS_v76_E1_corner}. The equilibrium chemistry retrieval delivers results for the C/O ratio and metallicity of the planet's atmosphere, in addition to the abundances of specific molecular constituents provided by the free chemistry approach. For {\ron the E1} retrieval case the C/O median value of $0.282$ is lower than the host star $C/O_*$ of $0.41 \pm 0.08$ \citep{polanskietal2022} while the metallicity at a median value of $\log(Z) \sim 1.57$ ($\sim 37$x Solar) is higher than the host star metallicity of $1.5$x Solar (inferred from the stellar [Fe/H] $\sim 0.19$; \citealt{Hellieretal2019}).

\begin{table*}
\caption{Free Parameter Posteriors from Equilibrium Chemistry Retrievals}

\begin{tabular}{lccccccc}
\hline \hline
ID\footnote{See Table~\ref{tab:retrieval_cases} for retrieval case definitions. Results shown here are based on \texttt{POSEIDON} and \texttt{PLATON} retrievals using equilibrium chemistry. For all retrieval cases listed, we are including a 1-D, opaque, cloud deck model. For \texttt{POSEIDON} retrievals, the bulk gases, H$_2$ and He, are present at an assumed primordial solar ratio of $X_\mathrm{He}/X_\mathrm{H_2}$ = 0.17.}
& $R_\mathrm{p,ref}$\footnote{We retrieve the $R_\mathrm{p,ref}$, the planet radius at the reference pressure ($1$ bar for \texttt{POSEIDON}, and \texttt{PLATON}).} & $M_p$ & $T_\mathrm{iso}$ & $C/O$ & $\log(Z)$ & $\log_{10}(P_\mathrm{cloud})$ & $\delta_\mathrm{rel}$\\
& R$_\mathrm{J}$ & M$_\mathrm{J}$ & K &  & $\log_{10}$ (xSolar) & $\log_{10}$(bar) & ppm \\
\hline
E1 & $0.5721^{+0.0041}_{-0.0042}$ & $0.0991^{+0.0044}_{-0.0039}$  & $1117^{+44}_{-39}$ & $0.282^{+0.078}_{-0.053}$ & $1.57^{+0.17}_{-0.18}$ & $-3.44^{+0.15}_{-0.15}$ & $35.9^{+6.3}_{-6.2}$ \\[2mm]
E2 & $0.6077^{+0.0029}_{-0.0056}$ & $0.0988^{+0.0047}_{-0.0052}$ & $973^{+137}_{-86}$ & $0.38^{+0.15}_{-0.11}$ & $2.475^{+0.088}_{-0.100}$ & $-0.3^{+2.1}_{-1.8}$ & $19^{+10}_{-10}$ \\[2mm]
E3 & $0.6092^{+0.0030}_{-0.0049}$ & $0.1011^{+0.0050}_{-0.0048}$ & $1030^{+150}_{-130}$ & $0.35^{+0.22}_{-0.19}$ & $2.50^{+0.12}_{-0.16}$ & $-0.3^{+2.1}_{-1.9}$ & - \\[2mm]
E4 & $0.6049^{+0.0055}_{-0.0115}$ & $0.0986^{+0.0049}_{-0.0049}$ & $1030^{+320}_{-150}$ & $0.39^{+0.16}_{-0.13}$ & $2.47^{+0.10}_{-0.11}$ & $-0.4^{+2.3}_{-2.2}$ & - \\[2mm]
E5 & $0.6058^{+0.0044}_{-0.0100}$ & $0.0989^{+0.0053}_{-0.0050}$ & $980^{+230}_{-120}$ & $0.34^{+0.22}_{-0.11}$ & $2.437^{+0.089}_{-0.126}$ & $-0.6^{+2.4}_{-1.9}$ & $16^{+14}_{-12}$ \\[2mm]
E6 & $0.5983^{+0.0020}_{-0.0026}$ & $0.1110^{+0.0049}_{-0.0057}$ & $811^{+41}_{-32}$ & $0.37^{+0.13}_{-0.14}$ & $1.57$\footnote{This metallicity ($\log(Z)$) preset to E1 retrieval value.} & $-2.42^{+1.82}_{-0.32}$ & $24^{+15}_{-15}$ \\[2mm]
\tableline

\label{tab:eq_chem_posteriors}
\end{tabular}
\end{table*}

{\ron Comparing the results of the Baseline free chemistry retrieval (B1; see Section~\ref{free_retrievals}) and the primary equilibrium chemistry retrieval (E1), we} see some similarities and some differences{\ron.} The retrieved median reference radius ($R_\mathrm{p,ref}$) in the equilibrium case has come down somewhat ($\sim 0.03$ R$_\mathrm{J}$), while the median retrieved mass ($M_p$) is about the same as for the free chemistry case. The median isothermal temperature ($T_\mathrm{iso}$) for the equilibrium case has increased significantly from {\ron $697^{+101}_{-59}$} K for free chemistry to {\ron $1117^{+44}_{-39}$} K for equilibrium. There is also a significant difference in the median retrieved cloud pressure ($\log_{10}(P_\mathrm{cloud})$). It has gone down substantially from $-1.88$ $\log_{10}$(bar) with free chemistry to $-3.44$ $\log_{10}$(bar) for equilibrium chemistry. It is not altogether surprising that we see these differences. The equilibrium chemistry approach to atmospheric characterization is very different than using free chemistry.

{\ron To verify our equilibrium chemistry results, we conducted a variety of equilibrium chemistry retrievals that are described in Appendix~\ref{ap:eq_retrievals}. First, in Section~\ref{pla_Fixed_Bins}, we present an equilibrium chemistry retrieval (E2) on the combined, fixed bin-width NIRISS and NIRSpec data using \texttt{PLATON}.} In Section~\ref{pla_single_instrument}, {\ron we again apply \texttt{PLATON} to fixed $0.01797 \ \mu$m bins} but we explore retrievals on each instrument separately (E3 for NIRISS and E4 for NIRSpec). Finally, in Section~\ref{pla_r100} we use \texttt{PLATON} to conduct a retrieval (E5) using the combined {\ron NIRISS and NIRSpec} data reduced to a constant spectral resolution ($R = 100$). All equilibrium chemistry retrieval inputs and results are available on Zenodo.\footnote{\url{https://doi.org/10.5281/zenodo.14503925}}

\section{Discussion} \label{results}

Analysis of our \textit{JWST} transmission spectroscopy observations of WASP-166b yielded detections of H$_2$O ($15.2\sigma$) and CO$_2$ ($14.7\sigma$) in the planet's atmosphere. We also found weak support for NH$_3$ ($2.3\sigma$) and an opaque cloud deck ($2.6\sigma$) at a pressure of $-1.88^{+0.70}_{-0.64}$ $\log_{10}$(bar).

From our \texttt{POSEIDON} {\ron (E1) equilibrium chemistry retrieval} we estimate a planetary C/O ratio of $C/O = 0.282^{+0.078}_{-0.053}$. By comparison, the host star WASP-166 has a stellar C/O ratio of $C/O_* = 0.41 \pm 0.08$ \citep{polanskietal2022}. Thus, the C/O ratio of WASP-166b is significantly lower than the Sun ($0.55$; \citealt{asplundetal2009}) and may also be substellar (low by $1.15\sigma$ compared to its host star). {\ron Additional observations may be needed to fully distinguish the planetary C/O ratio as either substellar or stellar.} Further, from \texttt{POSEIDON} {\ron (E1)} we also find that WASP-166b has a high atmospheric metallicity ($\log(Z) = 1.57^{+0.17}_{-0.18}$, $Z = 37^{+18}_{-13}$) relative to Solar.

In this section, we compare these results to prior analyses of WASP-166b, explain the broader context of similar hot Neptunes, and investigate how the atmospheric composition of WASP-166b informs our understanding of the planet's formation pathway and history.

\subsection{{\ron Comparison to} Prior Atmospheric Observations of WASP-166b}

{\ron Our results are not the first constraints on the atmospheric properties of WASP-166b. To start, sodium has been previously detected in the planet's atmosphere at $3.4\sigma$ confidence \citep{seideletal2020,seideletal2022}.} Our observations, collected with the \textit{JWST} NIRISS and NIRSpec instruments, provide wavelength coverage across $0.85-5.17$ $\mu$m and are thus insensitive to the {\ron core of the sodium absorption line at $0.6 \ \mu$m. However, we explore our sensitivity to the pressure broadened wings of sodium and potassium more fully in Section~\ref{Na_K_retrievals} and find that they also have a negligible impact on our observations even at the shortest wavelengths that we can access.}

Our results are consistent with that of \citet{seideletal2020} and \citet{seideletal2022} in that all three analyses point away from a flat transmission spectrum and towards an intermediate pressure cloud deck. The same ESPRESSO transmission spectroscopy observations (2 transits from 2020 Dec and 2021 Feb) used to confirm the sodium detection by \citet{seideletal2020} were also subsequently used to investigate the presence of water vapor and clouds in the atmosphere of WASP-166b \citep{lafargaetal2023}. They were able to strongly exclude scenarios with both high water content and high pressure (low altitude) clouds which would yield the strongest water absorption signals; they were also able to moderately exclude scenarios with low water content and low pressure (high altitude) clouds that would yield non-detections. They found good fits to intermediate absorption signal models with either low water content and high cloud pressure or high water content with low cloud pressure. Our results are consistent in that we make a strong detection of H$_2$O {\ron ($15.2\sigma$)} as well as show weak support for an intermediate pressure cloud deck {\ron ($2.6\sigma$)} dampening H$_2$O absorption signals.

\subsection{Comparison to Hot Neptune LTT 9779b} \label{ltt9779b}

In contrast to WASP-166b on the edge of the Hot Neptune Desert, there is a similar planet deep in the desert that already has atmospheric characterization with \textit{JWST}: LTT 9779b ($P = 0.792$ d, $R_p = 4.72 \pm 0.23$ R$_\oplus$, $M_p = 29.32 \pm 0.8$ M$_\oplus$; \citealt{jenkinsetal2020}). LTT 9779b is also the most irradiated Neptune planet known to date. The location of both LTT 9779b and WASP-166b in and near the edge of the Hot Neptune Desert can be seen in Fig~\ref{fig:HND_mass_v_insol} {\ron and their properties can be compared in Table~\ref{tab:similar_planets}}. \citet{radicaetal2024} collected and analyzed a full phase curve and transit of LTT 9779b with the NIRISS SOSS instrument ($0.6 - 2.85$ $\mu$m) but found only muted spectral features. Although their findings were discrepant from a flat transmission spectrum, they could not fully break the degeneracy between metallicity and cloud top pressure. By applying additional constraints on planetary metallicity through interior structure modeling, \citet{radicaetal2024} found that millibar pressure clouds in a H$_2$O or CH$_4$ dominated atmosphere was the most likely scenario. {\ron \citet{radicaetal2025} determined from \textit{HST} eclipse observations that LTT 9779b has a high geometric albedo consistent with atmospheric silicate condensates.} Not only must the atmosphere of LTT 9779b withstand much more irradiation than that of WASP-166b ($2500$ S$_\oplus$ versus $440$ S$_\oplus$), the planet is also far smaller than WASP-166b ($4.72 \pm 0.23$ R$_\oplus$ versus $6.9 \pm 0.3$ R$_\oplus$) despite similar planet masses and stellar radii, resulting in a smaller scale height and smaller spectral features. It is possible that LTT 9779b has only retained a thick atmosphere via the presence of a cloud deck that reduces the efficiency of atmospheric loss \citep{radicaetal2024}; meanwhile, the (still high but) relatively lower stellar irradiation for WASP-166b may allow for a puffier atmosphere with fewer or deeper clouds without extreme atmospheric mass loss.

\begin{table*}
    \scriptsize
    \caption{{\ron Planets Similar to WASP-166b in Bulk Properties or Atmospheric Detections. Detections are indicated with their detection significance or with a `Y' (where no detection significance is listed), non-detections (and inconclusive results) are indicated using `ND', upper limits are indicated using `$<$', and features excluded from a given atmospheric retrieval are indicated using `-'.}}
    \begin{center}
    \begin{tabular}{l|llll|cccccccl|cc}
        Name & Period & Insolation & Radius & Mass & \multicolumn{6}{c}{Detections} &   && Instrument & Ref. \\
         & (d) & Flux ($S_\oplus$) & ($R_\oplus$) & ($M_\oplus$) & $\rm{CH_4}$ & $\rm{SO_2}$ & CO & $\rm{NH_3}$ & $\rm{H_2O}$ & $\rm{CO_2}$  & HCN&Clouds & &  \\
        \hline
        \hline
        GJ 1214b & 1.580 & 17 & 2.74 & 8.19 & $2.0\sigma$ & ND& ND& ND& ND& $2.4\sigma$  &  -&ND& [g] & [1] \\
        HD 3167c & 29.838 & 17 & 2.86 & 9.79 & $<$& - & ND & $<$ & $3.17\sigma$ & $3.28\sigma$  &  -&ND & [b,d] & [2] \\
        GJ 3470b & 3.337 & 36 & 4.35 & 13.89 & ND & - & ND & ND & $5.2\sigma$ & ND & ND & ND & [b,c,d] & [3] \\
        & &  &  & &  $3.8\sigma$ &  $4.0\sigma$ &  $1.5\sigma$ & ND& $6.3\sigma$ & $7.3\sigma$  &  ND&ND& [b,d,e] & [4] \\
        HD 106315c & 21.057 & 103 & 4.35 & 15.19 & $<$& -& ND & $1.97\sigma$ & $5.68\sigma$ & ND  &  ND&ND& [b] & [2]\\
        & & & & & ND& -& ND& -& ND & ND & ND & ND & [b,d,i]&[5]\\
        HAT-P-11b & 4.888 & 98 & 4.36 & 23.39 & $2.6\sigma$ & -& -& $5\sigma$ & $3.4\sigma$ & $3.2\sigma$  & - &- & [a] & [6]\\ 
        &  &  & &  & ND& - & ND&  ND &  $5.1\sigma$  & ND & ND & ND & [b,d,i] & [7]\\ 
        LTT 9779b & 0.792 & 2500 & 4.72 & 29.32 & ND & - & ND & - & ND & ND & - & ND & [h] & [8] \\
        TOI-674b & 1.977 & 38 & 5.26 & 23.49 & ND& -& ND& ND& $2.9\sigma$ &  ND&  -&$2.2\sigma$& [b,d]& [9] \\
        HAT-P-26b & 4.235 & 160 & 6.39 & 18.59 & ND & - & ND & - & $8.8\sigma$ & $<$& - & Y & [b,c,d]& [10]\\
        & & & & & ND & - & ND & ND & 7.2$\sigma$ & ND& - & ND & [b,c,d,j] & [11]\\
        HD 89345b & 11.814 & 234 & 6.86 & 35.69 & -& - & - & - & - & - & - & ND & [a] & [12]\\
        \textbf{WASP-166b} & \textbf{5.444}& \textbf{440} & \textbf{6.90} & \textbf{32.1} &  \textbf{ND} &  \textbf{1.5$\sigma$} &  \textbf{1.1$\sigma$} & \textbf{2.3$\sigma$} & \textbf{15.2$\sigma$} & \textbf{14.7}$\sigma$  &  \textbf{ND} &  \textbf{2.6$\sigma$} & \textbf{[g,h]} & \textbf{[0]} \\
        HIP 67522b & 6.960 & 306 & 10.05 & 13.79 & ND& 1.8$\sigma$& 3.5$\sigma$& ND& $7\sigma$ & $11\sigma$  & - & ND & [g]& [13] \\
        WASP-107b & 5.721 & 51 & 10.54 & 38.14 & $5\sigma$ & $9\sigma$ & $7\sigma$ & $6\sigma$ & $21\sigma$ & $29\sigma$  & - &26$\sigma$& [b,e,f] & [14] \\
    \end{tabular}
    \end{center}
    {\ron \textit{Instruments}: [a] TNG/GIANO-B, [b] HST/WFC3, [c] HST/STIS, [d] Spitzer/IRAC, [e] JWST/NIRCam, [f] JWST/MIRI, [g] JWST/NIRSpec, [h] JWST/NIRISS SOSS, [i] Kepler/K2, [j] Magellan/LDSS-3C,. \\ \textit{References}: [0] This Work, [1] \citet{schlawinetal2024}, [2] \citet{guilluyetal2021}, [3] \citet{bennekeetal2019}, [4] \citet{beattyetal2024}, [5] \citet{kreidbergetal2022}, [6] \citet{basilicataetal2024}, [7] \citet{Fraine2014}, [8] \citet{radicaetal2024}, [9] \citet{brandeetal2022}, [10] \citet{wakefordetal2017}, [11] \citet{macdonaldandmadhusudhan2019}, [12] \citet{guilluyetal2023}, [13] \citet{thaoetal2024}, [14] \citet{welbanksetal2024}
    \label{tab:similar_planets}}
\end{table*}

\subsection{Comparison to Other Similar Planets} \label{other_planets}

While LTT 9779b is an interesting planet for comparison at an extremely high insolation flux, there are several additional planets that provide valuable context for the atmospheric results of WASP-166b, {\ron which can be seen in Table~\ref{tab:similar_planets}. (This table also tracks which instrument made the detection, allowing the reader to partially disentangle instrument sensitivity to a given molecule from its underlying presence in the atmosphere.)}

The most similar planet to WASP-166b in mass and radius with atmospheric characterization is HD 89345b \citep{guilluyetal2023}, which was observed with GIANO-B on the TNG and resulted in upper limits on the presence of helium.

WASP-107b ($P = 5.7215$ d, {\ron $R_p = 10.54$ R$_\oplus$, $M_p = 38.14$ M$_\oplus$}; \citealt{welbanksetal2024}) is especially comparable to WASP-166b given the atmospheric features detected by \citet{welbanksetal2024} {\ron and \citet{dyreketal2024}}: CH$_4$, SO$_2$, CO, NH$_3$, H$_2$O, CO$_2$, and a cloud deck, the latter four of which we also found evidence for in the atmosphere of WASP-166b. Additionally, they determined a planet metallicity of $10 - 18$x Solar ([M/H] = $1.09^{+0.17}_{-0.07}$) and a C/O ratio of $0.33^{+0.06}_{-0.05}$. WASP-107b has a metal-rich atmosphere within $2\sigma$ {\ron of the metallicity uncertainty of WASP-166b derived here}, and a substellar C/O ratio within $1\sigma$ {\ron that of WASP-166b}. \citet{welbanksetal2024} {\ron and \citet{singetal2024}} observe that tidal heating likely played an important role in the evolution of WASP-107b and may contribute to the inflated radius of the planet, while disequilibrium chemistry can result from interactions between the planet atmosphere and core. These lessons can be similarly applied to WASP-166b, as this planet too has evidence of disequilibrium processes and an inflated radius, {\ron resulting in a planetary density ($0.54$ g cm$^{-3}$) puffier than 94\% of other Neptunes ($1.5$ R$_\oplus < R_p < 7.5$ R$_\oplus$), according to the NASA Exoplanet Archive (accessed {\ron 2025 Mar 24}).}

Besides WASP-107b, there is only one other hot Neptune thus far known to have atmospheric ammonia: \citet{basilicataetal2024} detected ammonia in the atmosphere of HAT-P-11b as well as H$_2$O and evidence for CH$_4$ and CO$_2$. Additionally, there are several other planets similar to WASP-166b not only in bulk characteristics but also in atmospheric composition. H$_2$O has been detected in the atmospheres of HD 106315c \citep{guilluyetal2021,kreidbergetal2022}, TOI-674b \citep{brandeetal2022}, and HAT-P-26b \citep{wakefordetal2017,macdonaldandmadhusudhan2019}). Further, just like WASP-107b and WASP-166b, both H$_2$O and CO$_2$ have been detected in the atmospheres of HD 3167c \citep{guilluyetal2021}, GJ 3470b \citep{bennekeetal2019,beattyetal2024}, and HIP 67522b \citep{thaoetal2024}. Lastly, there is only one known hot Neptune with evidence of CO$_2$ but no detection at all of H$_2$O: GJ 1214b \citep{schlawinetal2024} has a $3.3\sigma$ detection of atmospheric CO$_2$.

{\ron Considering all these atmospheric detections (and more, as listed in Table~\ref{tab:similar_planets}), we can draw some very basic inferences about this planetary subset. Most of the Hot Neptunes and planets similar to WASP-166b that we have considered here have reported atmospheric H$_2$O detections, and a majority (though fewer) have reported CO$_2$ detections as well. Molecules like CH$_4$, SO$_2$, CO, and NH$_3$ have fewer reported detections, and nearly all of those correspond, as we might expect, to observations collected with \textit{JWST}.}

\subsection{The Formation of WASP-166b and the Hot Neptune Desert}

Our \texttt{POSEIDON} equilibrium chemistry retrieval indicated that WASP-166b has a high atmospheric metallicity ($\log(Z) = 1.57^{+0.17}_{-0.18}$, $Z = 37^{+18}_{-13}$), about $~\sim 25$x more metal-rich than the host WASP-166 (inferred from a stellar [Fe/H] of $0.19$). Our retrievals also yielded a substellar planetary atmospheric C/O ratio ($C/O = 0.282^{+0.078}_{-0.053}$) still consistent ($1.1\sigma$) with the host star ($C/O_* = 0.41 \pm 0.08$), which allows us to place some constraints on the formation pathways for this planet. For example, a discrepancy between stellar C/O and planetary C/O can be explained by planetary formation at a larger orbital separation, beyond one or several snow lines (e.g. H$_2$O, CO$_2$, CO) where the relative abundances of key molecules in gas or ice phase (and the corresponding C/O ratio) are distinct from the environment close to the host star. Planet formation beyond a given snow line can create a primordial C/O ratio that persists in the planet's atmosphere even following later mixing, modification, or inward migration across that snow line \citep{madhusudhan2019}.

On the other hand, as discussed by \citet{obergetal2011}, in situ planet formation and formation through gravitational instability should yield a stellar C/O ratio as the planet forms with the same chemical inventory as the host star; a higher metallicity and lower C/O ratio (as seen for WASP-166b) is indicative of ``pollution'' by planetesimal accretion modifying the C/O ratio away from its primordial value (regardless of formation pathway). \citet{thiabaudetal2015} similarly find that {\ron planetary accretion or mixing of accreted solids with the envelope can keep the C/O ratio substellar or stellar.} \citet{madhusudhanetal2017} echoed this result by finding that planetary core erosion can yield very high planetary metallicities and stellar or substellar C/O ratios, although photoevaporation could also play a role. Taking these results together, while it is still hard to clarify the exact formation pathway, WASP-166b fits the mold of a planet that attained a {\ron stellar or} substellar C/O ratio and high metallicity through a combination of planetesimal accretion, core erosion, or photoevaporation.

As WASP-166b is a planet residing at the edge of the Hot Neptune Desert, direct theories of how the planet formed necessarily also shed light on the formation mechanisms involved in the origin of the desert more broadly. Although the extent and paucity of the desert have been well documented, its origins and mechanisms are less clear. Hot Neptune Desert origin theories generally invoke either planet migration or in situ formation. \citet{mazehetal2016} proposed that the desert upper boundary is an inner migration limit due to either photoevaporation (atmospheric loss) at high insolation fluxes or a gap in the inner protoplanetary disk during early planetary formation. \citet{matsakosandkonigl2016} theorized that the desert forms as planets undergo high-eccentricity migration followed by tidal circularization near their Roche limit. As for in situ theories, the lower desert boundary could be the result of planet cores forming after protoplanetary disk gas has dispersed \citep{leeandchiang2016,rogersetal2011,helledetal2016}. \citet{batyginetal2016} suggested that in situ hot super-Earth formation can lead to runaway gas accretion and create hot Jupiters, with hot Neptunes only forming under fine-tuned accretion conditions.

Our constraints on the formation mechanisms for WASP-166b do not clearly reveal how much migration the planet underwent. However, our findings are consistent with planetesimal accretion followed by some degree of core erosion and/or photoevaporation. We cannot exclude any formation mechanisms for the Hot Neptune Desert through the analysis of a single planet, but theories that invoke photoevaporation \citep{mazehetal2016} or can otherwise explain a low C/O ratio with a high planetary metallicity are certainly worth exploring more thoroughly.

\section{Conclusion} \label{conclusion}

We present our atmospheric characterization of WASP-166b, {\ron a.k.a. Catalineta,} a low-density super-Neptune ($P = 5.44$ d, $R_p = 6.9 \pm 0.3$ R$_\oplus$, $M_p = 32.1 \pm 1.6$ M$_\oplus$) located at the edge of the Hot Neptune Desert. We collected two transit observations of the planet with \textit{JWST}, one transit with NIRSpec BOTS G395M/F290LP ($2.80-5.17$ $\mu$m) and one transit with NIRISS SOSS Order-1 ($0.85-2.81$ $\mu$m). This is the first publication thoroughly describing the NIRSpec G395M transmission spectroscopy data reduction process as well as demonstrating NIRSpec G395M and NIRISS SOSS used in combination. {\ron We found that the broader wavelength range of both instruments used in combination provides much greater sensitivity to the atmospheric composition that is otherwise lost to ambiguity and degeneracies in our single instrument retrievals, with NIRISS providing strong constraints on the H$_2$O abundance and cloud features (where NIRSpec performs poorly) and NIRSpec capturing the CO$_2$ and NH$_3$ features (where NIRISS performs poorly).}

We reduced these observations to planetary transmission spectra and conducted forward modeling and retrievals on the combined planetary spectrum in order to constrain the planet's atmospheric constituents. We find the planet's atmosphere contains H$_2$O with $15.2\sigma$ significance ($\log_{10}$(vmr) $= -1.42^{+0.20}_{-0.24}$), and CO$_2$ with $14.7\sigma$ significance ($\log_{10}$(vmr) $= -2.13^{+0.27}_{-0.39}$). We also show weak support ($2.3\sigma$) {\ron for} NH$_3$ ($\log_{10}$(vmr) $= -4.02^{+0.28}_{-0.34}$), and weak support ($2.6\sigma$) for the cloud deck pressure ($\log_{10}(P_\mathrm{cloud})$) $= -1.88^{+0.71}_{-0.63}$ $\log_{10}$(bar)). {\ron{Detections of CO, SO$_2$, and Na are inconclusive; and finally, we show non-detections of CH$_4$, C$_2$H$_2$, HCN, H$_2$S, and K.}}

Puffy planets like WASP-166b are prime targets for detailed atmospheric analysis. Our transmission spectroscopy of WASP-166b yields meaningful constraints on molecular abundances, metallicity, and C/O ratio, thereby providing insight into the formation {\ron process} and evolution of this low-density hot super-Neptune. The high planetary metallicity ($\log(Z) = 1.57^{+0.17}_{-0.18}$, $Z = 37^{+18}_{-13}$) and slightly substellar C/O ratio ($C/O = 0.282^{+0.078}_{-0.053}$) of WASP-166b can be explained through planetesimal accretion followed by core erosion or photoevaporation. This in turn provides evidence that photoevaporation, or other mechanisms that allow for {\ron stellar or} substellar C/O ratios and superstellar metallicities, {\ron are plausible pathways for} the formation of the Hot Neptune Desert. Collecting further atmospheric observations of similar hot planets at intermediate radii will give us the insights necessary to better discriminate between the various formation models for the origin and nature of the Hot Neptune Desert.


\section*{Acknowledgments}
The authors wish to thank the referee for their useful and enlightening feedback. The authors would also like to thank Taylor Bell for his helpful comments and suggestions related to the intricacies of \texttt{Eureka!}. Also, we would like to thank Ryan MacDonald for his quick responses and clear explanations to questions about \texttt{POSEIDON}. AWM acknowledges John Brewer's many useful discussions and feedback on the work and appreciates conversations with Munazza Alam about system observations. AWM would also like to thank the thesis committee that reviewed this work as a PhD thesis chapter, including Eugene Chiang and Josh Bloom.

This work is based on observations made with the NASA/ESA/CSA \textit{JWST}. The data were obtained from the Mikulski Archive for Space Telescopes at the Space Telescope Science Institute, which is operated by the Association of Universities for Research in Astronomy, Inc., under NASA contract NAS 5-03127 for \textit{JWST}. These observations are associated with program \#GO-2062. Support for program \#GO-2062 was provided by NASA through grant JWST-GO-02062.002-A from the Space Telescope Science Institute, which is operated by the Association of Universities for Research in Astronomy, Inc., under NASA contract NAS 5-03127.

This research has made use of the NASA Exoplanet Archive, which is operated by the California Institute of Technology, under contract with the National Aeronautics and Space Administration under the Exoplanet Exploration Program.

DRL acknowledges support by NASA Headquarters through an appointment to the NASA Postdoctoral Program at the Goddard Space Flight Center, administered by ORAU through a contract with NASA. Additionally, DRL acknowledges support from the GSFC Sellers Exoplanet Environments Collaboration (SEEC), which is supported by NASA's Planetary Science Division's Research Program, as well as support from NASA under award number 80GSFC24M0006. 

CKH acknowledges support from the National Science Foundation Graduate Research Fellowship Program under Grant No. DGE 2146752. AWM acknowledges support from the National Science Foundation Graduate Research Fellowship Program under Grant No. DGE 1752814.\\

{\ron \begin{large}\textit{Data availability:}\end{large} The specific observations analyzed can be accessed via the STScI MAST archive at \href{https://doi.org/10.17909/8gbq-gp71}{10.17909/8gbq-gp71}. All data products and models from our analyses are available at \href{https://doi.org/10.5281/zenodo.14503925}{10.5281/zenodo.14503925}.} 
{\ron \facilities{\textit{JWST} (NIRISS and NIRSpec)}}
\software{\texttt{Ahsoka} \citep[][]{Louie_2025}, 
\texttt{batman} \citep{Kreidbergetal2015},
\texttt{dynesty} \citep{Speagle2020},
\texttt{emcee} \citep{Foreman-Mackey2013, GoodmanWeare2010},
\texttt{Eureka!\hspace{0.2em}}\citep{belletal2022},
{\ron \texttt{exoTEDRF} \citep[formerly, \texttt{supreme-SPOON};][]{feinsteinetal2023,radicaetal2023,Radica2024JOSS}},
\texttt{ExoTiC-LD} \citep{david_grant_2022_ExoTiC-LD},
\texttt{ExoTransmit} \citep{kemptonetal2017},
\texttt{FastChem} \citep{Kitzmannetal2018},
{\ron \texttt{GGChem} \citep{woitkeetal2018,woitkeandhelling2021}},
\textit{JWST} \texttt{Science Calibration Pipeline} \citep{bushouse2023},
\texttt{MultiNest} \citep{ferozetal2009,ferozetal2013},
\texttt{nirHiss} \citep{feinsteinetal2023},
\texttt{PandExo} \citep{batalhaetal2017},
\texttt{PASTASOSS} \citep{baines2023traces,baines2023wavelength},
\texttt{PLATON} \citep{zhangetal2019,zhangetal2020},
\texttt{POSEIDON} \citep{macdonaldandmadhusudhan2017, macdonald2023},
\texttt{PyMultiNest} \citep{buchner2016},
\texttt{TauREx} \citep{alrefaieetal2021},
\texttt{transitspectroscopy} \citep{espinoza_nestor_2022_6960924}.
}

\clearpage
\appendix
\section{NIRSpec {\ron BOTS} G395M Data Reduction Details}\label{appA_g395m_reduction}

\subsection{\textit{JWST} Pipeline Stage 1: Detector-level processing}\label{subsubsec:NIRSpec_stage1}

Our analysis begins with the \textit{rateints.fits} files (i.e., ramp-fitting calibrated flux values) downloaded from the Mikulski Archive for Space Telescopes (MAST).\footnote{The data described here may be obtained from the MAST archive at \url{https://doi.org/10.17909/8gbq-gp71}.} The \textit{rateints.fits} files are output products of Stage 1 of the \textit{JWST} \texttt{Science Calibration Pipeline} (version 1.13.3),\footnote{\url{https://jwst-pipeline.readthedocs.io/en/latest/jwst/user_documentation/introduction.html}} and the \textit{JWST} Science Data Processing (SDP) subsystem (version 2023\_4a).

Stage 1 applies detector-level corrections to raw non-destructively read ``ramps'' from the uncalibrated (\textit{uncal.fits}) data in order to produce 2-dimensional count rate (slope) images per exposure. The corrections made in this stage include detector dark current subtraction, identifying known bad pixels using the Calibration Reference Data System map (CRDS v11.17.14),\footnote{\url{https://jwst-crds.stsci.edu/}} adjustments for detector non-linearity, removing effects of cosmic ray hits, and other effects.\footnote{\url{https://jwst-pipeline.readthedocs.io/en/latest/jwst/pipeline/calwebb_detector1.html}} We refer the reader to the footnotes in this section for a more detailed description of the definitions and processes in this stage.

One module of the \textit{JWST} \texttt{Science Calibration Pipeline} Stage 1 is the \texttt{jump} step, which identifies outliers such as cosmic ray hits. \citet{Kirketal2024} had suggested that the default \texttt{jump} detection step could be unreliable for small group numbers. Indeed, the \texttt{jump} step will automatically skip execution if the input data contain less than 3 groups per integration.\footnote{\url{https://jwst-pipeline.readthedocs.io/en/latest/jwst/jump/description.html}} As our NIRSpec observation only used 3 groups per integration, we conducted a test case feeding the uncalibrated \textit{uncal.fits} data through \texttt{Eureka!\hspace{0.2em}}Stage 1 (essentially a wrapper for Stage 1 of the \textit{JWST} \texttt{Science Calibration Pipeline}) while skipping the \texttt{jump} step. We detected no significant differences between this test case and the default \textit{JWST} Pipeline Stage 1 output \textit{rateints.fits} files, and saw no evidence of systematics or large scatter in the resultant transmission spectrum with the default \texttt{jump} step included. Having demonstrated that the default \textit{JWST} Pipeline Stage 1 was working properly, and to save local processing time, and avoid other unknowns, we chose to accept these output files and move forward. The remaining reduction stages for the NIRSpec observations after Stage 1 are processed through the \texttt{Eureka!\hspace{0.2em}}pipeline as described below.

\subsection{\texttt{Eureka!\hspace{0.2em}}Stage 2: Spectroscopic processing}\label{subsubsec:NIRSpec_stage2}

Starting with the \textit{rateints.fits} files from MAST, we used \texttt{Eureka!\hspace{0.2em}}for Stage 2 processing. This stage continues the calibration process initiated in Stage 1, with \texttt{Eureka!\hspace{0.2em}}serving again as a wrapper for the \textit{JWST} Science Calibration Pipeline (\texttt{calwebb\_spec2}) Stage 2 steps.\footnote{\url{https://jwst-pipeline.readthedocs.io/en/stable/jwst/pipeline/calwebb_spec2.html}} The outputs of Stage 2 for Time Series Observations (TSOs) are calibrated data (\textit{calints.fits} files) still on the native detector pixel grid where integrations remain distinct to capture the time-series information.

There are 16 processing steps in \texttt{Eureka!\hspace{0.2em}}(v0.10) Stage 2. For our instrument/mode (i.e., NIRSpec Fixed Slit, and TSO) 10 of these steps are skipped by default. Three of the remaining 6 steps: \texttt{extract\_2d}, \texttt{wavecorr}, and \texttt{srctype} are performed by default. 

First, the \texttt{extract\_2d}\footnote{\url{https://jwst-pipeline.readthedocs.io/en/stable/jwst/extract_2d/main.html}} step extracts 2D arrays from spectral images and computes an array of wavelengths to attach to the data. Next, the \texttt{wavecorr}\footnote{\url{https://jwst-pipeline.readthedocs.io/en/latest/jwst/wavecorr/index.html}} step adjusts the wavelength map for fixed slit sources not centered (in the dispersion direction) in the slit. Finally, the \texttt{srctype}\footnote{\url{https://jwst-pipeline.readthedocs.io/en/stable/jwst/srctype/index.html}} step addresses the question of whether a spectroscopic target is an extended object or a point source. This is then used in later processing to apply corrections dependent on the source type.

Additionally, based upon the Early Release Science (ERS) program literature (see Section~\ref{NIRSpec_Data_Reduction}), we skipped the last 3 steps that would otherwise be run by default. 
\texttt{Eureka!\hspace{0.1em}}Stage 2 typically performs flat-fielding and unit conversions; however, we skipped the \texttt{flat\_field},\footnote{\url{https://jwst-pipeline.readthedocs.io/en/latest/jwst/flatfield/index.html}} and \texttt{photom}\footnote{\url{https://jwst-pipeline.readthedocs.io/en/latest/jwst/photom/index.html}} steps as they result in a conversion to physical flux units that is not needed for our relative flux measurements. If included, these steps could potentially add noise depending on the accuracy of the latest detector flat fields provided by the CRDS. In addition, the small detector region that is affected by these steps can reduce the precision on background removal \citep{Mayetal2023}. None of the data-reduction pipelines used on the NIRSpec ERS program data described by \citet{Aldersonetal2023}, including \texttt{Eureka!\hspace{0.1em}}, performed a flat-field correction. As they explain: ``the available flat fields were of poor quality and unexpectedly removed portions of the spectral trace.''

Finally, we skipped the \texttt{extract\_1d}\footnote{\url{https://jwst-pipeline.readthedocs.io/en/latest/jwst/extract_1d/index.html}} step \citep{Rustamkulovetal2023} since we are using \texttt{Eureka!\hspace{0.2em}}to perform an Optimal spectral extraction \citep{Horne1986} in the next stage.

\subsection{\texttt{Eureka!\hspace{0.2em}}Stage 3: Spectral extraction}\label{subsubsec:NIRSpec_stage3}

The key steps of \texttt{Eureka!\hspace{0.2em}}Stage 3 are to correct the curvature of the spectral trace, perform column-by-column background subtraction, and to perform an Optimal spectral extraction \citep{Horne1986}. The main output products of Stage 3 are time series tables of 1-D stellar spectra (2-D light curves), first by pixel column, and then with the wavelength calibration applied. 

\begin{figure*}
    \centering
    \includegraphics[width=1.00\textwidth]{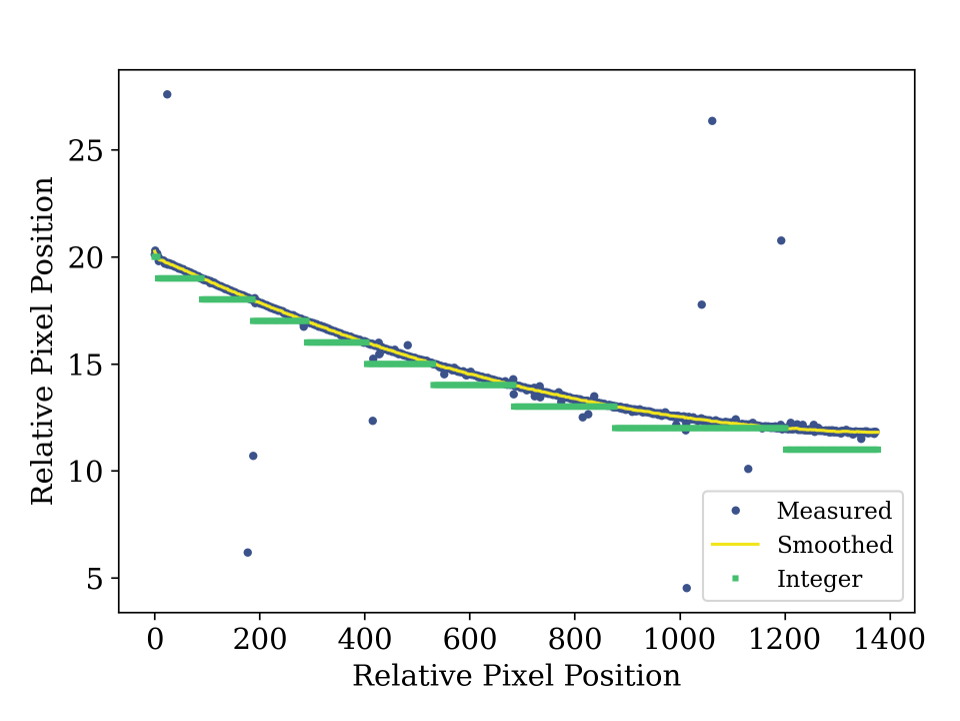}
    \caption{{\ron Trace Curvature.} The measured (blue {\ron points}), smoothed (yellow {\ron line}), and integer-rounded (green {\ron points}) relative position of the spectral trace on the NIRSpec G395M (NRS1) detector determined with the \texttt{Eureka!\hspace{0.2em}}data reduction and analysis pipeline \citep{belletal2022}. The majority of the measured {\ron (blue)} data agree closely with the smoothed {\ron (yellow)} line. Significant curvature is evident; however, the aspect ratio of the plot is significantly compressed along the horizontal/column axis.}
    \label{fig:3107}
\end{figure*}

The relative position of the spectral trace on the detector is shown in Figure~\ref{fig:3107}. \texttt{Eureka!\hspace{0.2em}}Stage 3 aligns the spectral trace by determining the effective center of light in each detector column with a Gaussian fit and integer-shifting along each column in order to bring this effective peak to the central row of the subarray field. 

Setting the various aperture limits and sigma clipping thresholds in Stages 3 and 4 is a process of trial and error. In setting the parameters in these stages, we considered the following five sources: (1) guidance from the \texttt{Eureka!\hspace{0.2em}}readthedocs Quickstart guide and tutorials,\footnote{\url{https://eurekadocs.readthedocs.io/en/stable/quickstart.html}} (2) notes and presentation materials from the 2023 Sagan Summer Workshop on Characterizing Exoplanet Atmospheres,\footnote{\url{https://nexsci.caltech.edu/workshop/2023/handson.shtml}} (3) data reduction procedure descriptions from the ERS WASP-39b papers \citep{JWSTERSTEAM2023,Ahreretal2023,Aldersonetal2023,Rustamkulovetal2023}, (4) direct conversations with the \texttt{Eureka!\hspace{0.2em}}developers, and finally (5) our own parametric study of the relevant settings/parameters affecting the white light curve residuals. 

For both the background subtraction and Optimal extraction steps, we performed a parametric analysis to optimize the values of the relevant \texttt{Eureka!\hspace{0.2em}}Stage 3 control parameters, which are listed in Table~\ref{tab:S3_parametric_study}. For each parameter, the table provides the name, description, trial values examined, and value used in our final analysis. We tested various combinations of these background and extraction parameters with the goal of determining the combination that would minimize the median absolute deviation (MAD) of the un-binned white light curve residuals along the time axis, as generated in \texttt{Eureka!\hspace{0.2em}}Stage 4 (Section~\ref{subsubsec:NIRSpec_stage4}). The MAD of the residuals in our parametric study ranged between 196 and 203 ppm. We selected the parameter configuration resulting in a MAD of 196 ppm. We next describe our application of background subtraction and Optimal spectral extraction using the final parameter values given in Table~\ref{tab:S3_parametric_study}.

\begin{table*}
\caption{NIRSpec \texttt{Eureka!\hspace{0.2em}}Stage 3 Background Subtraction and Spectral Extraction Parametric Analysis}
\begin{center}
\begin{tabular}{llccc}
\hline \hline
Parameter  &  Description                                              &  Trial Values  &  Final    \\
\hline
bg\_hw     &  background half-width (pixels from middle row)           &     6, 8, 9    &    8        \\[0.5ex]
bg\_thresh &  background threshold for temporal outliers               & [3,3], [10,10] & [10,10]   \\[0.5ex]

p3thresh   &  background spatial outlier rejection threshold           &   2.5, 3.0     &  2.5      \\[0.5ex]

spec\_hw   &  spectral extraction half-width                           &   1, 2, 4, 5     &    4      \\[0.5ex]

p7thresh   &  spectral extraction spatial outlier rejection threshold  &   10, 15       &   10      \\[0.5ex]

sigma      &  rolling median outlier threshold                         &     3, 4       &    3      \\[0.5ex]
\hline

\label{tab:S3_parametric_study}
\end{tabular}\\
Note: All parameters are those for \texttt{Eureka!\hspace{0.2em}}Stage 3, except for `sigma', which is a Stage 4 parameter. 
\end{center}
\end{table*}

\begin{figure*}
    \centering
    \includegraphics[width=1.00\textwidth]{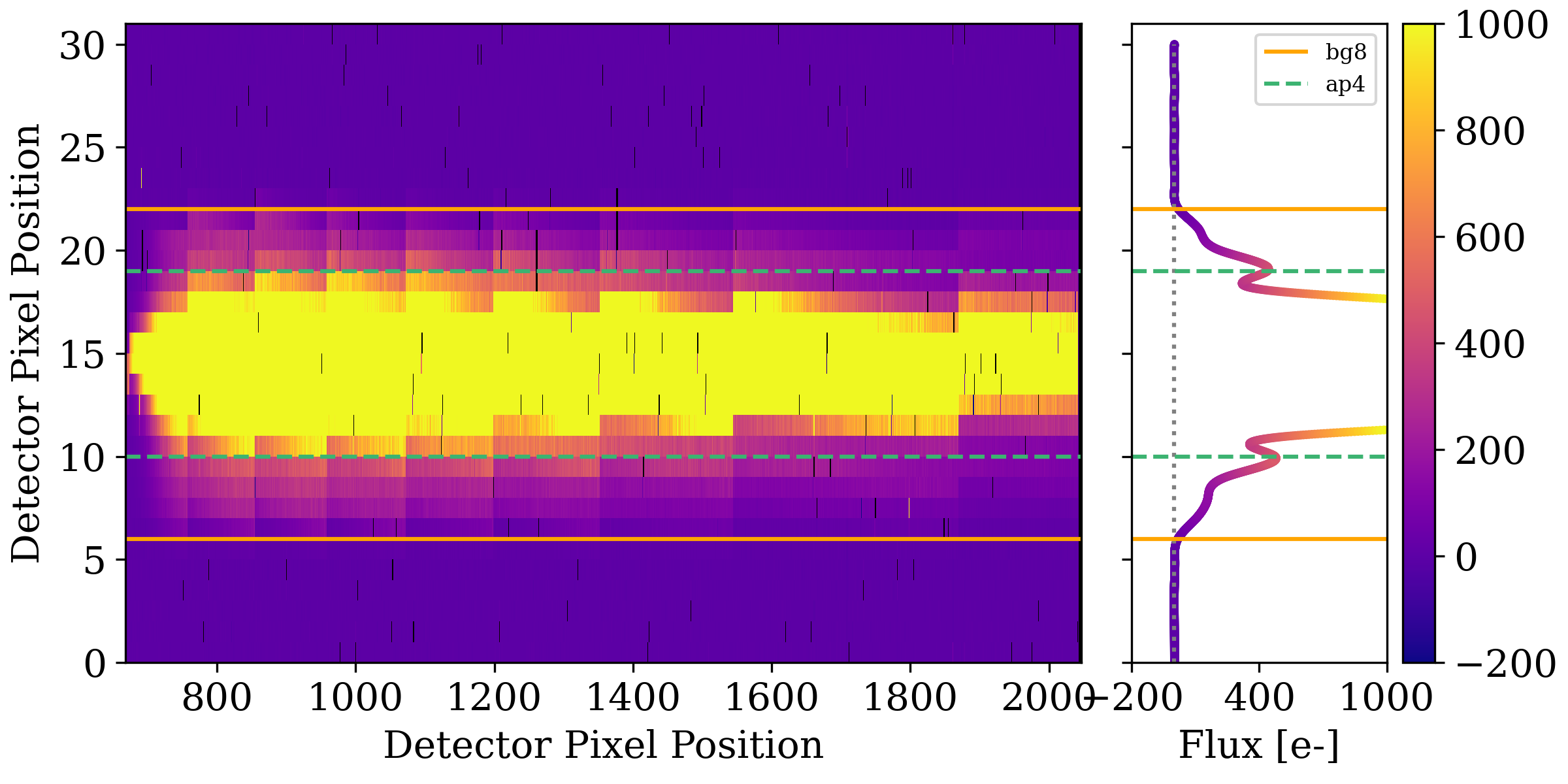}
    \caption{The aligned spectral trace and residual background are shown as part of the \texttt{Eureka!\hspace{0.2em}}Stage 3 reduction of the NIRSpec dataset. The background region is outside the bg8 lines (solid orange), and the spectral extraction aperture is inside the ap4 lines (dashed green). Outlier pixels can be seen with certain bad pixels identified via the CRDS shown in black. The color bar on the far right is a one-to-one mapping of the flux value (x-axis) in the right panel.}
    \label{fig:3304}
\end{figure*}

We used the frames corrected for curvature as inputs to the column-by-column background subtraction step. Here, we subtracted a median value from each column of these frames, with the background region (\texttt{bg\_hw} $= 8$ pixels) shown in Figure~\ref{fig:3304}. We also applied a two-step outlier removal process in both time and space, with a $10\sigma$ double-iteration rejection threshold for temporal outliers and a $2.5\sigma$ threshold for spatial outliers to eliminate the effects of cosmic ray events and prevent distortion from bad pixels in the background correction. 

\begin{figure*}
    \centering
    \includegraphics[width=1.00\textwidth]{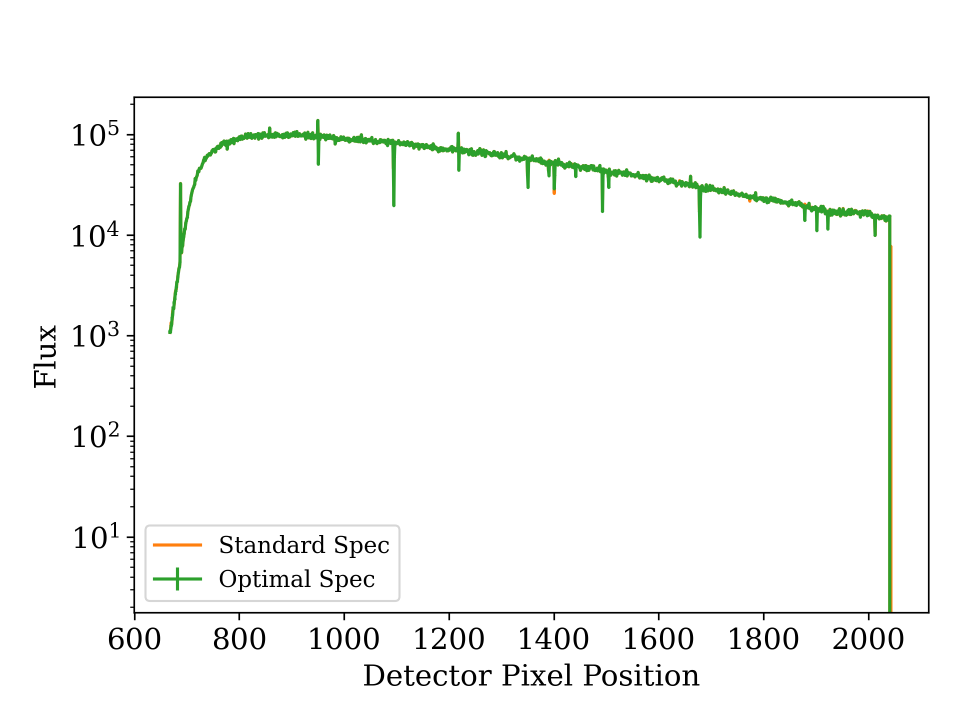}
    \caption{1-D spectrum, with flux versus absolute pixel position for an arbitrarily chosen early integration (in this case Integration 4) in our NIRSpec dataset. We use the ``Optimal'' spectrum \citep{Horne1986} shown in green, although it is almost an exact match to the Standard spectrum, shown in orange. The vertical spikes along the spectral profile indicate suspect pixel columns that were, in some cases, masked in Stage 4.}
    \label{fig:3302}
\end{figure*}

We constructed the median frame for Optimal spectral extraction using an outlier rejection threshold of $5\sigma$ (\texttt{median\_thresh}, default). Our spectral extraction half-width (\texttt{spec\_hw}) was 4 pixels on each side of the central pixel, for a total extraction width of 9 pixels. Figure~\ref{fig:3304} shows the aligned spectral trace and residual background, with the boundaries outlined. Following step 7 of Optimal spectral extraction described in \citet{Horne1986}, we used a spatial outlier rejection threshold (\texttt{p7thresh}) of $10\sigma$ (Table~\ref{tab:S3_parametric_study}). No spectral smoothing was applied to the Optimal extraction profile. Figure~\ref{fig:3302} shows an optimally extracted stellar spectrum for an arbitrary single integration.

\subsection{\texttt{Eureka!\hspace{0.2em}}Stage 4: Spectroscopic light curves}\label{subsubsec:NIRSpec_stage4}

Using outputs from the previous stage, \texttt{Eureka!\hspace{0.2em}}Stage 4 produces spectroscopic light curves by binning the time series of 1D spectra by wavelength. In addition, \texttt{Eureka!\hspace{0.2em}}generates a broadband (``white'') light curve. 

Based on careful consideration of the flux level for each pixel across the spectral trace, and, in particular, evidence of bad pixel columns in the 2D light curves generated in Stage 3, we manually constructed a pixel column mask for those situations where the existing CRDS Data Quality (bad pixel) mask or other unmasked outlier pixels could have significantly impacted the fit and level of the inferred peak flux along a column. The final mask included 10 columns ($\sim 0.8\%$ of active columns). Our goal in building this mask was to improve the results for the affected wavelength channels, while avoiding biases that could distort the final transmission spectrum.

The \texttt{Eureka!\hspace{0.2em}}package also performs sigma clipping of outlier integrations for each light curve at this stage. For example, in the case of a fixed spectral resolving power ($R =100$), we extracted the flux from $2.80$ to $5.17\ \mu$m, splitting the light into $62$ spectroscopic channels with bin width increasing in step with wavelength for equal resolution bins. For each light curve, we applied a box-car filter of $5$ integrations in length (\texttt{box\_width}, default) and clipped any $3\sigma$ outliers (\texttt{sigma}, see Table~\ref{tab:S3_parametric_study}). We performed $20$ iterations of sigma clipping (\texttt{maxiters}, default) in this manner. This procedure removed less than $0.5\%$ of integrations for any given light curve.

We ran three full (pre-binned) data reduction cases for each instrument. First, we fixed the bin size for NIRSpec to 10 pixels (corresponding to 18.6 pixel bins for NIRISS) and a bin width $\sim 0.01797 \ \mu$m, for $132$ bins, and $R \sim 221$ at $3.95\ \mu$m. The purpose of this case was to provide a consistent set of data matched to a similarly binned (in wavelength) case for NIRISS to move forward into our atmospheric analysis phase. Second, we ran a full reduction for $R = 100$, which resulted in a total of $62$ bins with bin size varying with wavelength. Finally, we ran a data reduction case with a fixed bin size of $\sim 2.2$ pixels (bin width $\sim 0.003954\ \mu$m), for $600$ bins, and $R \sim 1000$ at $3.95\ \mu$m. $\sim 2.2$ pixels is the Nyquist sampling limit for NIRSpec G395M \citep{jakobsenetal2022} and thus the minimum spectral resolution element\footnote{\url{https://jwst-docs.stsci.edu/jwst-near-infrared-spectrograph/nirspec-instrumentation/nirspec-dispersers-and-filters}}. 

{\ron Because we binned to the highest instrument resolution (i.e., the Nyquist limit), going to a smaller bin width (e.g., $1$ pixel) would not provide meaningful higher resolution information. However, for our NIRISS reduction we chose to go down to the $1$ pixel level. Although this was beyond the $\sim 2$ pixel Nyquist limit for NIRISS \citep{albertetal2023}, we felt that this would leave no doubt that we had captured all of the available information in the signal. (See Section~\ref{subsubsec:NIRISS_Stages456} for more information about the NIRISS binning scheme.)}

\citet{Mayetal2023} have argued that using pre-binned data at the working resolution of the transmission spectral analysis is the best approach, rather than using the high-resolution data and separately binning it down to the working resolution (i.e., post-binning). We chose to adopt this approach and use the pre-binned constant-wavelength-bin ($\sim 0.01797 \ \mu$m) spectra for nearly all our downstream analysis.

\subsection{\texttt{Eureka!\hspace{0.2em}}Stages 5 {\ron and 6}: Light curve fitting {\ron and transmission spectrum plotting}}\label{subsubsec:NIRSpec_stage56}

Stage 5 of \texttt{Eureka!\hspace{0.2em}}is generally concerned with fitting light curves, both broadband and narrowband (spectroscopic). The focus is on determining the value of the transit depth with errors, across the usable wavelength range. 

The orbital parameters of WASP-166b (see Table~\ref{tab:system_params}), period (\textit{P}), inclination ($i_p$), semi-major axis (\textit{$a/R_*$}), eccentricity (\textit{e}), and argument of periapsis (\textit{$\omega$}), taken from recent literature were calculated from a very high-quality dataset based on multiple transits, with multiple instruments, both ground- and space-based \citep{Hellieretal2019,Doyleetal2022}.

Throughout our analysis of the white-light and spectroscopic light curves, we fit for the coefficients of our limb darkening model ($u_1$, $u_2$), following the guidance of \citet{EspinozaJordan2016}. We chose the \texttt{Eureka!\hspace{0.2em}}implementation of the widely used quadratic limb darkening model first described by \citet{Kopal1950} and further discussed by \citet{EspinozaJordan2015}.

For all light curve fits, we kept the following as additional free parameters: the planet-star radius ratio ($R_p$/$R_*$), a constant and linear photometric polynomial coefficient ($c_0$, $c_1$), the centroid decorrelation parameters (coefficients for linear decorrelation against drift/jitter and PSF width in the spatial direction, $ypos$, $ywidth$), and the scatter multiple ($scatter\_mult$), a multiplier to rescale errors and produce a reduced chi-squared value of $\sim1.0$. Values for the fixed parameters, and prior bounds for the free parameters are shown in Table~\ref{tab:wlc_fitting}.

\begin{figure*}
    \centering
    \includegraphics[width=1.00\textwidth]{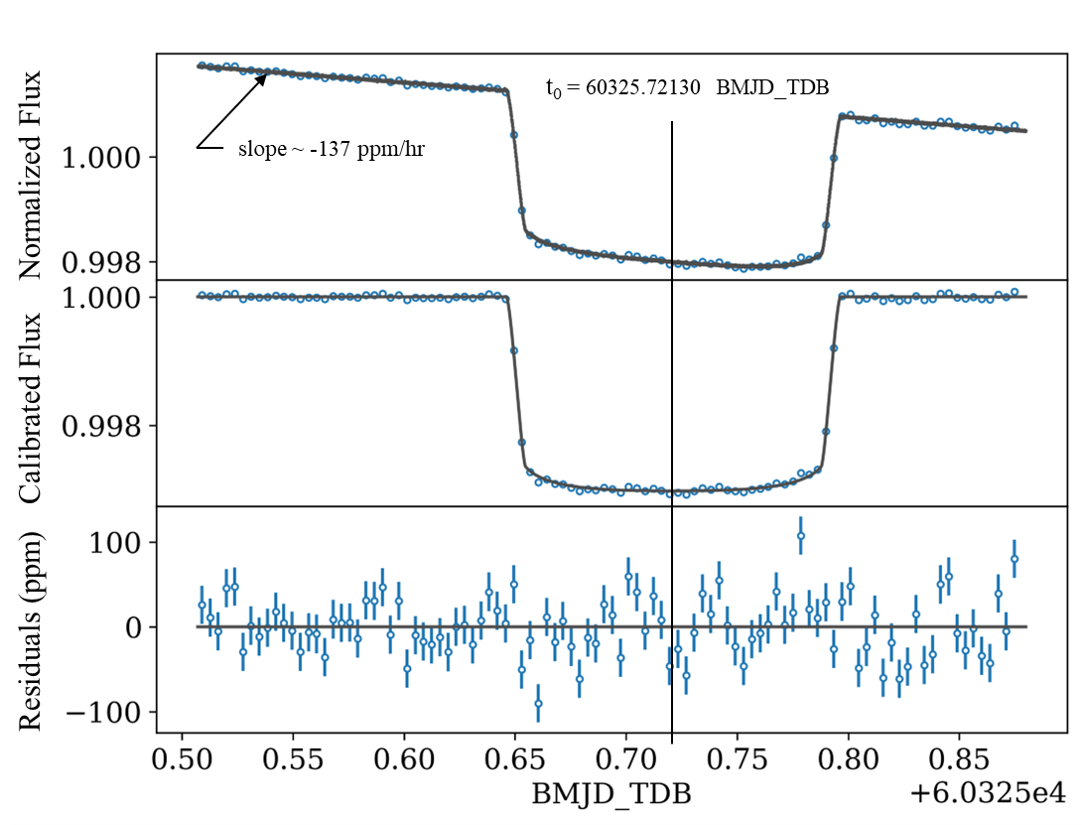}
    \caption{Broadband (white) light curve from our NIRSpec observation of WASP-166b, showing the fitted central transit time ($t_0$), and showing the exposure-long downward linear trend in normalized flux (-137 ppm/hr). We saw no other indication of unusual events or trends in the light curve.}
    \label{fig:5101}
\end{figure*}

We used \texttt{batman} \citep{Kreidbergetal2015} within the \texttt{Eureka!\hspace{0.2em}}framework to model the transit light curves (see Table~\ref{tab:wlc_fitting}). We fit the white light curve of our NIRSpec observation to determine the central transit time (see Figure~\ref{fig:5101}). The result was $60325.72130^{+0.00004}_{-0.00003}$ BMJD\_TDB (note uncertainty of $<$ 4 seconds) which we fixed to the median value for the analysis of our spectroscopic light curves. 

We also fit the white-light curve to refine the values for $i_p$ and $a/R_*$ (see Table~\ref{tab:wlc_fitting}). The $1\sigma$ bounds of these parameters from our fit fell within the $1\sigma$ bounds of the values determined by \citet{Hellieretal2019} and \citet{Doyleetal2022}, but were more tightly constrained. Again, we fixed the parameters $i_p$ and $a/R_*$ to the new fit median values for the analysis of our spectroscopic light curves.

\begin{table*}[htpb]
{\footnotesize
  	\caption{WASP-166b Light Curve Fitting Parameter Information. }         
  	\label{tab:wlc_fitting}
        \begin{flushleft}
  	\begin{tabular}{ l c c c c c }        
  	\hline\hline
   & & \multicolumn{2}{c}{NIRISS SOSS White Light Curve (WLC) } & \multicolumn{2}{c}{NIRSpec G395M WLC} \\ 
    \cmidrule(l){3-6}
    Parameters\tablenotemark{a} & Default WLC Prior\tablenotemark{b} &  Prior &  Fit\tablenotemark{c} &  Prior &  Fit\tablenotemark{c} \\
    \hline
    $R_p/R_\star$ &  0.05177$^{+0.00063}_{-0.00035}$\tablenotemark{d}  &  $\mathcal{N}(\rm{Default}, 0.005)$ & $0.053303^{+9.5\times 10^{-5}}_{-9.8\times 10^{-5}}$  & $\mathcal{N}(\rm{Default}, 0.01)$ & $0.053691^{+7.2 \times 10^{-5}}_{-7.1\times 10^{-5}}$ \\
    $P$ (days) & 5.44354215 (fixed)\tablenotemark{d}  & Default & Default & Default & Default \\
    $t_0$ (BMJD$_{\rm TDB}$) & \nodata & $\mathcal{N}(60309.39, 0.05)$ & $60309.390933^{+3.5\times 10^{-5}}_{-3.5\times 10^{-5}}$ & $\mathcal{N}(60325.7213, 4.0$$\times$$10^{-4})$ & $60325.72130^{+4.0 \times 10^{-5}}_{-3.0 \times 10^{-5}}$ \\
    $i$ (degrees) & 88.85$^{+0.74}_{-0.94}$\tablenotemark{d} & $\mathcal{N}(\rm{Default}, 3.0)$ & $87.78^{+0.13 }_{-0.12}$  & $\mathcal{N}(\rm{Default}, 0.94)$ & $88.39^{+0.16 }_{-0.15}$  \\    
    $a/R_{\star}$ & 11.83$^{+0.29}_{-0.68}$\tablenotemark{d} & $\mathcal{N}(\rm{Default}, 2.1)$  & $11.07^{+0.12}_{-0.11}$ &  $\mathcal{N}(\rm{Default}, 0.68)$ & $11.57^{+0.12}_{-0.11}$ \\
    $e$ & 0 (fixed)\tablenotemark{e} & Default & Default & Default & Default \\
    $\omega$ (degrees) & 90 (fixed)\tablenotemark{e} & Default & Default & Default & Default \\
    \\ [-2.5 ex]
    \hline \\ [-2.5 ex]
    \multicolumn{6}{l}{Quadratic Limb Darkening Coefficients}  \\
    \\ [-2.5 ex]
    \hline \\ [-2.5 ex]
    $u_{1}$ & $\mathcal{U}(0, 1)$  & Default & $0.149^{+0.030}_{-0.030}$ & Default & $0.044^{+0.026}_{-0.024}$ \\
    $u_{2}$ &  $\mathcal{U}(0, 1)$  & Default & $0.209^{+0.046}_{-0.046}$ & Default & $0.204^{+0.038}_{-0.041}$ \\
    \\ [-2.5 ex]
    \hline \\ [-2.5 ex]
    \multicolumn{6}{l}{Systematic Polynomial Coefficients}  \\
    \\ [-2.5 ex]
    \hline \\ [-2.5 ex]
    $c_{0}$ & \nodata  & $\mathcal{N}(1.0, 0.01)$ & $1.0010994^{+2.8\times 10^{-6} }_{-2.8\times 10^{-6} }$  & $\mathcal{N}(1.0, 0.01)$ & $1.0011134^{+3.0 \times 10^{-6} }_{-2.9 \times 10^{-6} }$ \\
    $c_{1}$ & \nodata  & $\mathcal{N}(0.0, 0.01)$ & $-0.000355^{+2.1\times 10^{-5} }_{-2.1\times 10^{-5} }$  & $\mathcal{N}(0.0, 0.01)$ & $-0.003297^{+2.2 \times 10^{-5} }_{-2.1 \times 10^{-5} }$ \\
    \\ [-2.5 ex]
    \hline
    \multicolumn{6}{l}{Centroid Decorrelation Parameters}  \\
    \\ [-2.5 ex]
    \hline \\ [-2.5 ex]
    ypos & \nodata  & \nodata & \nodata  & $\mathcal{N}(0.0, 0.10)$ & $-0.0029^{+1.0 \times 10^{-3} }_{-1.0 \times 10^{-3} }$ \\
    ywidth & \nodata  & \nodata & \nodata  & $\mathcal{N}(0.0, 10.0)$ & $-0.0015^{+2.0 \times 10^{-3} }_{-2.0 \times 10^{-3} }$ \\
    \\ [-2.5 ex]
    \hline
    \multicolumn{6}{l}{White Noise Parameter}  \\
    \\ [-2.5 ex]
    \hline \\ [-2.5 ex]
    scatter\_mult  & \nodata  & $\mathcal{N}(1.1, 1.0)$ & $0.3636^{+0.0037 }_{-0.0037  }$  & $\mathcal{N}(1.4, 0.4)$ & $1.737^{+0.013}_{-0.013}$ \\
    \\ [-2.5 ex]
    \hline
    \end{tabular}
    \tablenotetext{a}{Parameter definitions: $R_p/R_\star = $ planet radius in units of stellar radii; $P = $ orbital period; $t_0 = $ time of transit center, where BMJD$_{\rm TDB} =$ BJD$_{\rm TDB}$ - 2400000.5; $i = $ inclination; $a/R_{\star}= $ semi-major axis in units of stellar radii; $e = $ eccentricity; $\omega =$ argument of periastron; $u_{1}$ and $u_{2} = $ quadratic limb darkening coefficients; $c_{0}$ and $c_{1} = $ systematics polynomial coefficients (constant and linear, respectively); $ypos$ and $ywidth = $ linear decorrelation coefficients against the drift/jitter and against changes in the PSF width, respectively, for spatial direction of spectroscopic trace; $scatter\_mult = $ white noise parameter, a multiplier to the expected noise from Stage 3.}
    \tablenotetext{b}{For our prior distributions, $\mathcal{N}$($\mu$, $\sigma$) represents a normal distribution with mean $\mu$ and standard deviation $\sigma$; $\mathcal{U}$(a, b) represents a uniform distribution between a and b. In specifying priors for NIRISS and NIRSpec, the word \textit{Default} refers to either the default median value, fixed value, or prior specified in this column.}
    \tablenotetext{c}{We note the median WLC fit values for $t_0$, $i$, and $a/R_{\star}$ were set as fixed values for the spectroscopic light curve fits. However, for NIRSpec G395M, the median value for $i$ was rounded to $88.4$ deg. All other parameters used the same prior bounds used in the WLC fits with the exception of the scatter\_mult parameter for NIRSpec, which used a prior distribution of $\mathcal{N}(1.74, 0.75)$. Also, NIRISS did not use the Centroid Decorrelation Parameters.}
    \tablenotetext{d}{\citet{Doyleetal2022} values.}
    \tablenotetext{e}{\citet{Hellieretal2019} values.}
    \end{flushleft}
}
\end{table*}

In our inspection of the 2-D light curves generated in Stages 3 and 4, we did not see any evidence of failed or degraded integrations, saturated wavelength regions, disruptions caused by high gain antennae moves, mirror tilt, or any other large excursions in drift and jitter. As a result, we did not manually clip any integrations from the analysis in this stage.

Model fits were performed by the dynamic nested sampling package, \texttt{dynesty} \citep{Speagle2020} within the \texttt{Eureka!\hspace{0.2em}}pipeline framework and were run until convergence (using 2000 live points and a $\Delta ln Z$ evidence tolerance of $0.001$). Figure~\ref{fig:EurekaNIRSpecFitSpectroscopicLCs} depicts a sample of spectroscopic light curves spaced across the NIRSpec bandpass, overplotted with the corresponding \texttt{dynesty} fits.

We saw an indication of correlated noise in the spectroscopic light curves, but it is only significant in two channels ($4.166 $ to $ 4.184\ \mu$m, and $4.427 $ to $ 4.490\ \mu$m) for our fixed ($0.01797 \ \mu$m) bin width reduction.\footnote{This assessment is based on inspection of Allan deviation plots, a diagnostic tool generated by \texttt{Eureka!\hspace{0.2em}}that shows the level of RMS error in the spectroscopic light curves for increasing time averaging bins ($N$). They show the deviation from white (random) $\sqrt{1/N}$ noise and give an indication of whether (or not) correlated noise is present.} It may be a result of 1/f noise, the modeling of systematics, or some other unidentified source, but further investigation would be needed to isolate the root cause (a similar situation with NIRSpec Prism was described by \citealt{Sarkaretal2024}).

We saw evidence of an exposure-long slope ($-137 \pm 1$ ppm per hour) in the normalized transit light curves (Figure~\ref{fig:5101}). This is very similar behavior to what was seen with the NRS1 detector during the NIRSpec commissioning campaign reported by \citet{Espinozaetal2023}. Evidence points to this being a detector-level effect. As of this writing, STScI has not determined the source of this trend, and is continuing to investigate. This slope is removed/flattened in the calibration process of Stage 5 and does not appear to affect our results.

We note the possible presence of periodic behavior at approximately $P = 0.035$ d, which we identified visually in the NIRSpec white light curve. We did not see evidence of this behavior in the NIRISS SOSS white light curve. We constructed a Lomb-Scargle periodogram that confirmed the most substantial sub-day periodicity is around $0.035$ d, but this behavior falls well short of even a $10\%$ false alarm probability, the lowest FAP threshold we explored, suggesting that this periodicity is likely due to random statistical noise.

One final possible source of concern for the reliability of our reduction is the ``transit light source effect''. The stellar surface along the transit chord that the planet passes through may have a different abundance and arrangement of starspots than the full stellar disk; therefore, using the disk-integrated stellar spectrum as a baseline to subtract from the spectra collected during transit can introduce biases into the inferred planetary transmission spectrum \citep{rackhametal2018}. However, this effect is most pronounced for M dwarfs and becomes less significant with larger stellar mass. {\ron \citet{rackhametal2019} found that for an F-type star such as WASP-166, the expected spot covering fraction is only $\sim 0.1\%$. They further noted that because the transit light source effect becomes more pronounced at shorter wavelengths, there is a concern with F dwarfs that unocculted faculae can cause stellar contamination in the transmission spectrum at UV wavelengths. Fortunately, our \textit{JWST} observations did not cover UV wavelengths. UV transit observations of WASP-166 and WASP-166b have recently been collected with \textit{HST} as part of the HUSTLE treasury program (ID 17183, Alam et al. in prep) which may be able to detect and better constrain the presence of any stellar contamination and its impact on transmission spectroscopy.}

{\ron In spite of the various systematic effects and noise sources that have been discussed here, we still} showed excellent spectral precision with our reduced NIRSpec data. For the case of a constant 10 pixel bin size, we measured transit depth errors from 36 to 111 ppm over the wavelength range. A simulation using \texttt{PandExo} \citep{batalhaetal2017} with this binning shows from 3\% to 12\% better performance than we show; however, according to \citet{Espinozaetal2023}, \texttt{PandExo} likely is underestimating the actual errors by as much as $20$\%. Our realized precision exceeds the adjusted \texttt{PandExo} results.

Stage 6, the final stage of \texttt{Eureka!\hspace{0.1em}}, creates and displays the planet transmission spectrum in figure and table form using results from Stage 5.

\section{NIRISS SOSS Data Reduction Details}\label{appB_soss_reduction}

\subsection{\texttt{Ahsoka} Stage 1: Detector-level processing}\label{subsubsec:NIRISS_Stage1}

We began by downloading the \textit{uncal.fits} (uncalibrated, pixel-level data) files from MAST, and then performed the following \texttt{JWST} \texttt{Science Calibration Pipeline}\footnote{\url{https://jwst-pipeline.readthedocs.io/en/latest/jwst/user_documentation/introduction.html}} Stage 1 detector-level\footnote{\texttt{calwebb\_detector1}, see \url{https://jwst-pipeline.readthedocs.io/en/stable/jwst/pipeline/calwebb_detector1.html}} steps: \texttt{dq\_init, saturation, superbias,} and \texttt{refpix.} For this initial stage of the NIRISS data reduction we used the \textit{JWST} \texttt{Science Calibration Pipeline} (version 1.11.4) and the \textit{JWST} Science Data Processing (SDP) subsystem (version 2023\_3b).

We next applied the {\ron \texttt{exoTEDRF} (formerly, \texttt{supreme-SPOON})} group-level background subtraction and 1/f noise removal steps \citep{radicaetal2023,Radica2024JOSS}. The 1/f noise \citep[see, e.g.,][]{rauscher2014,schlawin2020} is introduced during detector readout. As explained by \citet{albertetal2023} and \citet{radicaetal2023}, 1/f noise is one of the last noise sources affecting \textit{JWST} near infrared (NIR) detector data, and should therefore be one of the first noise sources removed, which in turn requires removal at the group level. During the 1/f noise removal step, sources of flux that could bias our results must be masked or removed, which include field star contaminants, target star flux, and Zodiacal light. The location of field star contaminants can be masked using the F277W exposure, while target star flux is masked by using the spectral trace.\footnote{See Section \ref{subsubsec:NIRISS_Stage3} for a description of our procedures determining spectral trace location and width.} 

The {\ron \texttt{exoTEDRF}} background subtraction step operates by scaling the STScI background model\footnote{See SOSS Background Observations at \href{https://jwst-docs.stsci.edu/jwst-near-infrared-imager-and-slitless-spectrograph/niriss-observing-strategies/niriss-soss-recommended-strategies\#NIRISSSOSSRecommendedStrategies-SOSSBackgroundObservations}{https://jwst-docs.stsci.edu}.} to group-level median frames of our observations. The 2 median frames (one for each group) are created from the out-of-transit integrations for each of the 2 groups \citep{radicaetal2023,Radica2024JOSS}. Scaling was performed in a small region in the lower left of our median frames unaffected by other flux sources, located at pixel locations: x $\in$ [5,401], y $\in$ [5,21]. In applying this step to our data, we found that Zodiacal light subtraction was unnecessary in the Stage 1 group-level integrations. (The \texttt{exoTEDRF} background subtraction step yielded a scale factor of 0 for both group-level median frames.) 
In other words, our bright (J$\sim$8.35 mag) target star WASP-166 overwhelms any contributing background sources.

\begin{figure*}
    \centering
    \includegraphics[width=1.00\textwidth]{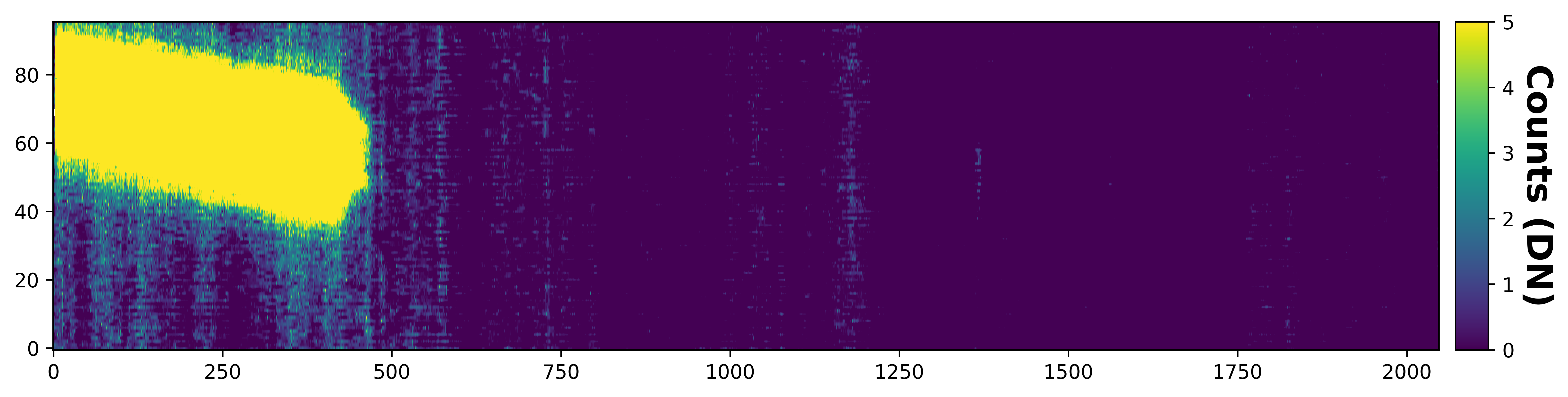}
    \caption{ Median image of our NIRISS SOSS F277W integrations. The F277W filter blocks those wavelengths $\lesssim$2.6 $\mu$m, and thus allows only the longest wavelengths of the order 1 spectrum to be dispersed upon the subarray. These longest wavelengths are visible left of column 500. The F277W exposure may be used to find the locations of any order 0 field star contaminants, which would appear as bright chevron shapes if present. Our observations have no significant order 0 contaminants.}
    \label{fig:NIRISS_F277W}
\end{figure*}

We next applied the {\ron \texttt{exoTEDRF} (formerly, \texttt{supreme-SPOON})} group-level 1/f noise subtraction algorithm, masking any field star contaminants (negligible for our data, see Figure~\ref{fig:NIRISS_F277W}) and spectral traces. Here, the noise-weighted average of each column in the group-level median frames is computed, and this is subtracted from each column of the raw image frames for each group. The final step in the 1/f noise subtraction algorithm is to add the background noise back into the image frames to ensure that additional steps in the \texttt{Ahsoka} pipeline are applied to the as-observed astrophysical images.\footnote{Refer to \cite{radicaetal2023} for additional details.} 

The F277W filter blocks those wavelengths $\lesssim$2.6 $\mu$m, and thus allows only the longest wavelengths of the order 1 spectrum to be dispersed upon the subarray. Additionally, the F277W exposure may be used to find the locations of any order 0 field star contaminants, since
order 0 field star contaminants only appear at column pixel indices higher than $\sim$700 \citep{albertetal2023}. The median image of our F277W integrations in Figure~\ref{fig:NIRISS_F277W} indicates that our observations have no significant order 0 contaminants.

The \texttt{Ahsoka} detector-level reduction concludes with the following \texttt{jwst} pipeline steps: \texttt{linearity, jump, ramp\_fitting,} and \texttt{gain\_scale.} Since our NIRISS SOSS data only included 2 groups, the \texttt{jump} step was automatically skipped, as discussed in Section \ref{subsubsec:NIRSpec_stage1}.

\subsection{\texttt{Ahsoka} Stage 2: Spectroscopic processing}\label{subsubsec:NIRISS_Stage2}

We next applied Stage 2 spectroscopic processing\footnote{\texttt{calwebb\_spec2}, see \url{https://jwst-pipeline.readthedocs.io/en/latest/jwst/pipeline/calwebb_spec2.html}} to our Stage 1 output files, which are similar to the \textit{rateints.fits} files used as inputs to our NIRSpec Stage 2 processing (see Section~\ref{subsubsec:NIRSpec_stage2}). We began with the following \texttt{jwst} pipeline steps: \texttt{assign\_wcs, srctype,} and \texttt{ flat\_field.} We then again employed the {\ron \texttt{exoTEDRF}} background subtraction algorithm, using the same procedures and scaling region as with the detector-level process, except that only one median frame need be constructed during Stage 2. (Here, the computed background scaling factor was $0.76582$.)

We concluded Stage 2 with the {\ron \texttt{exoTEDRF} (formerly, \texttt{supreme-SPOON})} \texttt{BadPix} custom cleaning step to flag and correct outlying/hot pixels \citep{radicaetal2023,Radica2024JOSS}. The \texttt{BadPix} step first creates a median frame using the out-of-transit integrations from the background subtraction step. Then, each pixel of the median frame is compared to surrounding pixels. Any pixel with a NaN or negative value, or that differs from surrounding pixels by more than $5\sigma$, is flagged. A mask records the locations of the flagged pixels on the NIRISS SOSS subarray, and flagged pixels are then replaced by the median value of surrounding pixels. Finally, the outlying/hot pixels (indicated by the mask) in each integration frame from the Stage 2 background subtraction step are replaced by the corresponding pixel values on the corrected median frame, which is scaled to the transit white light curve.

\begin{figure*}
    \centering
    \includegraphics[width=1.00\textwidth]{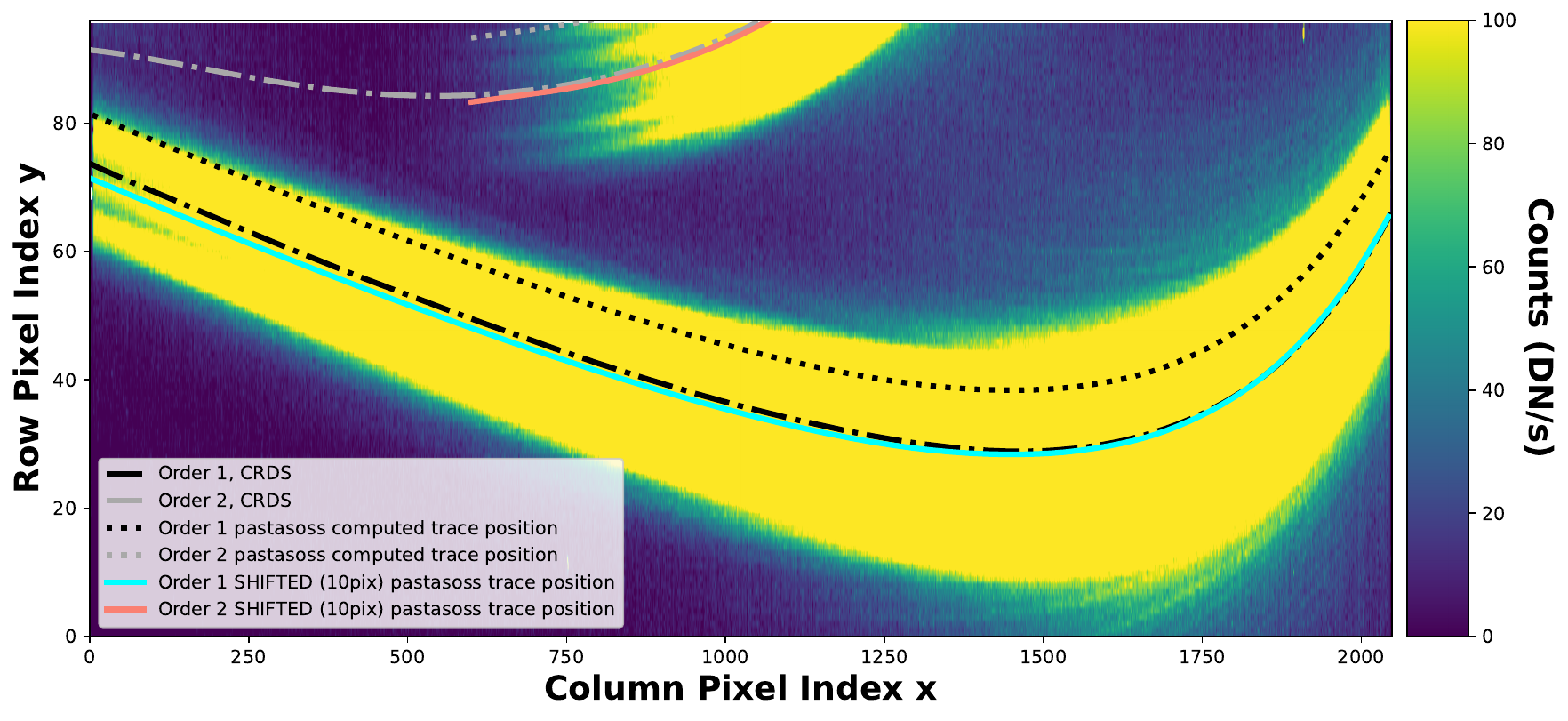}
    \caption{Comparison of the NIRISS SOSS CRDS spectral trace, the \texttt{PASTASOSS}-derived spectral trace, and the \texttt{PASTASOSS} trace shifted by 10 pixels, all overplotted on our \texttt{BadPix} output frame from integration 0. A portion of Order 2 is visible near the top center of SUBSTRIP96. The \texttt{PASTASOSS} trace shifted vertically by 10 pixels aligns best with our data. }
    \label{fig:NIRISStracesWithBadPixData}
\end{figure*}

\subsection{\texttt{Ahsoka} Stage 3: Spectral extraction}\label{subsubsec:NIRISS_Stage3}

The \texttt{Ahsoka} pipeline described by \cite{Louie_2025} produces a time series of 1D (flux versus wavelength) stellar spectra by applying the box extraction algorithm from \texttt{nirHiss} to the \texttt{BadPix} step output frames derived in Stage 2. \texttt{Ahsoka} employs the STScI-developed \texttt{PASTASOSS} package\footnote{https://github.com/spacetelescope/pastasoss} \citep{baines2023traces,baines2023wavelength} to determine the spectral trace position and wavelength solution. During observations, both the NIRISS SOSS spectral trace and wavelength solution are known to vary by a few pixels depending upon the precise position of the pupil wheel, which aligns the GR700XD grism with the optical path. The \texttt{PASTASOSS} package takes the pupil wheel position into account to determine the order 1 spectral trace position and wavelength solution to sub-pixel level accuracy. 

Direct application of both the \texttt{PASTASOSS} package and \texttt{nirHiss} box extraction algorithm to our SUBSTRIP96 data was problematic. Below, we describe in turn how we overcame problems encountered with these two algorithms.

The \texttt{PASTASOSS} package was developed and tested on the SUBSTRIP256 subarray. We found that the \texttt{PASTASOSS}-derived trace was not centered upon the spectral trace of our SUBSTRIP96 data. Upon further investigation, we discovered that the order 1 spectrum for SUBSTRIP96 is shifted 10 pixels vertically from the trace on SUBSTRIP256, as described by \citet{albertetal2023}. Figure~\ref{fig:NIRISStracesWithBadPixData} compares the CRDS spectral trace, the \texttt{PASTASOSS}-derived spectral trace, and the \texttt{PASTASOSS} trace shifted by 10 pixels, all overplotted on our \texttt{BadPix} output frame from integration 0. The \texttt{PASTASOSS} trace shifted by 10 pixels (as documented in \citealt{albertetal2023}) aligns best with our data, and we used that trace moving forward.

\begin{figure*}
    \centering
    \includegraphics[width=1.00\textwidth]{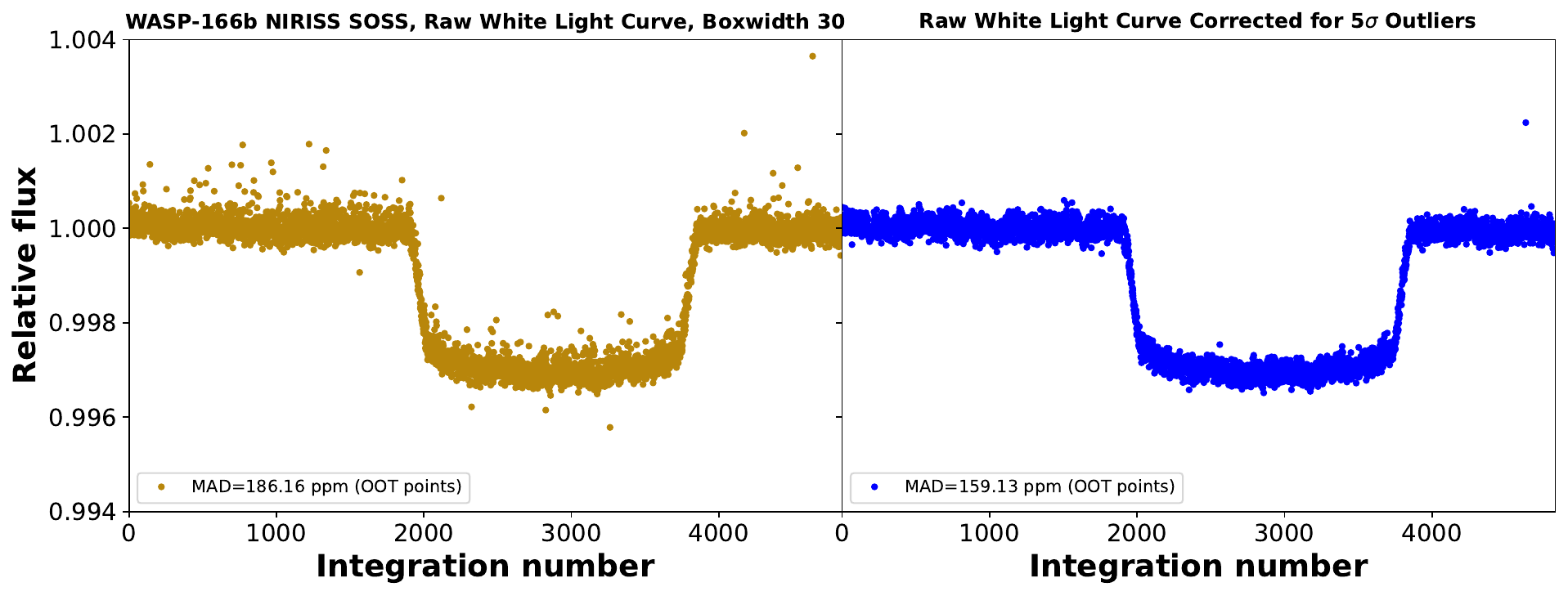}
    \caption{NIRISS SOSS raw white light curve for the 30 pixel width \texttt{nirHiss} box extraction before (\textbf{\underline{left}}) and after (\textbf{\underline{right}}) correcting the stellar spectra for $5\sigma$ outliers. We normalized the flux for each integration by the median flux of the out-of-transit integrations. For this computation, we used integration indices [0,1750] and [4000,4835] as the out-of-transit points. }
    \label{fig:NIRISSboxExtRawWhiteLight}
\end{figure*}

To apply the \texttt{nirHiss} box extraction algorithm to our data, we tested a variety of spectral extraction widths varying between 24 and 36 pixels, ultimately choosing a width of 30 pixels. To select the best extraction width, we computed the median absolute deviation for out-of-transit points in the raw white light curve produced from our data. We created the raw white light curve by summing the flux within the extraction width for all wavelengths, computing total flux for each of the 4836 integrations. We then normalized the flux for each integration by the median flux of the out-of-transit integrations. For this computation, we used integration indices [0,1750] and [4000,4835] as the out-of-transit points. Our extraction width of 30 pixels minimized out-of-transit median absolute deviation. Figure~\ref{fig:NIRISSboxExtRawWhiteLight} (left panel) shows the raw white light curve for the 30 pixel width box extraction.

\begin{figure*}
    \centering
    \includegraphics[width=1.00\textwidth]{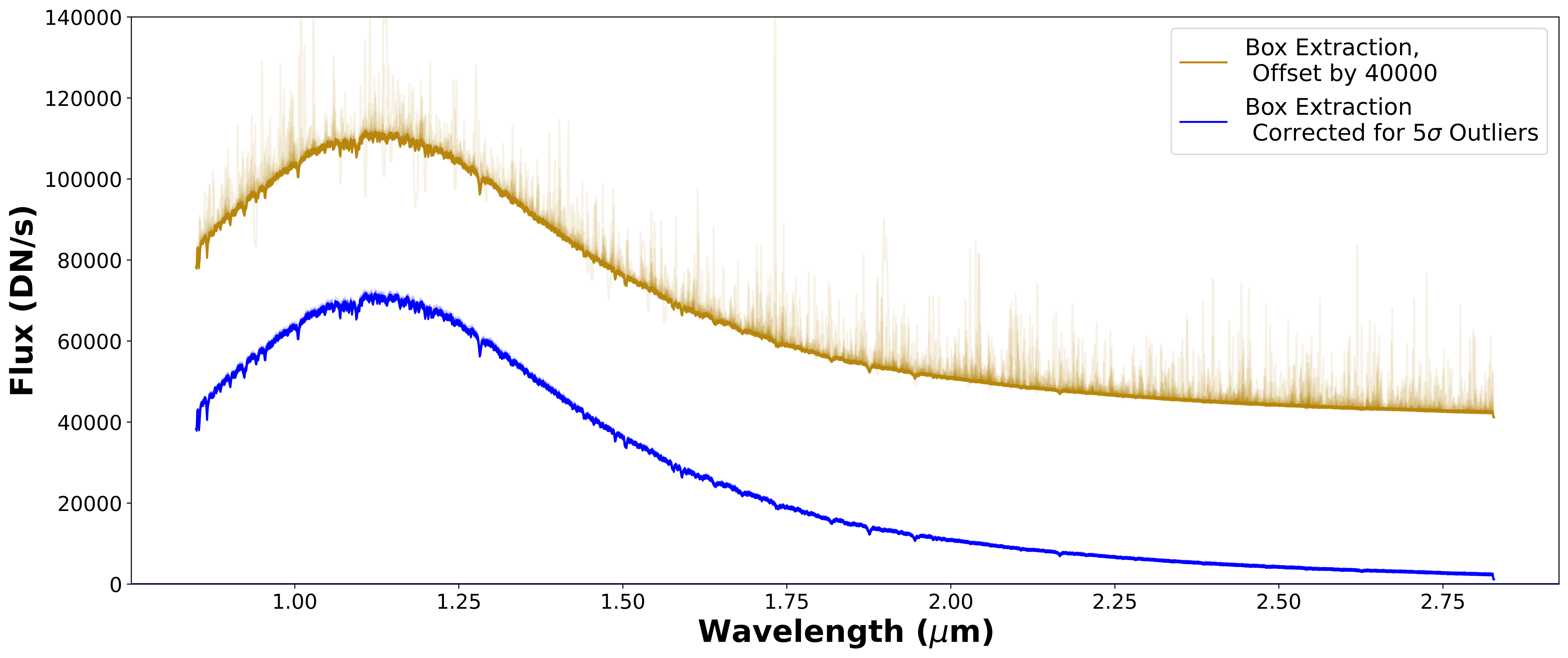}
    \caption{Comparison of NIRISS SOSS 1D stellar spectra (flux versus wavelength) before (top) and after (bottom) correcting for $5\sigma$ outliers. Stellar spectra for all integrations are overplotted, with each individual stellar spectrum plotted in partially transparent color, such that the most opaque region corresponds to the expected stellar spectrum, while the more transparent regions indicate outlying values of flux. An excessive number of outliers were evident in the stellar spectra from our initial box extraction (top). These outliers are no longer visible after our correction (bottom). }
    \label{fig:NIRISSboxCORRECTEDStellarSpectraAllInts}
\end{figure*}

Once the extraction width is determined, the \texttt{nirHiss} box extraction algorithm creates one stellar spectrum for each integration by summing the flux across the extraction width in each column. This provides a time series of 1D stellar spectra (flux versus wavelength). The extracted stellar spectra for all integrations are overplotted in Figure~\ref{fig:NIRISSboxCORRECTEDStellarSpectraAllInts} (top spectra). Examination of the plot reveals an excessive number of outliers in our stellar spectra. These outliers would normally be flagged during the \textit{JWST} pipeline \texttt{jump} step, which was automatically skipped for our data since we only have 2 groups.

To correct our stellar spectra, we removed $5\sigma$ outliers using a technique demonstrated by \cite{feinsteinetal2023}, \cite{Coulombe_2023NatureERSW18b}, and \cite{Gressier_2025} in their application of the \texttt{transitspectroscopy} pipeline \citep{espinoza_nestor_2022_6960924}. 
{\ron We began by following two steps to produce a \textit{median master spectrum.} First, for each integration, the stellar spectrum is \textit{normalized} by the median flux value across all wavelengths (all columns) for that integration. Then, the \textit{median master spectrum} is created by setting the flux value at each wavelength (in each column) equal to the \textit{normalized} median flux value for that wavelength across all \textit{normalized} stellar spectra (i.e., across all integrations). For each integration, the \textit{normalized} stellar spectrum is then compared to this \textit{median master spectrum} to identify $5\sigma$ outliers. Any identified $5\sigma$ outlier value is replaced by the corresponding \textit{median master spectrum} value at the corresponding wavelength (or column), which is re-scaled in the final corrected stellar spectrum by multiplying by the median flux value across all wavelengths (all columns) for that integration.}
{\ron Note that} we used the median absolute deviation at each wavelength multiplied by a statistical scale factor to estimate the associated standard deviation $\sigma$ for each column. 

In Figure~\ref{fig:NIRISSboxCORRECTEDStellarSpectraAllInts}, we compare the corrected stellar spectra for all integrations (bottom spectra) to the uncorrected box extraction (top, offset spectra). Figure~\ref{fig:NIRISSboxExtRawWhiteLight}, right panel, shows the white light curve derived from the corrected stellar spectra. \texttt{Eureka!\hspace{0.2em}}Stages 4, 5, and 6 are then applied to the corrected time series of 1D stellar spectra, as described in the next section.

\subsection{\texttt{Ahsoka} Stages 4, 5, 6: \texttt{Eureka!\hspace{0.2em}}application}\label{subsubsec:NIRISS_Stages456}

Our application of the \texttt{Eureka!\hspace{0.2em}}pipeline \citep[v0.10,][]{belletal2022} to NIRISS SOSS is similar to that described for NIRSpec G395M Stages 4, 5, and 6 (see Section~\ref{subsubsec:NIRSpec_stage4} and Section~\ref{subsubsec:NIRSpec_stage56}). In this section, we highlight the differences in our NIRISS SOSS analysis compared to NIRSpec.

We applied \texttt{Eureka!\hspace{0.2em}}Stage 4 to extract both white light and spectroscopic light curves from the NIRISS SOSS time series of 1D stellar spectra. We used three binning schemes across the NIRISS SOSS bandpass ($0.85$ to $2.81\ \mu$m) for the spectroscopic light curves. First, we created 109 constant 0.01797 $\mu$m width bins, which match the constant bin width used for NIRSpec G395M. Next, we created a constant $R = 100$ spectrum with 119 channels. For each light curve, we applied a box-car filter of 10 integrations in length and masked any $5\sigma$ outliers, performing 5 iterations of sigma clipping, masking no more than 0.2\% of all integrations for any one light curve. During \texttt{Eureka!\hspace{0.2em}}Stage 4, we exercised the option to compute quadratic limb darkening coefficients for each spectral channel using the \texttt{ExoTiC-LD} package\footnote{\url{https://exotic-ld.readthedocs.io/en/latest/}} \citep{david_grant_2022_ExoTiC-LD}, making use of the \citet{magic2015stagger} 3D stellar models. Finally, we ran a final full resolution data reduction case, where we fit light curves at the pixel-level (i.e., one fit per column on the detector).

We used \texttt{Eureka!\hspace{0.2em}}Stage 5 to fit both white light and spectroscopic light curves. Like our NIRSpec Stage 5 analysis, we modeled the transit light curves using the \texttt{batman} package \citep{Kreidbergetal2015}, used a polynomial model with constant ($c_{0}$) and linear ($c_{1}$) coefficients to model systematics, and employed a scatter multiplier to model white noise. Period (\textit{P}), eccentricity (\textit{e}), and argument of periapsis (\textit{$\omega$}) were fixed to the values shown in Table~\ref{tab:wlc_fitting} throughout our analysis. We used the \texttt{ExoTiC-LD} coefficients computed in Stage 4 as prior limb darkening values for both white light and spectroscopic light curves. We used the \texttt{emcee} \citep{Foreman-Mackey2013} affine invariant Markov Chain Monte Carlo package \citep{GoodmanWeare2010} within the \texttt{Eureka!\hspace{0.2em}}pipeline for NIRISS SOSS Stage 5 light curve fitting. As with the NIRSpec analysis, we did not manually clip any integrations, since we saw no evidence of failed or degraded integrations in the Stage 4 2D light curves.

\begin{figure}
    \includegraphics[width=0.9\textwidth]{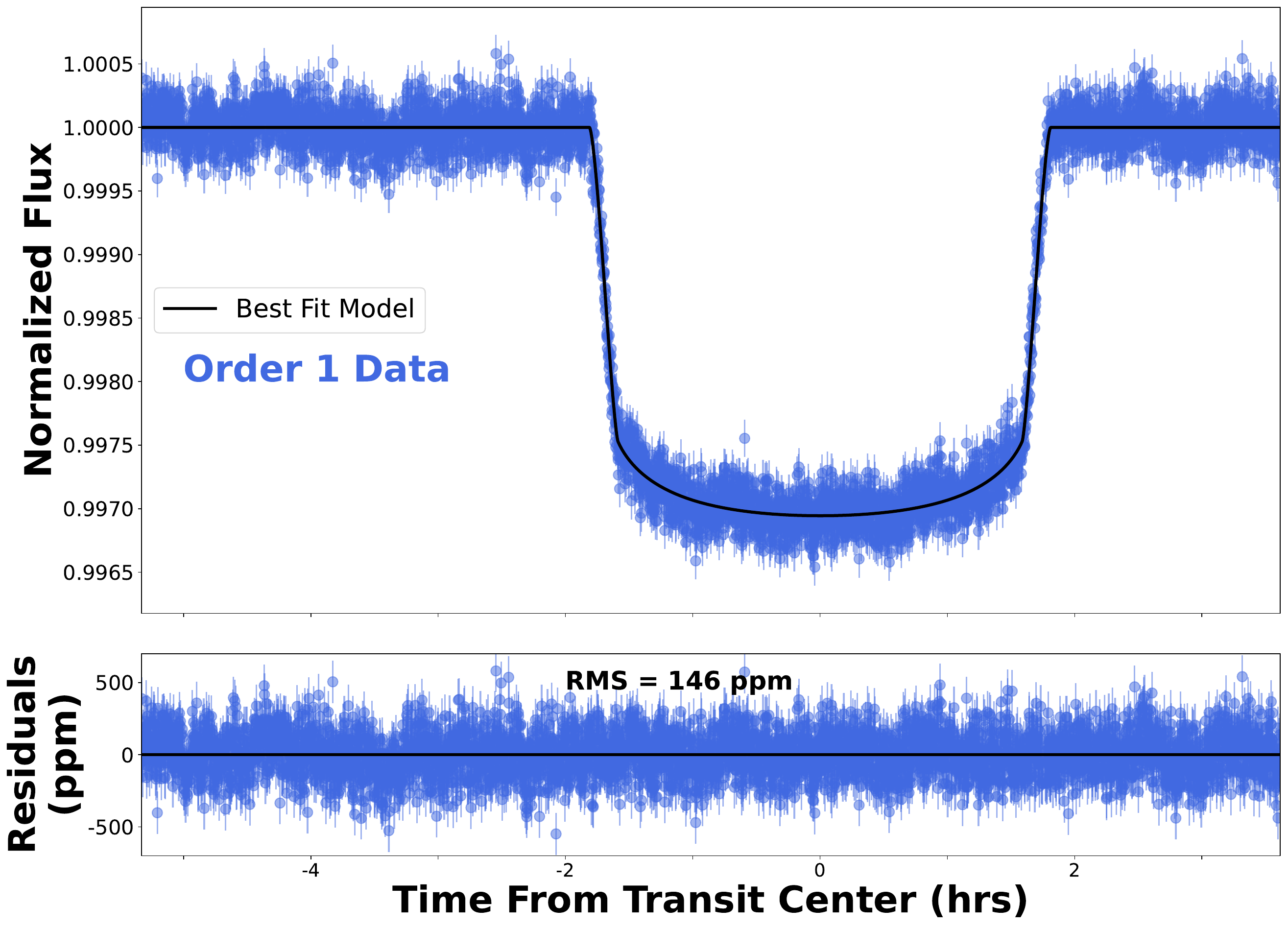}
    \includegraphics[width=0.9\textwidth]{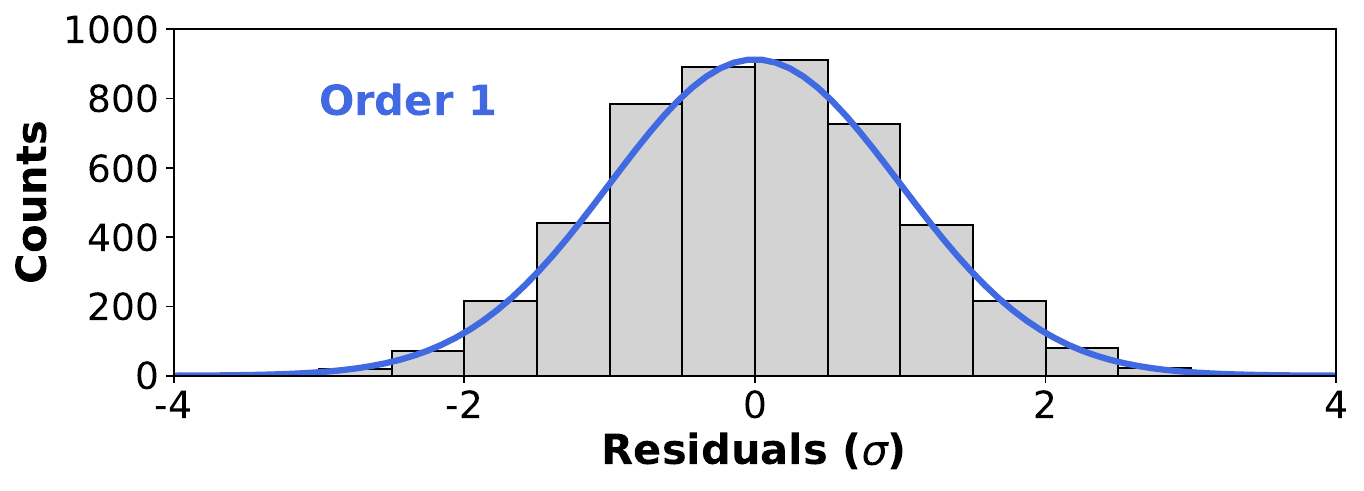}
    \caption{\texttt{Eureka!\hspace{0.2em}}\citep{belletal2022} Stage 5 white light curve \texttt{emcee} \citep{Foreman-Mackey2013} fit to the NIRISS SOSS \texttt{Ahsoka} data reduction. The \underline{\textbf{Top}} panel shows the Order 1 data corrected with the polynomial systematics model (blue points), and overplotted with the best fit transit model (black curve). The \underline{\textbf{Center}} panel indicates residuals with RMS scatter in black text. Mean error bars in the top and center plots are 146 ppm. The \underline{\textbf{Bottom}} panel shows the histogram distribution of residuals in terms of number of standard deviations $\sigma$ overplotted with a Gaussian curve of the same mean and standard deviation as the residuals. The Gaussian trend of our distribution indicates the systematics are well managed. 
    }
    \label{fig:EurekaFitWLCs}
\end{figure}

For white light curves, we fit for the planet radius ($R_p/R_*$), time of transit center ($t_0$), inclination (\textit{$i$}), and semi-major axis (\textit{$a/R_*$}), as well as quadratic limb darkening coefficients $u_{1}$ and $u_{2}$, systematic polynomial coefficients $c_{0}$ and $c_{1}$, and the white noise parameter using the prior distributions listed in Table~\ref{tab:wlc_fitting}. For our \texttt{emcee} fit, we ran 20000 steps, discarding the first 15000 as burn-in, using 500 walkers. From the white light curve we saw no unusual events or substantial systematic trends (e.g., saturation, mirror tilt, High Gain Antenna move disturbance, etc.). As can be computed from the $c_{1}$ fit in Table~\ref{tab:wlc_fitting}, the polynomial systematics model shows a minor slope of -15 ppm/hr across the transit. This effect was somewhat similar to what was seen with the NIRSpec G395M response, however an order of magnitude smaller. The source of this behavior is unknown at the time of this writing. Figure~\ref{fig:EurekaFitWLCs} shows the NIRISS SOSS white light curve overplotted with the \texttt{emcee} fit.

For spectroscopic light curves, we fit for $R_p/R_*$ on a normal distribution using as priors the median value from the white light curve fit, with a width of 0.005. We fit for quadratic limb darkening coefficients, systematic polynomial coefficients, and the white noise parameter using the same prior distributions as for our white light curve fits. We fixed $t_0$, \textit{$i$}, and \textit{$a/R_*$} to the median values from our white light curve fits (Table~\ref{tab:wlc_fitting}). For each of our \texttt{emcee} spectroscopic fits, we ran 7500 steps, discarding the first 1500 as burn-in, using 200 walkers. Figure~\ref{fig:EurekaFitSpectroscopicLCs} depicts a sample of spectroscopic light curves spaced across the NIRISS SOSS bandpass, overplotted with the corresponding \texttt{emcee} fits.

The key output product of \texttt{Eureka!\hspace{0.2em}}Stage 6 is a transmission spectrum, which can be displayed as a plot of transit depths with errors across wavelength, where wavelengths correspond to those covered by each spectral channel. \texttt{Eureka!\hspace{0.2em}}Stage 6 computes transit depths and corresponding errors using output values of $R_p/R_\star$ from Stage 5.

\section{Additional Comparative Atmospheric Retrievals with Free Chemistry} \label{ap:free_retrievals}
{\ron We present our primary Reference free chemistry retrieval (R1) and our Baseline free chemistry retrieval (B1) in Section~\ref{free_retrievals}. In this section, we explore several different approaches to analyzing the data to ensure the robustness of our free chemistry analysis. Unless otherwise stated, these retrievals also assume isothermal atmospheres and employ \texttt{POSEIDON} to analyze the combined NIRISS and NIRSpec data binned to equal bin widths ($0.01797 \ \mu$m).} 

{\ron First, in Section~\ref{Na_K_retrievals}, we discuss another Reference case (R2) that added Na and K to the constituents used in the B1 case. Then, we describe a Bayesian model comparison removing Na (R2a) and K (R2b) in turn. Next, in Section~\ref{CO_SO2_retrievals}, we present another Reference case (R3) that added SO$_2$ and CO to the B1 constituents. This allows us to make a Bayesian model comparison between the B1 case and the R3 case (with the combined SO$_2$+CO constituents). For both the Na and K study, and the SO$_2$+CO study, we find non-detections or inconclusive support for the presence of the constituents.}

{\ron The remaining sections present additional retrievals using the same atmospheric constituents as B1 (i.e., no CO, SO$_2$, Na, or K). In Section~\ref{Non-isothermal}, we discuss B2, which relaxes the assumption of an isothermal atmosphere and instead uses a gradient P-T profile. In Section~\ref{pos_single_instrument}, we return to an isothermal atmosphere and analyze the NIRISS and NIRSpec data independently. These retrievals are denoted as B3 (Section~\ref{pos_niriss}) and B4 (Section~\ref{pos_nirspec}), respectively. In Section~\ref{pos_r100}, we then assess the possible ramifications of our chosen binning by analyzing the full NIRISS and NIRSpec data set binned to a fixed resolution ($R = 100$) rather than fixed-width bins (B5). Next, in Section~\ref{TauREx_Retrieval}, we check for any dependence on our specific analysis framework by using the \texttt{TauREx} retrieval code instead of \texttt{POSEIDON} to complete retrieval B6. Finally, in Section~\ref{deck_haze_retrieval}, we perform retrieval B7, which tests the effects of using a more sophisticated cloud model that incorporates the effects of hazes as well as clouds.}

\subsection{{\ron \texttt{POSEIDON} Na and K Models and Bayesian Model Comparison}} \label{Na_K_retrievals}

\begin{figure*}[p]
    \centering
    \includegraphics[width=\textwidth]{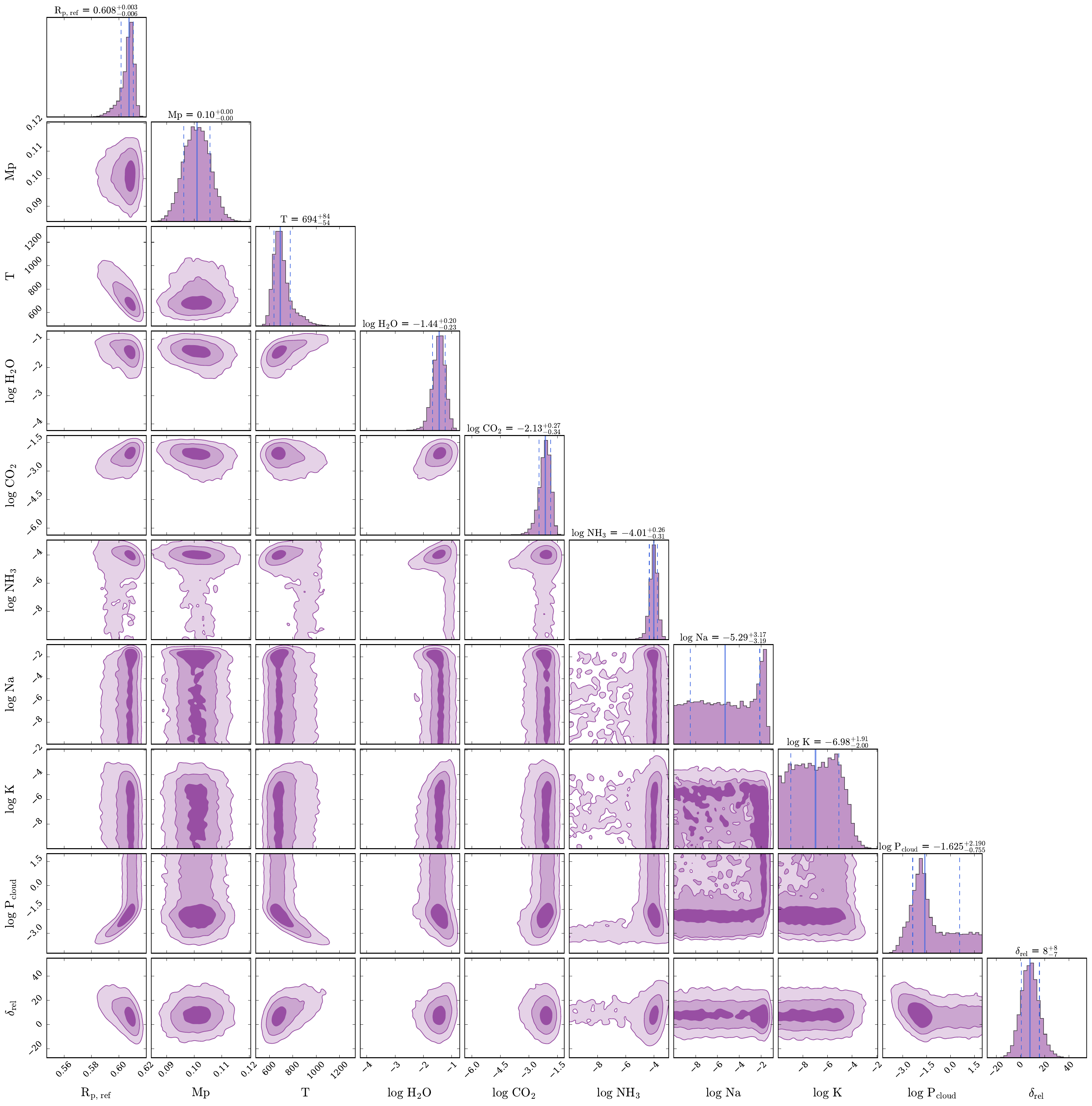}
    \caption{{\ron{Posterior distributions of free parameters (corner plot) for \texttt{POSEIDON} Reference retrieval case, R2, with Na and K added to the Baseline model. We see some faint hints of Na, but it is very poorly constrained.}}}
    \label{fig:POS_v83_R2_corner}
\end{figure*}

{\ron
In the region between $0.85$ and $1.2 \ \mu$m the extended, pressure broadened wings of the spectral features of both sodium (Na) and potassium (K) could potentially form part of the continuum underlying the water absorption features that we see (\citealt{welbanksetal2019}; Figure 1). However, given the cooler temperatures ($< 1000$ K), and lower pressures ($\sim10^{-3}$ bar) consistent with our transmission spectroscopy of WASP-166b, the opacity cross-sections and related absorption feature strengths are likely to be very small; cross-sections at $0.85 \ \mu$m are $\sim 9$ and $\sim 12$ orders of magnitude below core line peak values for K and Na, respectively (\citealt{welbanksetal2019}; Figure 1).

To explore whether or not the pressure broadened wings of Na and/or K are contributing to our observed spectrum, we conducted another Reference retrieval sequence. First, we ran a Reference retrieval (R2) that added Na and K to the Baseline (B1) constituents. The posterior distribution (see corner plot, Figure~\ref{fig:POS_v83_R2_corner}) from this retrieval showed hints of Na, but its abundance was very poorly constrained. Then we ran a retrieval (R2a) that only removed Na from the constituent list. Finally, we ran a retrieval (R2b) that only removed K. The results of this Bayesian model comparison are listed in Table~\ref{tab:BMC}. They show an inconclusive detection ($1.7\sigma$) of Na and a non-detection ($0.9\sigma$) of K. 
}

\subsection{{\ron\texttt{POSEIDON} CO and SO$_2$ Models and Bayesian Model Comparison}} \label{CO_SO2_retrievals}

\begin{figure*}[p]
    \centering
    \includegraphics[width=\textwidth]{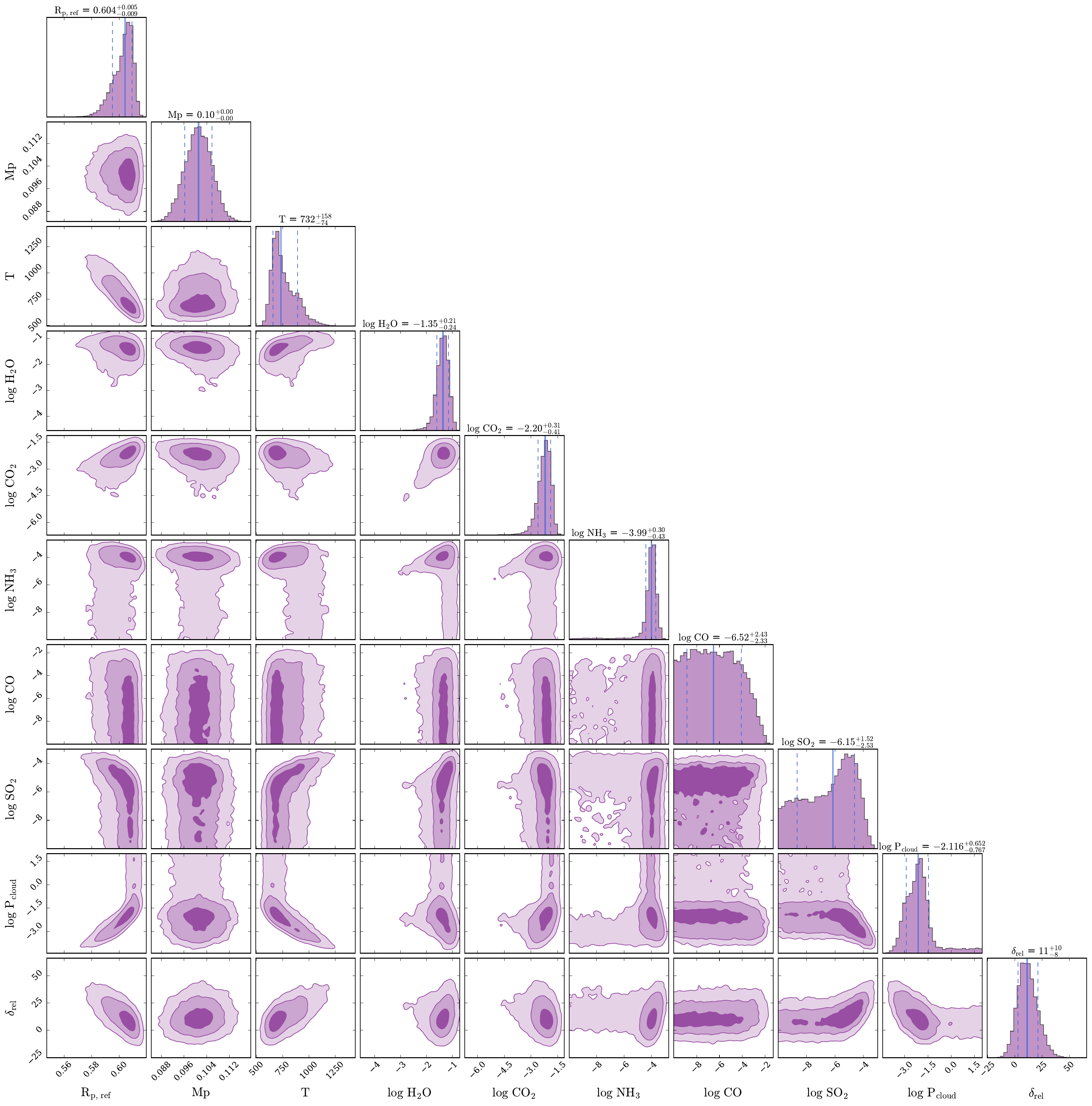}
    \caption{{\ron{Posterior distributions of free parameters (corner plot) for \texttt{POSEIDON} Reference retrieval case, R3, with CO and SO$_2$ added to the Baseline model. We see some hints of SO$_2$, but it is very poorly constrained.}}}
    \label{fig:POS_v81_R3_corner}
\end{figure*}

{\ron In our R1 retrieval (Section~\ref{free_retrievals}), the abundance of CO is very poorly constrained (see Figure~\ref{fig:POS_v62_R1_corner} and Table~\ref{tab:reference_retrieval_posteriors}), with a very small median value ($-6.36^{+2.35}_{-2.40}$ $\log_{10}$(vmr)). Likewise, the measured abundance of SO$_2$ is very small ($-5.69^{+1.37}_{-2.77}$ $\log_{10}$(vmr)) and poorly constrained. The detection significance (see Table~\ref{tab:BMC}) of CO alone is only $1.1\sigma$, and the detection significance of SO$_2$ alone is $1.5\sigma$; both are inconclusive detections.

To test whether adding CO+SO$_2$ in combination would show a higher detection significance, we ran another Reference retrieval case (R3), adding CO+SO$_2$ together to the Baseline (B1) case (see Figure~\ref{fig:POS_v81_R3_corner} and Table~\ref{tab:free_chem_posteriors}). The R3 free parameter posteriors showed very minor changes from the Baseline. Abundances for CO and SO$_2$ were both still poorly constrained and showed median values below 1 ppm ($\log_{10}$(vmr)). We compared this R3 case to the Baseline, effectively removing CO+SO$_2$. The result was $1.5\sigma$ (inconclusive) support for CO+SO$_2$ combined. There was no substantial difference in this detection significance value and the result using the primary R1 retrieval sequence (Table~\ref{tab:BMC}).
}

\subsection{\texttt{POSEIDON} (Non-isothermal Model) Retrieval} \label{Non-isothermal}

\begin{figure*}
    \centering
    \includegraphics[width=0.6\textwidth]{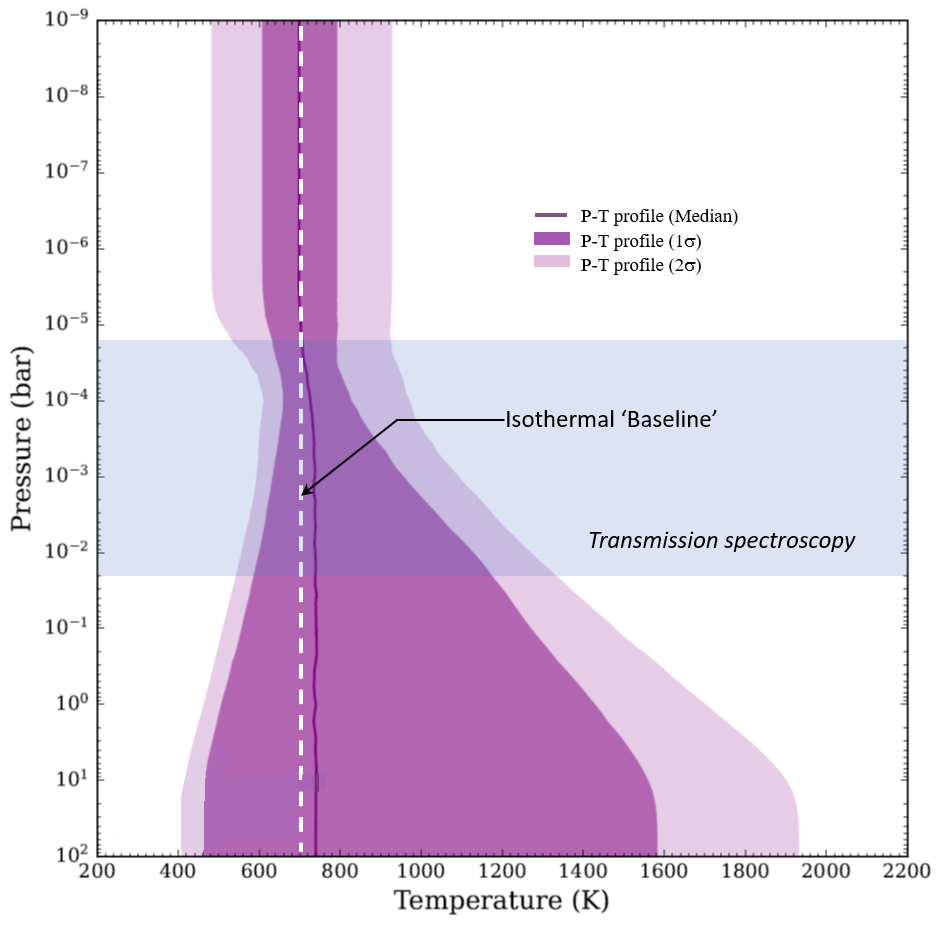}
    \caption{Retrieved P-T profile for the atmosphere of WASP-166b using the \texttt{POSEIDON} 2-parameter 'gradient' model (B2). The observational data for this case has been reduced to fixed $0.01797 \ \mu$m bins. As with the Baseline (B1) case, the fill gases, H$_2$ and He are present at the primordial solar ratio. Trace gas abundances for H$_2$O, CO$_2$, and NH$_3$, as well as cloud top pressure are free parameters in the retrieval. There is a retrieved transit depth offset between the instruments of $\sim 11$ ppm. The median temperature (dark purple line) is consistent with the Baseline isothermal temperature of $697$ K (white dashed line) at low pressures, but there is a small increase in temperature at pressures above $\sim 0.01$ mbar. The temperature is very poorly constrained in the deep atmosphere.}
    \label{fig:POS_Non_isothermal_Retrieval_v72_mod2}
\end{figure*}

For nearly all of our forward modeling and retrieval analyses, we have made the assumption that the atmosphere is isothermal over the full pressure range. Here we considered the case {\ron{(B2)}} of a non-isothermal atmosphere using the simple 2-parameter \texttt{POSEIDON} ``gradient'' P-T model. {\ron{The pressures associated with the high ($T_\mathrm{high}$) and deep ($T_\mathrm{deep}$) temperature points are $P_\mathrm{high} = 10^{-5}$ bar and $P_\mathrm{deep} = 10$ bar, respectively (\texttt{POSEIDON} default). These points serve as anchors outside the pressure range probed by low-resolution transmission spectroscopy. Figure~\ref{fig:POS_Non_isothermal_Retrieval_v72_mod2} shows the P-T profile for the full atmosphere grid from $10^{-9}$ bar to $10^{2}$ bar. \texttt{POSEIDON} smooths each profile, avoiding discontinuities in the temperature gradient at $P_\mathrm{high}$ and $P_\mathrm{deep}$ \citep{macdonaldlewis2022}.}} We set a broad prior range for both the deep atmosphere (high pressure) and high altitude (low pressure) temperature parameters (see prior ranges in Table~\ref{tab:retrieval_priors}). Otherwise this retrieval was the same as the Baseline {\ron{case (B1)}} described {\ron in Section~\ref{free_retrievals}}. The retrieved temperatures and other free parameter posteriors can be easily compared to the {\ron B1} values in Table~\ref{tab:free_chem_posteriors}.

As shown in our retrieved P-T profile (Figure~\ref{fig:POS_Non_isothermal_Retrieval_v72_mod2}) the median line is not far from isothermal, particularly in the pressure region probed by transmission spectroscopy. This region is generally considered to be in the range from $\sim 0.01$ to $100$ mbar \citep[e.g.,][]{Kemptonetal2014, macdonaldandmadhusudhan2017, Fortney2005, singetal2016}

The pressure region above $1$ mbar has a median temp of $\sim740$ K which is {\ron consistent ($0.43\sigma$) with} the {\ron B1} isothermal temperature ($697^{+101}_{-59}$ K). At lower pressures the temperature trends back to the same level as {\ron B1}. In general, the temperature is very poorly constrained, particularly in the deep atmosphere; this region is not probed by transmission spectroscopy. Further characterization of WASP-166b's P-T profile may await future emission spectroscopy observations.

\subsection{\texttt{POSEIDON} Single Instrument Retrievals} \label{pos_single_instrument}
To assess the quality of our retrievals, and investigate which molecular features and planetary properties are primarily constrained by NIRSpec observations versus NIRISS observations, we conducted retrievals with \texttt{POSEIDON} on each of the two instrument datasets separately. Overall, the results of the single instrument retrievals are generally consistent with expectations, {\ron but the broader wavelength range of the full data set provides much greater sensitivity to the atmospheric composition. Specifically, the shorter wavelength NIRISS data provide valuable constraints on the H$_2$O abundance and cloud features but tend to miss CO$_2$ and NH$_3$, while the longer wavelength NIRSpec data capture the CO$_2$ and NH$_3$ but show a lower abundance for H$_2$O and appear to miss the cloud features entirely.} Without the broad spectrum coverage of the combined instruments we would have a very different view of the makeup of the atmosphere of WASP-166b. 

\subsubsection{NIRISS SOSS Only}
\label{pos_niriss}
\begin{figure*}
    \centering
    \includegraphics[width=0.9\textwidth]{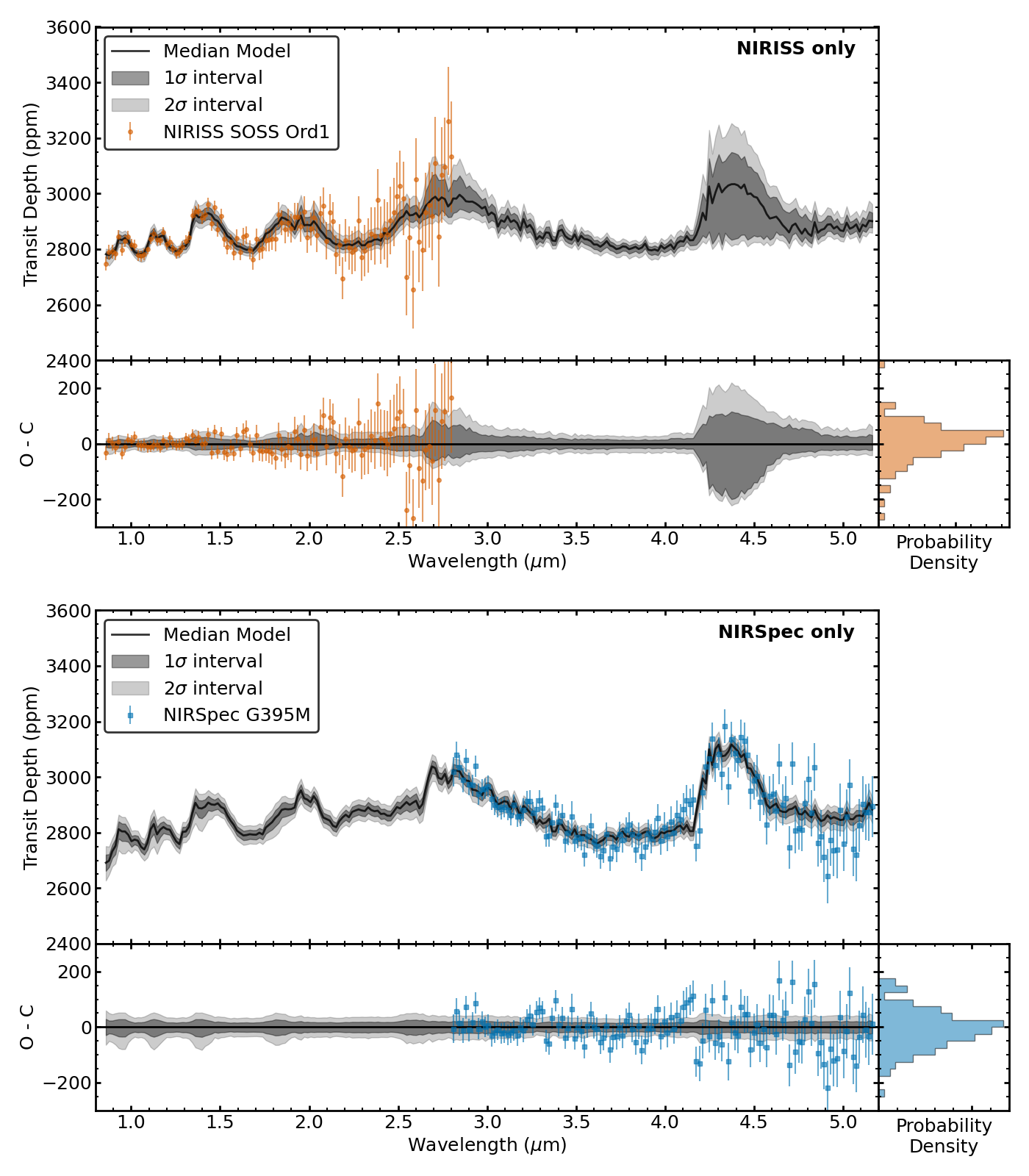}
    \caption{WASP-166b transmission spectra from \texttt{POSEIDON} free chemistry retrievals, {\ron based on the NIRISS SOSS dataset alone (B3; \textbf{Top panels}) and based on the NIRSpec dataset alone (B4; \textbf{Bottom panels}). \textbf{Top:} observational data shown for NIRISS SOSS (orange circles) for the case of fixed ($0.01797 \ \mu$m) bins.} The retrieval is using an isothermal P-T profile with an opaque cloud deck. Fill gases (H$_2$ and He) are present at the primordial solar ratio. The trace gases H$_2$O, CO$_2$, and NH$_3$ are included in the model. \ron{The median retrieved spectrum for this case is shown as a fine black line with the $1\sigma$ and $2\sigma$ confidence intervals shown in dark and light grey shading respectively. Below the primary top panel is a smaller panel with residuals between observational data and retrieval model, with an orange histogram of NIRISS residuals on the right.} Retrieval results are as expected; they show that the water abundance and cloud top pressure are similar to the Baseline (B1) values, but that the CO$_2$ abundance is understated, and NH$_3$ is missed altogether. \ron{\textbf{Bottom:} Same as top, but the observational data shown is for NIRSpec G395M (blue squares) for the case of fixed ($0.01797 \ \mu$m) bins.} Retrieval results are as expected. Water is seen but at a lower abundance than the Baseline (B1). Since we are missing the shorter wavelength data, the cloud effects are not captured, leaving the cloud top pressure essentially unconstrained. The CO$_2$ is reasonably well captured, and NH$_3$ appears slightly higher in abundance than the Baseline. \ron{Again, below the primary bottom panel is a smaller panel with residuals between observational data and retrieval model, with a blue histogram of NIRSpec residuals on the right.}}
    \label{fig:POS_single_inst_v73_v74}
\end{figure*}

{\ron First, we ran a free chemistry retrieval B3, considering only} NIRISS SOSS data. The data and the retrieved spectrum are shown in Figure~\ref{fig:POS_single_inst_v73_v74}. \texttt{POSEIDON} extends the retrieval modeling for the full $0.85$ to $5.17\ \mu$m wavelength range, out beyond the range of the single-instrument data.

The free parameter posterior results from this retrieval case are shown in Table~\ref{tab:free_chem_posteriors}, and can be compared directly with the {\ron B1} retrieval. It is apparent that the main H$_2$O and cloud features are captured within the NIRISS SOSS Order-1 bandpass ($0.85$ to $2.81\ \mu$m), while the CO$_2$ and NH$_3$ features are not as well covered, or are missed altogether. The largest absorption feature in the combined instrument spectrum is due to CO$_2$ at $\sim 4.4 \ \mu$m. This feature is outside the NIRISS bandpass, which (in this single instrument case) results in a {\ron lower and much less precise, but still consistent} abundance estimate for CO$_2$. {\ron Specifically, B3 returns $X_{\rm CO_2} =-3.2^{+1.3}_{-4.6}$ versus $X_{\rm CO_2} = -2.13^{+0.27}_{-0.39}$ for B1. } 

{\ron Compared to B1, the} NIRISS only retrieval indicates somewhat more water {\ron ($X_{\rm H_2O} =-0.96^{+0.24}_{-0.43}$ for B3 versus $X_{\rm H_2O} = -1.42^{+0.20}_{-0.24}$ for B1). This may be due to the lower signal from CO$_2$ changing the relative mixing ratios. The subtle NH$_3$ feature within the NIRISS bandpass at $\sim 2.3 \ \mu$m does not seem to have an impact. The other very subtle feature of NH$_3$ at $\sim 3.9 \ \mu$m is outside the NIRISS bandpass. {\ron Similar to the impact on CO$_2$}, the} net effect {\ron is a lower and much less precise but still consistent NH$_3$} abundance {\ron ($X_{\rm NH_3} =-6.5^{+1.9}_{-2.1}$ for B3 versus $X_{\rm NH_3} = -4.02^{+0.28}_{-0.34}$ for B1).} The retrieved cloud pressure is lower for the NIRISS alone case. This may be due to subtle effects being missed on the red side of the $2.81 \ \mu$m limit.

\subsubsection{NIRSpec G395M Only}
\label{pos_nirspec}
{\ron Next, we ran a free chemistry retrieval B4 considering only NIRspec data.} The data and the retrieved spectrum are shown in Figure~\ref{fig:POS_single_inst_v73_v74}. As in the NIRISS only case, the retrieval modeling for the full $0.85$ to $5.17 \ \mu$m wavelength range, is displayed. The free parameter posterior results from this retrieval case are also shown in Table~\ref{tab:free_chem_posteriors}, and can be compared directly with the {\ron B1} retrieval. The retrieved isothermal temperature for {\ron B4} is very consistent with {\ron B1} (slightly cooler than the {\ron B1} value by $\sim 15$ K).

The effect on the spectrum due to clouds, clipping the lower extent of the features in the $0.85$ to $1.5 \ \mu$m region as seen in Figure~\ref{fig:POS_Spectral_Decomp_v42} is not seen with this instrument alone, leading to essentially an unconstrained, and much higher result for the cloud deck pressure. {\ron Unlike the NIRISS data, the NIRSpec data does not sample strong water features. B4 does have some sensitivity to water because of the} broad water feature {\ron between} $2.3$ to $3.5 \ \mu$m,{\ron but this broad feature provides less stringent constraints than the more pronounced water peaks detectable in the NIRISS data (see Figure~\ref{fig:POS_Spectral_Decomp_v42}.). Accordingly, B4 finds a lower water abundance than B3 or B1}. 

{\ron The wavelength range probed by NIRSpec includes a} subtle NH$_3$ feature at $\sim 3.9 \ \mu$m {\ron and a strong} CO$_2$ feature at $\sim 4.4 \ \mu$m. {\ron Compared to B1, B4 measures a NH$_3$ abundance that is slightly higher but still consistent ($X_{\rm NH_3} = -3.58^{+0.44}_{-0.49}$ for B4 versus $X_{\rm NH_3} = -4.02^{+0.28}_{-0.34}$ for B1). The inferred CO$_2$ abundance is} very near the {\ron B1} value ($X_{\rm CO_2} =-2.33^{+0.34}_{-0.46}$ for B4 versus $X_{\rm CO_2} =-2.13^{+0.27}_{-0.39}$ for B1). 

\begin{figure*}
    \centering
    \includegraphics[width=\textwidth]{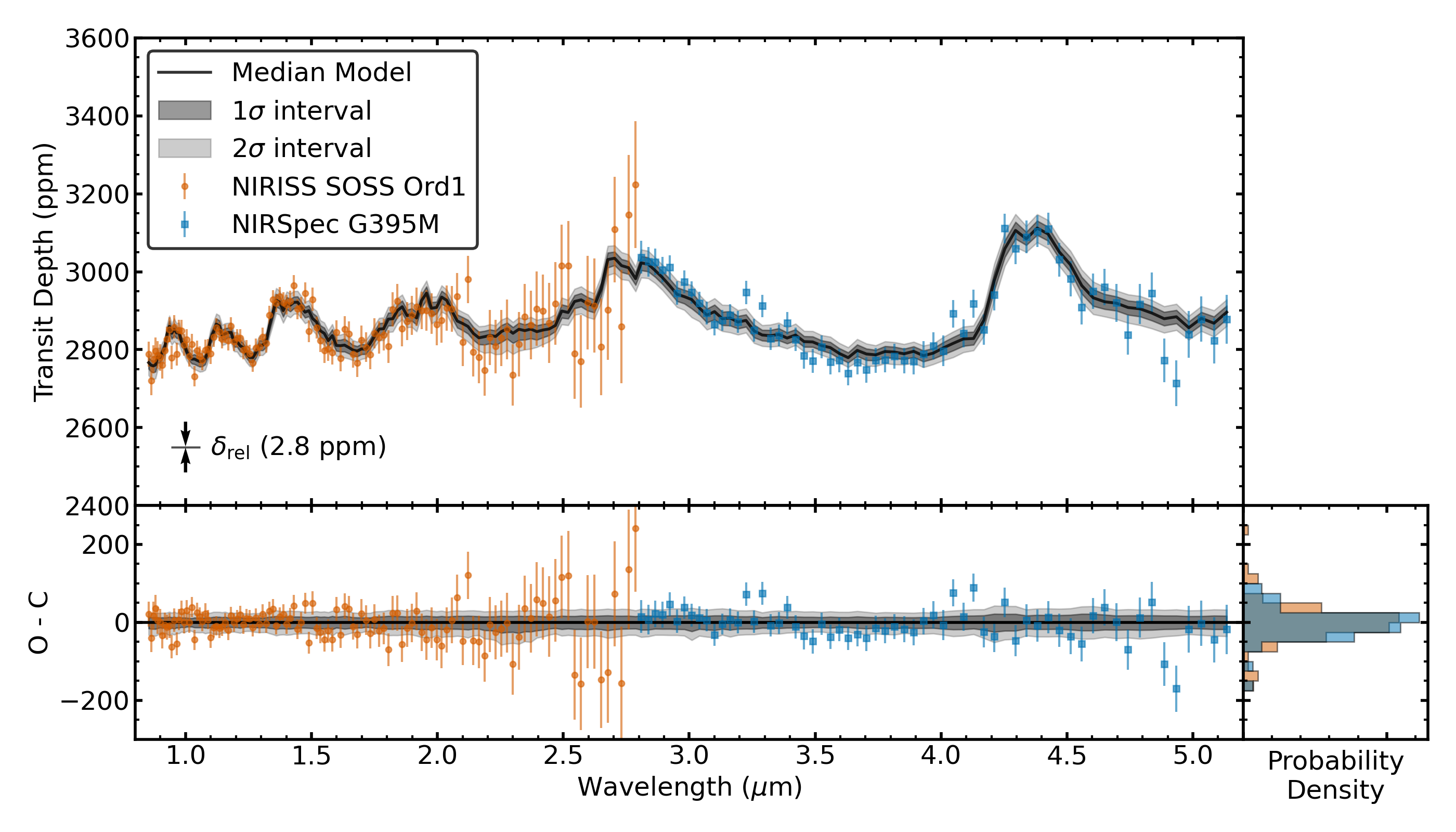}
    \caption{WASP-166b transmission spectrum from \texttt{POSEIDON} free chemistry retrieval; {\ron the observational data are for the case of constant $R = 100$ binning (B5). \textbf{Top:} observational data shown for NIRISS SOSS (orange circles) and NIRSpec G395M (blue squares)}. The retrieval is using an isothermal P-T profile with an opaque cloud deck. Fill gases (H$_2$ and He) are present at the primordial solar ratio. The trace gases H$_2$O, CO$_2$, and NH$_3$ are included in the model. \ron{The plotted NIRISS data have been adjusted down (post-retrieval) by a small retrieved instrument offset (see plotted $\delta_\mathrm{rel}$ value). The median retrieved spectrum for this case is shown as a fine black line with the $1\sigma$ and $2\sigma$ confidence intervals shown in dark and light grey shading respectively. \textbf{Bottom:} residuals between observational data and retrieval model from Top panel plot. \textbf{Bottom Right:} histogram of residuals for each instrument wavelength range, with orange for NIRISS data and blue for NIRSpec data.}}
    \label{fig:POS_Free_Chem_R100_v75}
\end{figure*}

\subsection{\texttt{POSEIDON} Combined Instrument Retrieval, with $R = 100$ Binning} \label{pos_r100}

In order to test whether the binning approach with our data reduction led to significant differences in our retrieval results, we performed a free chemistry retrieval {\ron{(B5)}} using the reduced instrument dataset holding a constant $R = 100$ value (pre-binned) for both instruments. The reduced data and the retrieved spectrum are shown plotted in Figure~\ref{fig:POS_Free_Chem_R100_v75}, with retrieval results included in Table~\ref{tab:free_chem_posteriors}. Only very minor differences are apparent between this and our Baseline (B1) case, with fixed ($0.01797 \ \mu$m) bins. 

The median isothermal temperatures are within $1$ K ($0.01\sigma$) of each other. We also see differences between the median $\log_{10}$(vmr) of only $0.02$ ($0.07\sigma$) for H$_2$O, $0.28$ ($0.71\sigma$) for CO$_2$, and $0.27$ ($0.54\sigma$) for NH$_3$. The median cloud-top pressure $\log_{10}(P_\mathrm{cloud})$ is within $0.18$ ($0.21\sigma$), in units of $\log_{10}$(bar). These differences are all well within the $1 \sigma$ error bounds.

\subsection{\texttt{TauREx} Retrieval}\label{TauREx_Retrieval}

As an independent {\ron check} on the validity of our \texttt{POSEIDON} Baseline {\ron{(B1)}} results, we ran a free chemistry retrieval {\ron{(case B6)}} with another well-known retrieval code, \texttt{TauREx} \citep{alrefaieetal2021}. We used the same system parameters and the same priors (Table~\ref{tab:retrieval_priors}) as were used in the \texttt{POSEIDON} Baseline case. The {\ron observed NIRISS transit depth values} were shifted down by the retrieved offset ($\delta_\mathrm{rel}$ $\sim 8.4$ ppm) from the \texttt{POSEIDON} Baseline, since {\ron unlike \texttt{POSEIDON} and \texttt{PLATON},} \texttt{TauREx} does not retrieve {\ron an offset parameter}.

\begin{figure*}
    \centering
    \includegraphics[width=\textwidth]{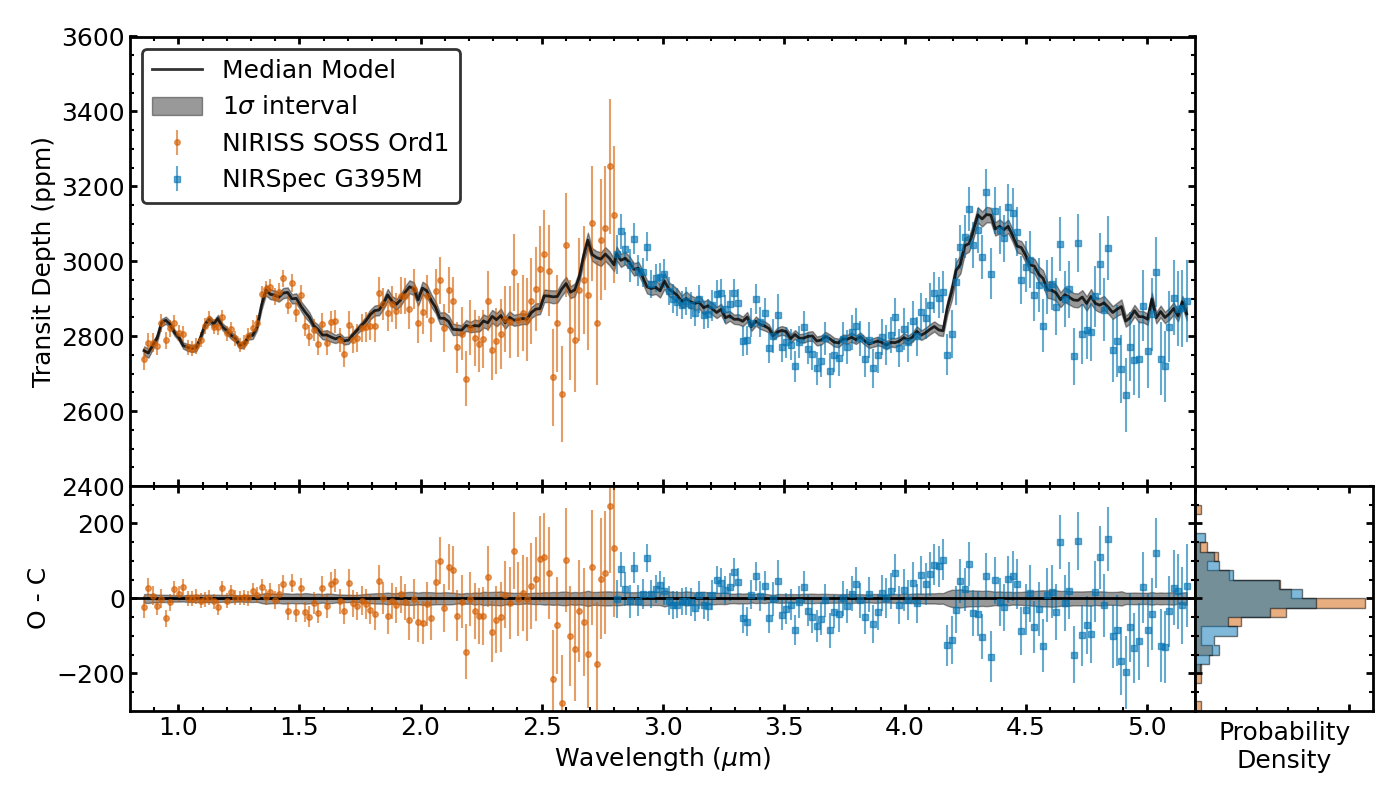}
    \caption{WASP-166b transmission spectrum from \texttt{TauREx} free chemistry retrieval {\ron (B6). We used the same free parameter priors (see Table~\ref{tab:retrieval_priors}) here as in the \texttt{POSEIDON} Baseline retrieval case (B1). \textbf{Top:} observational data shown for NIRISS SOSS (orange circles) and NIRSpec G395M (blue squares)}, for the case of fixed ($0.01797 \ \mu$m) bins. The retrieval is using the native \texttt{TauREx} ``isothermal'' P-T profile with the \texttt{TauREx} opaque cloud deck model. Fill gases (H$_2$ and He) are present at the primordial solar ratio (same ratio as used by \texttt{POSEIDON}, $X_\mathrm{He}/X_\mathrm{H_2}$ = 0.17). The trace gases H$_2$O, CO$_2$, and NH$_3$ are included in the model. Since \texttt{TauREx} does not directly retrieve an instrument offset, the actual data was adjusted here (all NIRISS transit depth values had {\ron the estimated B1 retrieval offset of} $\sim 8.4$ ppm subtracted). Posteriors of the free parameters for this case are compiled in Table~\ref{tab:free_chem_posteriors}. \ron{The median retrieved spectrum for this case is shown as a fine black line with the $1\sigma$ confidence interval shown in dark grey shading. \textbf{Bottom:} residuals between observational data and retrieval model from Top panel plot. \textbf{Bottom Right:} histogram of residuals for each instrument wavelength range, with orange for NIRISS data and blue for NIRSpec data.}}
    \label{fig:TRx_Retrieval_v8}
\end{figure*}

The combined instrument data and the retrieved spectrum are shown in Figure~\ref{fig:TRx_Retrieval_v8}. The results (free parameter posteriors) shown in Table~\ref{tab:free_chem_posteriors} are consistent with the \texttt{POSEIDON} Baseline results. The median isothermal temperatures are within $11$ K of each other, {\ron well within the error bounds}. We also see negligible differences between the median volume mixing ratios ($\log_{10}$(vmr)): only $0.05$ ($0.17\sigma$) for H$_2$O, $0.04$ ($0.09\sigma$) for CO$_2$, $0.09$ ($0.21\sigma$) for NH$_3$, and $0.11$ ($0.13\sigma$) for the cloud-top pressure $\log_{10}(P_\mathrm{cloud})$ (in $\log_{10}$(bar)). In summary, these differences are well within the $1\sigma$ error bounds of each respective parameter. Clearly, \texttt{POSEIDON} and \texttt{TauREx} are delivering consistent retrieval results for the key atmospheric parameters of WASP-166b.

{\ron
\subsection{\texttt{POSEIDON} Cloud Deck Haze Retrieval}\label{deck_haze_retrieval}

So far our retrievals have employed a simple (single free parameter, opaque deck) cloud model. This approach does not allow for spectral variation in the cloud scattering/absorption properties. Clouds with a variable scattering slope could potentially alter our conclusions.
We conducted a free chemistry retrieval (B7) with the Baseline (B1) constituents, and a 3-parameter, `deck\_haze' cloud model (an extension of the single parameter, \texttt{POSEIDON} `MacMad17' model). This `deck\_haze' model includes a `log\_a' parameter (Rayleigh enhancement factor of a power-law haze), and `gamma' (scattering slope of a power-law haze). This model still includes the `log\_P\_cloud' parameter for the opaque deck pressure.} 

{\ron As shown in Table~\ref{tab:free_chem_posteriors}, the B7 results are very near the B1 results with differences in the spectral fit nearly imperceptible. We show that the log\_P\_cloud posterior has shifted from the B1 value of $-1.88^{+0.70}_{-0.64}$ $\log_{10}$(bar) to a slightly higher (but consistent) pressure of $-1.64^{+2.14}_{-0.71}$ $\log_{10}$(bar). Both the log\_a and gamma values are poorly constrained, which is another indication that Rayleigh scattering may not be a particularly significant factor in the wavelength range that we can access. The retrieved log\_a of $0.70^{+2.78}_{-2.94}$ could indicate that the Rayleigh scattering inflection point is pushed far to the blue end of the spectrum (probably to lower wavelengths than our cut-off of $0.85 \ \mu$m); while the retrieved gamma of $-9.91^{+7.19}_{-6.94}$ could indicate a moderately steep slope.} 

{\ron The net result of testing this more complex cloud/haze model is that our simple (single free parameter) opaque cloud deck model seems to be doing a reasonable job of representing the WASP-166b cloud situation, at least to first order. The more complex `deck\_haze' model does not appear to alter our conclusions to any significant degree. Future work could include consideration of more sophisticated cloud/haze models, but this may need to wait for new observations that deliver high quality data at shorter wavelengths. For example, in 2023 the HUSTLE \textit{HST} treasury program \citep{wakefordetal2022} collected UV transit observations of WASP-166b that may provide insights into the planet's clouds.}

\section{Additional Comparative Atmospheric Retrievals with Equilibrium Chemistry} \label{ap:eq_retrievals}

{\ron As discussed in Section~\ref{equilibrium_retrievals}, we conducted our primary equilibrium chemistry retrieval (E1) with \texttt{POSEIDON}.} In order to {\ron corroborate the} \texttt{POSEIDON} equilibrium chemistry results we used the Python package \texttt{PLATON} v5.3 \citep{zhangetal2019,zhangetal2020}, based on the well-known atmospheric modeling code, \texttt{ExoTransmit} \citep{kemptonetal2017}. While \texttt{POSEIDON} allows the flexibility of including or excluding equilibrium chemistry, \texttt{PLATON} v5.3 operates on the assumption of equilibrium chemistry. \texttt{PLATON} models an atmosphere using a grid of pressure and temperature. We use a $250$-layer grid, spaced uniformly in $\log_{10}$ pressure space, with a maximum pressure of $1000$ bar, a minimum pressure of $10^{-9}$ bar, and a reference pressure (at the reference radius) of $1$ bar. \texttt{PLATON} includes an opaque cloud deck model and parameterizes clouds through a cloud-top pressure (log\_cloudtop\_P), a scattering slope, and an amplitude parameter (log\_scatt\_factor). {\ron We also retrieved a systematic transit depth offset ($\delta_\mathrm{rel}$) when modeling the combined data from both instruments.} We fixed the scattering slope to $4$ corresponding to pure Rayleigh scattering (test retrievals that left scattering slope free could not place any constraints on the parameter). With this arrangement, the scattering factor multiplies the scattering absorption coefficient by the same factor across all wavelengths. Priors for the \texttt{PLATON} free parameters are listed in Table~\ref{tab:retrieval_priors}. 

{\ron In this section, we present the results of a series of retrievals with \texttt{PLATON}. Unless otherwise stated, the retrievals consider both the NIRISS and NIRSpec data using bins with a uniform width of $0.01797 \ \mu$m. In Section~\ref{pla_Fixed_Bins}, we describe retrieval E2, which is the \texttt{PLATON} equivalent of our primary \texttt{POSEIDON} equilibrium chemistry retrieval E1 (see Section~\ref{equilibrium_retrievals}). This section also includes a retrieval (E6) that explores a metallicity discrepancy between E1 and E2. Next, in Section~\ref{pla_single_instrument}, we apply \texttt{PLATON} to the NIRISS and NIRSpec data individually; the NIRISS retrieval E3 is discussed in Section~\ref{pla_niriss} and the NIRSpec retrieval E4 is discussed in Section~\ref{pla_nirspec}. Finally, in Section~\ref{pla_r100}, we use \texttt{PLATON} to perform an equilibrium chemistry retrieval (E5) on the combined NIRISS and NIRSpec data using a fixed spectral resolution ($R=100$) rather than a fixed bin width. 
}

{\ron 
It should be noted that the basic equilibrium chemistry codes, \texttt{FastChem\_2} \citep{stocketal2022} used by \texttt{POSEIDON} v1.2.1, and \texttt{GGChem} \citep{woitkeetal2018,woitkeandhelling2021} used by \texttt{PLATON} v5.3, do not model the photochemical formation pathway for SO$_2$ (see \citealt{stocketal2022}). There are various add-on codes available, e.g., \texttt{VULCAN} \citep{tsaietal2017,tsaietal2021}, that can give \texttt{FastChem} the ability to model disequilibrium photochemical processes; but the implementation of this or any other ancillary code was beyond the scope of our work in this paper.

Without including this photochemical formation mechanism for SO$_2$, our equilibrium chemistry estimates for the abundance of SO$_2$ may be understated. This could change our assessment of the overall atmospheric metallicity and C/O ratio. However, our free chemistry abundance estimates for SO$_2$ and H$_2$S, while only relevant in the lower pressure and temperature regime (above the cloud deck and in the region probed by transmission spectroscopy), would suggest that these constituents only exist (if at all) at very low abundances.
}

\subsection{\texttt{PLATON} {\ron{Equilibrium Chemistry Retrieval}}}
\label{pla_Fixed_Bins}
\begin{figure*}
    \centering
    \includegraphics[width=\textwidth]{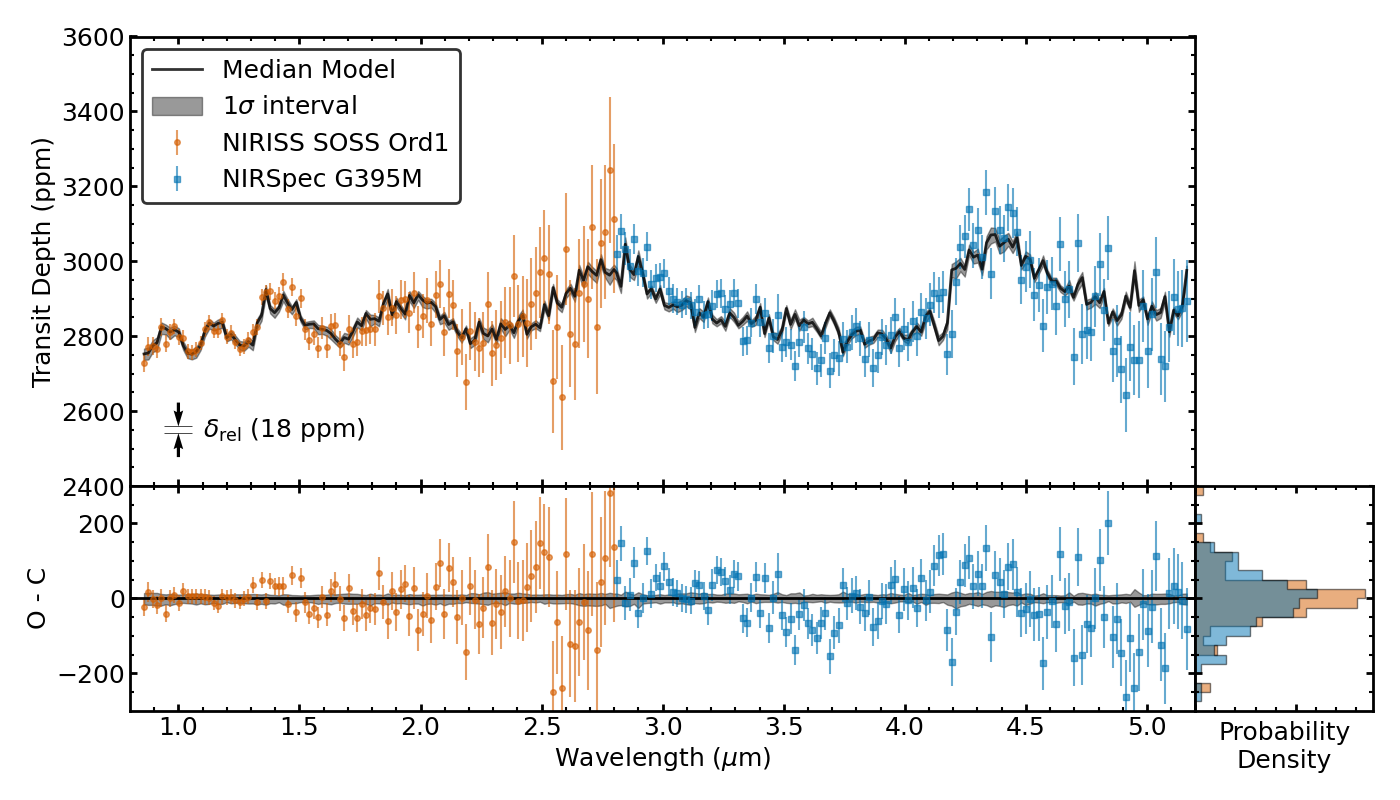}
    \caption{WASP-166b transmission spectrum from \texttt{PLATON} equilibrium chemistry retrieval {\ron (E2). \textbf{Top:} observational data shown for NIRISS SOSS (orange circles) and NIRSpec G395M (blue squares)}, for the case of fixed ($0.01797 \ \mu$m) bins. The retrieval is using an isothermal P-T profile with an opaque cloud deck, and 31 molecular species \citep{zhangetal2020}. \ron{The plotted NIRISS data have been adjusted down (post-retrieval) by a small retrieved instrument offset (see plotted $\delta_\mathrm{rel}$ value). The median retrieved spectrum for this case is shown as a fine black line with the $1\sigma$ confidence interval shown in dark grey shading. \textbf{Bottom:} residuals between observational data and retrieval model from Top panel plot. \textbf{Bottom Right:} histogram of residuals for each instrument wavelength range, with orange for NIRISS data and blue for NIRSpec data.}}
    \label{fig:pla_Eq_Chem}
\end{figure*}

\begin{figure*}
    \centering
    \includegraphics[width=\textwidth]{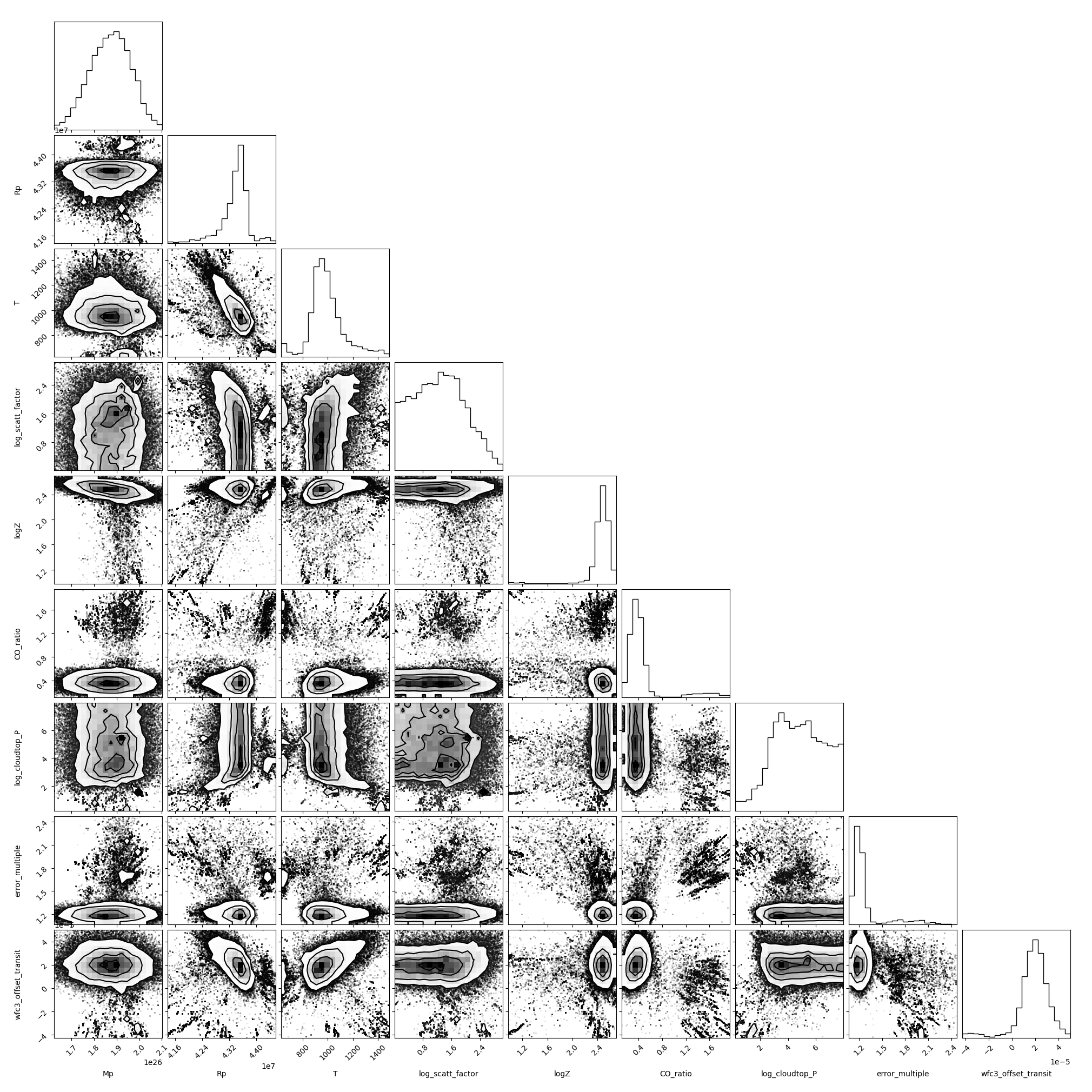}
    \caption{{\ron{Posterior distributions of free parameters (corner plot) for \texttt{PLATON} Equilibrium Chemistry Case, E2. (Note: the 'wfc\_offset\_transit' parameter is equivalent to $\delta_\mathrm{rel}$ in our \texttt{POSEIDON} retrievals; we adjust the wavelength range for 'wfc\_offset\_transit' from $1-1.7$ $\mu$m (for \textit{HST} WFC3) to $0.8-2.8$ $\mu$m (for NIRISS SOSS).)}}}
    \label{fig:POS_v31_E2_corner}
\end{figure*}

Running \texttt{PLATON} on both instruments {\ron (case E2)} yielded good agreement {\ron with the primary equilibrium chemistry case E1} across most parameters. We present the retrieved spectrum for the {\ron E2} case in Figure~\ref{fig:pla_Eq_Chem}. {\ron The E2} posterior values are shown in table form in Table~\ref{tab:eq_chem_posteriors} and in corner plot form in Figure~\ref{fig:POS_v31_E2_corner}.

We find excellent agreement between {\ron the} retrieved {\ron E1 and E2} values for the planet mass $M_p$ ($0.05\sigma$) and the carbon-to-oxygen ratio $C/O$ ($0.73\sigma$), as well as reasonable agreement for the planetary isothermal temperature $T$ ($1.01\sigma$), the cloud deck pressure $\log_{10}(P_\mathrm{cloud})$ ($1.7\sigma$), and the instrumental offset $\delta_\mathrm{rel}$ ($1.4\sigma$).

There is however a noticeable mismatch between \texttt{PLATON} and \texttt{POSEIDON} for the reference radius $R_\mathrm{p,ref}$ ($5.1\sigma$) and metallicity $\log(Z)$ ($4.6\sigma$). Degeneracies between several exoplanet parameters in transmission spectroscopy have been well documented and discussed \citep{welbanksandmadhusudhan2019}, especially between planet radius, reference pressure, cloud-top pressure, and metallicity. It is possible that \texttt{POSEIDON} settled on a low-metallicity, low-radius, intermediate-pressure cloud deck solution to this degeneracy while \texttt{PLATON} settled instead on a high-metallicity, high-radius, high-pressure (and poorly constrained) cloud deck solution. 

{\ron Furthermore, the differences in the reference radius and metallicity could be due to discrepancies} in the equilibrium chemistry assumptions between \texttt{POSEIDON} and \texttt{PLATON} such as a difference in the opacity line lists (see Section~\ref{free_retrievals}) or the radiative transfer calculations. {\ron In addition, while \texttt{POSEIDON} and \texttt{PLATON} consider very similar sets of molecules, there are some differences. For instance, \texttt{PLATON} includes HF, N$_2$, H$_2$CO, and HCl, while \texttt{POSEIDON} does not; similarly, \texttt{POSEIDON} includes C$_2$, CaH, CrH, FeH, SO, and TiH while \texttt{PLATON} does not. Also, as discussed in Section~\ref{retrievals} and as shown in Table~\ref{tab:opacity_data}, the various packages use absorption line lists from different combinations of sources.}

{\ron We explored the planet radius and metallicity discrepancy by performing an additional \texttt{PLATON} retrieval (E6) with metallicity fixed to the lower value determined by the \texttt{POSEIDON} (E1) retrieval ($\log(Z) = 1.57^{+0.17}_{-0.18}$). This adjustment to \texttt{PLATON} resulted in forcing the planet radius down from $R_\mathrm{p,ref} = 0.6077^{+0.0029}_{-0.0056}$ R$_J$ (E2) to $0.5983^{+0.0020}_{-0.0026}$ R$_J$ (closer to the \texttt{POSEIDON} E1 value of $R_\mathrm{p,ref} = 0.5721^{+0.0041}_{-0.0042}$ R$_J$). Planet mass also increased to $M_p = 0.1110^{+0.0049}_{-0.0057}$ M$_\oplus$. Decreased planet radius and increased planet mass both contribute to increased surface gravity, cooling the isothermal temperature significantly down to $811^{+41}_{-32}$ K, and pushing the high pressure and poorly constrained cloud deck from E2 ($\log_{10}(P_\mathrm{cloud})$) $= -0.3^{+2.1}_{-1.8}$ $\log_{10}$(bar)) down to lower pressures and a tighter constraint ($-2.4^{+1.8}_{-0.3}$ $\log_{10}$(bar)).}

{\ron The results of our fixed metallicity fit are consistent with the work of \citet{kemptonetal2017}, who} found that the strength of transmission features tends to peak around $5$x Solar metallicity. {\ron Accordingly,} forcibly decreasing metallicity from $Z \sim 300$x Solar to $Z \sim 40$x Solar would increase the strength of spectral features. Increasing surface gravity, decreasing temperature, and raising the cloud deck to higher altitudes (lower pressures) all serve to suppress spectral features, so these parameter shifts all appear to be counterbalancing the stronger features that would result from the lower metallicity.

We encourage further investigation of this issue. We also stress that regardless of the solution to this discrepancy, our conclusion holds that WASP-166b is quite metal-rich compared to its host star.

\subsection{\texttt{PLATON} Single Instrument Retrievals} \label{pla_single_instrument}

In order to assess the quality of our retrievals as well as investigate which molecular features and planetary properties are primarily constrained by NIRSpec observations versus NIRISS SOSS observations, we conducted equilibrium chemistry retrievals with \texttt{PLATON} on each of the two instrument data sets separately. We performed the same procedure with \texttt{POSEIDON} assuming free chemistry in Section~\ref{pos_single_instrument}. {In all cases, we found that the results from single-instrument analyses are consistent but much less precise than those from analyses of the joint NIRISS and NIRSpec data set.} 

\subsubsection{NIRISS SOSS Alone}
\label{pla_niriss}

Running \texttt{PLATON} on NIRISS SOSS alone {\ron (case E3)} yielded good agreement across all parameters against the benchmark \texttt{PLATON} equilibrium retrieval E2. Free parameter priors are given in Table~\ref{tab:retrieval_priors}, and posterior results from this retrieval case (and the combined instrument retrieval case) are shown in Table~\ref{tab:eq_chem_posteriors}. All parameters agree to within $1\sigma$ except the error multiple parameter (a scaling parameter used to account for over- or underestimation of transit depth error bars by the user). 

Although the NIRISS SOSS alone retrieval agrees with the results from both instruments, we still find greatly improved precision using both instruments. We find that key parameters are more tightly constrained when both instruments are included. Uncertainties on C/O ratio and planetary metallicity are $56\%$ and $48\%$ larger respectively when retrieved with just NIRISS SOSS compared to both instrument data sets.

\subsubsection{NIRSpec G395M Alone}
\label{pla_nirspec}

Similar to NIRISS SOSS, running \texttt{PLATON} on NIRSpec alone {\ron (case E4)} yielded good agreement {\ron with E2} across all parameters. Free parameter priors are given in Table~\ref{tab:retrieval_priors}, and posterior results from this retrieval case (and the combined instrument retrieval case) are shown in Table~\ref{tab:eq_chem_posteriors}. Again, all parameters agree to within $1\sigma$ except the error multiple parameter.

Although the NIRSpec alone retrieval agrees with the results from both instruments, we still find greatly improved precision with both instruments. We find that key parameters are more tightly constrained when both instruments are included. Uncertainties on C/O ratio, metallicity, and the isothermal temperature are $11\%$, $11\%$, and $110\%$ larger, respectively, when retrieved with just NIRSpec compared to the combined instrument data sets.

\subsection{\texttt{PLATON} Combined Instrument Retrieval, with $R = 100$ Binning} \label{pla_r100}
{\ron Next, we ran \texttt{PLATON} on the same $R = 100$ spectrum we analyzed with \texttt{POSEIDON} in Section~\ref{pos_r100}. The fixed resolution \texttt{PLATON} analysis (case E5) is consistent with E2.} All parameters agree to within $1\sigma$. In other words, our choice of equal wavelength binning ($0.01797 \ \mu$m bins) over equal resolution binning ($R = 100$) has a negligible impact on our retrieved results.

\bibliography{main}{}

\begin{thebibliography}{}
\expandafter\ifx\csname natexlab\endcsname\relax\def\natexlab#1{#1}\fi
\providecommand{\url}[1]{\href{#1}{#1}}
\providecommand{\dodoi}[1]{doi:~\href{http://doi.org/#1}{\nolinkurl{#1}}}
\providecommand{\doeprint}[1]{\href{http://ascl.net/#1}{\nolinkurl{http://ascl.net/#1}}}
\providecommand{\doarXiv}[1]{\href{https://arxiv.org/abs/#1}{\nolinkurl{https://arxiv.org/abs/#1}}}

\bibitem[{{Ahrer} {et~al.}(2023){Ahrer}, {Stevenson}, {Mansfield}, {Moran}, {Brande}, {Morello}, {Murray}, {Nikolov}, {Petit dit de la Roche}, {Schlawin}, {Wheatley}, {Zieba}, {Batalha}, {Damiano}, {Goyal}, {Lendl}, {Lothringer}, {Mukherjee}, {Ohno}, {Batalha}, {Battley}, {Bean}, {Beatty}, {Benneke}, {Berta-Thompson}, {Carter}, {Cubillos}, {Daylan}, {Espinoza}, {Gao}, {Gibson}, {Gill}, {Harrington}, {Hu}, {Kreidberg}, {Lewis}, {Line}, {L{\'o}pez-Morales}, {Parmentier}, {Powell}, {Sing}, {Tsai}, {Wakeford}, {Welbanks}, {Alam}, {Alderson}, {Allen}, {Anderson}, {Barstow}, {Bayliss}, {Bell}, {Blecic}, {Bryant}, {Burleigh}, {Carone}, {Casewell}, {Changeat}, {Chubb}, {Crossfield}, {Crouzet}, {Decin}, {D{\'e}sert}, {Feinstein}, {Flagg}, {Fortney}, {Gizis}, {Heng}, {Iro}, {Kempton}, {Kendrew}, {Kirk}, {Knutson}, {Komacek}, {Lagage}, {Leconte}, {Lustig-Yaeger}, {MacDonald}, {Mancini}, {May}, {Mayne}, {Miguel}, {Mikal-Evans}, {Molaverdikhani}, {Palle}, {Piaulet}, {Rackham}, {Redfield}, {Rogers}, {Roy}, {Rustamkulov},
  {Shkolnik}, {Sotzen}, {Taylor}, {Tremblin}, {Tucker}, {Turner}, {de Val-Borro}, {Venot}, \& {Zhang}}]{Ahreretal2023}
{Ahrer}, E.-M., {Stevenson}, K.~B., {Mansfield}, M., {et~al.} 2023, \nat, 614, 653, \dodoi{10.1038/s41586-022-05590-4}

\bibitem[{{Al-Refaie} {et~al.}(2021){Al-Refaie}, {Changeat}, {Waldmann}, \& {Tinetti}}]{alrefaieetal2021}
{Al-Refaie}, A.~F., {Changeat}, Q., {Waldmann}, I.~P., \& {Tinetti}, G. 2021, \apj, 917, 37, \dodoi{10.3847/1538-4357/ac0252}

\bibitem[{{Al-Refaie} {et~al.}(2015){Al-Refaie}, {Yachmenev}, {Tennyson}, \& {Yurchenko}}]{al-refaieetal2015}
{Al-Refaie}, A.~F., {Yachmenev}, A., {Tennyson}, J., \& {Yurchenko}, S.~N. 2015, \mnras, 448, 1704, \dodoi{10.1093/mnras/stv091}

\bibitem[{{Albert} {et~al.}(2023){Albert}, {Lafreni{\`e}re}, {Ren{\'e}}, {Artigau}, {Volk}, {Goudfrooij}, {Martel}, {Radica}, {Rowe}, {Espinoza}, {Roy}, {Filippazzo}, {Darveau-Bernier}, {Talens}, {Sivaramakrishnan}, {Willott}, {Fullerton}, {LaMassa}, {Hutchings}, {Rowlands}, {Vila}, {Zhou}, {Aldridge}, {Maszkiewicz}, {Beaulieu}, {Cook}, {Piaulet}, {Roy}, {Lamontagne}, {Morel}, {Frost}, {Salhi}, {Coulombe}, {Benneke}, {MacDonald}, {Johnstone}, {Turner}, {Fournier-Tondreau}, {Allart}, \& {Kaltenegger}}]{albertetal2023}
{Albert}, L., {Lafreni{\`e}re}, D., {Ren{\'e}}, D., {et~al.} 2023, \pasp, 135, 075001, \dodoi{10.1088/1538-3873/acd7a3}

\bibitem[{{Alderson} {et~al.}(2023){Alderson}, {Wakeford}, {Alam}, {Batalha}, {Lothringer}, {Adams Redai}, {Barat}, {Brande}, {Damiano}, {Daylan}, {Espinoza}, {Flagg}, {Goyal}, {Grant}, {Hu}, {Inglis}, {Lee}, {Mikal-Evans}, {Ramos-Rosado}, {Roy}, {Wallack}, {Batalha}, {Bean}, {Benneke}, {Berta-Thompson}, {Carter}, {Changeat}, {Col{\'o}n}, {Crossfield}, {D{\'e}sert}, {Foreman-Mackey}, {Gibson}, {Kreidberg}, {Line}, {L{\'o}pez-Morales}, {Molaverdikhani}, {Moran}, {Morello}, {Moses}, {Mukherjee}, {Schlawin}, {Sing}, {Stevenson}, {Taylor}, {Aggarwal}, {Ahrer}, {Allen}, {Barstow}, {Bell}, {Blecic}, {Casewell}, {Chubb}, {Crouzet}, {Cubillos}, {Decin}, {Feinstein}, {Fortney}, {Harrington}, {Heng}, {Iro}, {Kempton}, {Kirk}, {Knutson}, {Krick}, {Leconte}, {Lendl}, {MacDonald}, {Mancini}, {Mansfield}, {May}, {Mayne}, {Miguel}, {Nikolov}, {Ohno}, {Palle}, {Parmentier}, {Petit dit de la Roche}, {Piaulet}, {Powell}, {Rackham}, {Redfield}, {Rogers}, {Rustamkulov}, {Tan}, {Tremblin}, {Tsai}, {Turner}, {de Val-Borro},
  {Venot}, {Welbanks}, {Wheatley}, \& {Zhang}}]{Aldersonetal2023}
{Alderson}, L., {Wakeford}, H.~R., {Alam}, M.~K., {et~al.} 2023, \nat, 614, 664, \dodoi{10.1038/s41586-022-05591-3}

\bibitem[{{Asplund} {et~al.}(2009){Asplund}, {Grevesse}, {Sauval}, \& {Scott}}]{asplundetal2009}
{Asplund}, M., {Grevesse}, N., {Sauval}, A.~J., \& {Scott}, P. 2009, \araa, 47, 481, \dodoi{10.1146/annurev.astro.46.060407.145222}

\bibitem[{{Azzam} {et~al.}(2016){Azzam}, {Tennyson}, {Yurchenko}, \& {Naumenko}}]{azzam_et_al2016}
{Azzam}, A. A.~A., {Tennyson}, J., {Yurchenko}, S.~N., \& {Naumenko}, O.~V. 2016, \mnras, 460, 4063, \dodoi{10.1093/mnras/stw1133}

\bibitem[{{Baines} {et~al.}(2023{\natexlab{a}}){Baines}, {Espinoza}, {Filippazzo}, \& {Volk}}]{baines2023traces}
{Baines}, T., {Espinoza}, N., {Filippazzo}, J., \& {Volk}, K. 2023{\natexlab{a}}, arXiv e-prints, arXiv:2311.07769, \dodoi{10.48550/arXiv.2311.07769}

\bibitem[{{Baines} {et~al.}(2023{\natexlab{b}}){Baines}, {Espinoza}, {Filippazzo}, \& {Volk}}]{baines2023wavelength}
---. 2023{\natexlab{b}}, {Characterization of the visit-to-visit Stability of the GR700XD Wavelength Calibration for NIRISS/SOSS Observations}, Technical Report JWST-STScI-008571, 12 pages

\bibitem[{{Barber} {et~al.}(2014){Barber}, {Strange}, {Hill}, {Polyansky}, {Mellau}, {Yurchenko}, \& {Tennyson}}]{barber_et_al2014}
{Barber}, R.~J., {Strange}, J.~K., {Hill}, C., {et~al.} 2014, \mnras, 437, 1828, \dodoi{10.1093/mnras/stt2011}

\bibitem[{{Barton} {et~al.}(2017){Barton}, {Hill}, {Yurchenko}, {Tennyson}, {Dudaryonok}, \& {Lavrentieva}}]{bartonetal2017}
{Barton}, E.~J., {Hill}, C., {Yurchenko}, S.~N., {et~al.} 2017, \jqsrt, 187, 453, \dodoi{10.1016/j.jqsrt.2016.10.024}

\bibitem[{{Barton} {et~al.}(2013){Barton}, {Yurchenko}, \& {Tennyson}}]{bartonetal2013}
{Barton}, E.~J., {Yurchenko}, S.~N., \& {Tennyson}, J. 2013, \mnras, 434, 1469, \dodoi{10.1093/mnras/stt1105}

\bibitem[{{Basilicata} {et~al.}(2024){Basilicata}, {Giacobbe}, {Bonomo}, {Scandariato}, {Brogi}, {Singh}, {Di Paola}, {Mancini}, {Sozzetti}, {Lanza}, {Cubillos}, {Damasso}, {Desidera}, {Biazzo}, {Bignamini}, {Borsa}, {Cabona}, {Carleo}, {Ghedina}, {Guilluy}, {Maggio}, {Mainella}, {Micela}, {Molinari}, {Molinaro}, {Nardiello}, {Pedani}, {Pino}, {Poretti}, {Southworth}, {Stangret}, \& {Turrini}}]{basilicataetal2024}
{Basilicata}, M., {Giacobbe}, P., {Bonomo}, A.~S., {et~al.} 2024, \aap, 686, A127, \dodoi{10.1051/0004-6361/202347659}

\bibitem[{{Batalha} {et~al.}(2017){Batalha}, {Mandell}, {Pontoppidan}, {Stevenson}, {Lewis}, {Kalirai}, {Earl}, {Greene}, {Albert}, \& {Nielsen}}]{batalhaetal2017}
{Batalha}, N.~E., {Mandell}, A., {Pontoppidan}, K., {et~al.} 2017, \pasp, 129, 064501, \dodoi{10.1088/1538-3873/aa65b0}

\bibitem[{{Batygin} {et~al.}(2016){Batygin}, {Bodenheimer}, \& {Laughlin}}]{batyginetal2016}
{Batygin}, K., {Bodenheimer}, P.~H., \& {Laughlin}, G.~P. 2016, \apj, 829, 114, \dodoi{10.3847/0004-637X/829/2/114}

\bibitem[{{Beatty} {et~al.}(2024){Beatty}, {Welbanks}, {Schlawin}, {Bell}, {Line}, {Murphy}, {Edelman}, {Greene}, {Fortney}, {Henry}, {Mukherjee}, {Ohno}, {Parmentier}, {Rauscher}, {Wiser}, \& {Arnold}}]{beattyetal2024}
{Beatty}, T.~G., {Welbanks}, L., {Schlawin}, E., {et~al.} 2024, \apjl, 970, L10, \dodoi{10.3847/2041-8213/ad55e9}

\bibitem[{Bell {et~al.}(2022)Bell, Ahrer, Brande, Carter, Feinstein, {Guzman Caloca}, Mansfield, Zieba, Piaulet, Benneke, Filippazzo, May, Roy, Kreidberg, \& Stevenson}]{belletal2022}
Bell, T.~J., Ahrer, E.-M., Brande, J., {et~al.} 2022, Journal of Open Source Software, 7, 4503, \dodoi{10.21105/joss.04503}

\bibitem[{{Benneke} \& {Seager}(2013)}]{BennekeandSeager2013}
{Benneke}, B., \& {Seager}, S. 2013, \apj, 778, 153, \dodoi{10.1088/0004-637X/778/2/153}

\bibitem[{{Benneke} {et~al.}(2019){Benneke}, {Knutson}, {Lothringer}, {Crossfield}, {Moses}, {Morley}, {Kreidberg}, {Fulton}, {Dragomir}, {Howard}, {Wong}, {D{\'e}sert}, {McCullough}, {Kempton}, {Fortney}, {Gilliland}, {Deming}, \& {Kammer}}]{bennekeetal2019}
{Benneke}, B., {Knutson}, H.~A., {Lothringer}, J., {et~al.} 2019, Nature Astronomy, 3, 813, \dodoi{10.1038/s41550-019-0800-5}

\bibitem[{{Bernath}(2020)}]{Bernath2020}
{Bernath}, P.~F. 2020, \jqsrt, 240, 106687, \dodoi{10.1016/j.jqsrt.2019.106687}

\bibitem[{{Birkmann} {et~al.}(2022){Birkmann}, {Ferruit}, {Giardino}, {Nielsen}, {Garc{\'\i}a Mu{\~n}oz}, {Kendrew}, {Rauscher}, {Beck}, {Keyes}, {Valenti}, {Jakobsen}, {Dorner}, {Alves de Oliveira}, {Arribas}, {B{\"o}ker}, {Bunker}, {Charlot}, {de Marchi}, {Kumari}, {L{\'o}pez-Caniego}, {L{\"u}tzgendorf}, {Maiolino}, {Manjavacas}, {Marston}, {Moseley}, {Prizkal}, {Proffitt}, {Rawle}, {Rix}, {te Plate}, {Sabbi}, {Sirianni}, {Willott}, \& {Zeidler}}]{BirkmanetalIV2022}
{Birkmann}, S.~M., {Ferruit}, P., {Giardino}, G., {et~al.} 2022, \aap, 661, A83, \dodoi{10.1051/0004-6361/202142592}

\bibitem[{{B{\"o}ker} {et~al.}(2023){B{\"o}ker}, {Beck}, {Birkmann}, {Giardino}, {Keyes}, {Kumari}, {Muzerolle}, {Rawle}, {Zeidler}, {Abul-Huda}, {Alves de Oliveira}, {Arribas}, {Bechtold}, {Bhatawdekar}, {Bonaventura}, {Bunker}, {Cameron}, {Carniani}, {Charlot}, {Curti}, {Espinoza}, {Ferruit}, {Franx}, {Jakobsen}, {Karakla}, {L{\'o}pez-Caniego}, {L{\"u}tzgendorf}, {Maiolino}, {Manjavacas}, {Marston}, {Moseley}, {Ogle}, {Perna}, {Pe{\~n}a-Guerrero}, {Pirzkal}, {Plesha}, {Proffitt}, {Rauscher}, {Rix}, {Rodr{\'\i}guez del Pino}, {Rustamkulov}, {Sabbi}, {Sing}, {Sirianni}, {te Plate}, {{\'U}beda}, {Wahlgren}, {Wislowski}, {Wu}, \& {Willott}}]{Bokeretal2023}
{B{\"o}ker}, T., {Beck}, T.~L., {Birkmann}, S.~M., {et~al.} 2023, \pasp, 135, 038001, \dodoi{10.1088/1538-3873/acb846}

\bibitem[{{Brady} {et~al.}(2024){Brady}, {Yurchenko}, {Tennyson}, \& {Kim}}]{Bradyetal2024}
{Brady}, R.~P., {Yurchenko}, S.~N., {Tennyson}, J., \& {Kim}, G.-S. 2024, \mnras, 527, 6675, \dodoi{10.1093/mnras/stad3508}

\bibitem[{{Brande} {et~al.}(2022){Brande}, {Crossfield}, {Kreidberg}, {Oklop{\v{c}}i{\'c}}, {Polanski}, {Barman}, {Benneke}, {Christiansen}, {Dragomir}, {Foreman-Mackey}, {Fortney}, {Greene}, {Howard}, {Knutson}, {Lothringer}, {Mikal-Evans}, \& {Morley}}]{brandeetal2022}
{Brande}, J., {Crossfield}, I. J.~M., {Kreidberg}, L., {et~al.} 2022, \aj, 164, 197, \dodoi{10.3847/1538-3881/ac8b7e}

\bibitem[{{Brooke} {et~al.}(2016){Brooke}, {Bernath}, {Western}, {Sneden}, {Af{\c{s}}ar}, {Li}, \& {Gordon}}]{brookeetal2016}
{Brooke}, J. S.~A., {Bernath}, P.~F., {Western}, C.~M., {et~al.} 2016, \jqsrt, 168, 142, \dodoi{10.1016/j.jqsrt.2015.07.021}

\bibitem[{{Bryant} {et~al.}(2020){Bryant}, {Bayliss}, {McCormac}, {Wheatley}, {Acton}, {Anderson}, {Armstrong}, {Bouchy}, {Belardi}, {Burleigh}, {Tilbrook}, {Casewell}, {Cooke}, {Gill}, {Goad}, {Jenkins}, {Lendl}, {Pollacco}, {Queloz}, {Raynard}, {Smith}, {Vines}, {West}, \& {Udry}}]{bryantetal2020}
{Bryant}, E.~M., {Bayliss}, D., {McCormac}, J., {et~al.} 2020, \mnras, 494, 5872, \dodoi{10.1093/mnras/staa1075}

\bibitem[{{Buchner}(2016)}]{buchner2016}
{Buchner}, J. 2016, {PyMultiNest: Python interface for MultiNest}, Astrophysics Source Code Library, record ascl:1606.005

\bibitem[{{Bushouse} {et~al.}(2023){Bushouse}, {Eisenhamer}, {Dencheva}, {Davies}, {Greenfield}, {Morrison}, {Hodge}, {Simon}, {Grumm}, {Droettboom}, {Slavich}, {Sosey}, {Pauly}, {Miller}, {Jedrzejewski}, {Hack}, {Davis}, {Crawford}, {Law}, {Gordon}, {Regan}, {Cara}, {MacDonald}, {Bradley}, {Shanahan}, {Jamieson}, {Teodoro}, {Williams}, \& {Pena-Guerrero}}]{bushouse2023}
{Bushouse}, H., {Eisenhamer}, J., {Dencheva}, N., {et~al.} 2023, {JWST Calibration Pipeline}, 1.12.0, Zenodo,  Zenodo, \dodoi{10.5281/zenodo.6984365}

\bibitem[{{Cabot} {et~al.}(2020){Cabot}, {Madhusudhan}, {Welbanks}, {Piette}, \& {Gandhi}}]{cabotetal2020}
{Cabot}, S. H.~C., {Madhusudhan}, N., {Welbanks}, L., {Piette}, A., \& {Gandhi}, S. 2020, \mnras, 494, 363, \dodoi{10.1093/mnras/staa748}

\bibitem[{{Casasayas-Barris} {et~al.}(2017){Casasayas-Barris}, {Palle}, {Nowak}, {Yan}, {Nortmann}, \& {Murgas}}]{casasayas-barrisetal2017}
{Casasayas-Barris}, N., {Palle}, E., {Nowak}, G., {et~al.} 2017, \aap, 608, A135, \dodoi{10.1051/0004-6361/201731956}

\bibitem[{{Casasayas-Barris} {et~al.}(2019){Casasayas-Barris}, {Pall{\'e}}, {Yan}, {Chen}, {Kohl}, {Stangret}, {Parviainen}, {Helling}, {Watanabe}, {Czesla}, {Fukui}, {Monta{\~n}{\'e}s-Rodr{\'\i}guez}, {Nagel}, {Narita}, {Nortmann}, {Nowak}, {Schmitt}, \& {Zapatero Osorio}}]{casasayas-barrisetal2019}
{Casasayas-Barris}, N., {Pall{\'e}}, E., {Yan}, F., {et~al.} 2019, \aap, 628, A9, \dodoi{10.1051/0004-6361/201935623}

\bibitem[{{Cegla} {et~al.}(2016){Cegla}, {Lovis}, {Bourrier}, {Beeck}, {Watson}, \& {Pepe}}]{ceglaetal2016}
{Cegla}, H.~M., {Lovis}, C., {Bourrier}, V., {et~al.} 2016, \aap, 588, A127, \dodoi{10.1051/0004-6361/201527794}

\bibitem[{{Charbonneau} {et~al.}(2002){Charbonneau}, {Brown}, {Noyes}, \& {Gilliland}}]{charbonneauetal2002}
{Charbonneau}, D., {Brown}, T.~M., {Noyes}, R.~W., \& {Gilliland}, R.~L. 2002, \apj, 568, 377, \dodoi{10.1086/338770}

\bibitem[{{Chen} {et~al.}(2020){Chen}, {Casasayas-Barris}, {Pall{\'e}}, {Welbanks}, {Madhusudhan}, {Luque}, \& {Murgas}}]{chenetal2020}
{Chen}, G., {Casasayas-Barris}, N., {Pall{\'e}}, E., {et~al.} 2020, \aap, 642, A54, \dodoi{10.1051/0004-6361/202038661}

\bibitem[{{Chen} {et~al.}(2018){Chen}, {Pall{\'e}}, {Welbanks}, {Prieto-Arranz}, {Madhusudhan}, {Gandhi}, {Casasayas-Barris}, {Murgas}, {Nortmann}, {Crouzet}, {Parviainen}, \& {Gandolfi}}]{chenetal2018}
{Chen}, G., {Pall{\'e}}, E., {Welbanks}, L., {et~al.} 2018, \aap, 616, A145, \dodoi{10.1051/0004-6361/201833033}

\bibitem[{{Chubb} {et~al.}(2020){Chubb}, {Tennyson}, \& {Yurchenko}}]{chubb_et_al2020}
{Chubb}, K.~L., {Tennyson}, J., \& {Yurchenko}, S.~N. 2020, \mnras, 493, 1531, \dodoi{10.1093/mnras/staa229}

\bibitem[{{Coles} {et~al.}(2019){Coles}, {Yurchenko}, \& {Tennyson}}]{coles_et_al2019}
{Coles}, P.~A., {Yurchenko}, S.~N., \& {Tennyson}, J. 2019, \mnras, 490, 4638, \dodoi{10.1093/mnras/stz2778}

\bibitem[{{Coulombe} {et~al.}(2023){Coulombe}, {Benneke}, {Challener}, {Piette}, {Wiser}, {Mansfield}, {MacDonald}, {Beltz}, {Feinstein}, {Radica}, {Savel}, {Dos Santos}, {Bean}, {Parmentier}, {Wong}, {Rauscher}, {Komacek}, {Kempton}, {Tan}, {Hammond}, {Lewis}, {Line}, {Lee}, {Shivkumar}, {Crossfield}, {Nixon}, {Rackham}, {Wakeford}, {Welbanks}, {Zhang}, {Batalha}, {Berta-Thompson}, {Changeat}, {D{\'e}sert}, {Espinoza}, {Goyal}, {Harrington}, {Knutson}, {Kreidberg}, {L{\'o}pez-Morales}, {Shporer}, {Sing}, {Stevenson}, {Aggarwal}, {Ahrer}, {Alam}, {Bell}, {Blecic}, {Caceres}, {Carter}, {Casewell}, {Crouzet}, {Cubillos}, {Decin}, {Fortney}, {Gibson}, {Heng}, {Henning}, {Iro}, {Kendrew}, {Lagage}, {Leconte}, {Lendl}, {Lothringer}, {Mancini}, {Mikal-Evans}, {Molaverdikhani}, {Nikolov}, {Ohno}, {Palle}, {Piaulet}, {Redfield}, {Roy}, {Tsai}, {Venot}, \& {Wheatley}}]{Coulombe_2023NatureERSW18b}
{Coulombe}, L.-P., {Benneke}, B., {Challener}, R., {et~al.} 2023, \nat, 620, 292, \dodoi{10.1038/s41586-023-06230-1}

\bibitem[{{D'Angelo} {et~al.}(2010){D'Angelo}, {Durisen}, \& {Lissauer}}]{DAngeloetal2010}
{D'Angelo}, G., {Durisen}, R.~H., \& {Lissauer}, J.~J. 2010, in Exoplanets, ed. S.~{Seager}, 319--346, \dodoi{10.48550/arXiv.1006.5486}

\bibitem[{{D'Angelo} \& {Lissauer}(2018)}]{DAngeloetal2018}
{D'Angelo}, G., \& {Lissauer}, J.~J. 2018, in Handbook of Exoplanets, ed. H.~J. {Deeg} \& J.~A. {Belmonte}, 140, \dodoi{10.1007/978-3-319-55333-7_140}

\bibitem[{{Deibert} {et~al.}(2019){Deibert}, {de Mooij}, {Jayawardhana}, {Fortney}, {Brogi}, {Rustamkulov}, \& {Tamura}}]{deibertetal2019}
{Deibert}, E.~K., {de Mooij}, E. J.~W., {Jayawardhana}, R., {et~al.} 2019, \aj, 157, 58, \dodoi{10.3847/1538-3881/aaf56b}

\bibitem[{{Doyle} {et~al.}(2022){Doyle}, {Cegla}, {Bryant}, {Bayliss}, {Lafarga}, {Anderson}, {Allart}, {Bourrier}, {Brogi}, {Buchschacher}, {Kunovac}, {Lendl}, {Lovis}, {Moyano}, {Roguet-Kern}, {Seidel}, {Sosnowska}, {Wheatley}, {Acton}, {Burleigh}, {Casewell}, {Gill}, {Goad}, {Henderson}, {Jenkins}, {Tilbrook}, \& {West}}]{Doyleetal2022}
{Doyle}, L., {Cegla}, H.~M., {Bryant}, E., {et~al.} 2022, \mnras, 516, 298, \dodoi{10.1093/mnras/stac2178}

\bibitem[{{Doyon} {et~al.}(2023){Doyon}, {Willott}, {Hutchings}, {Sivaramakrishnan}, {Albert}, {Lafreni{\`e}re}, {Rowlands}, {Bego{\~n}a Vila}, {Martel}, {LaMassa}, {Aldridge}, {Artigau}, {Cameron}, {Chayer}, {Cook}, {Cooper}, {Darveau-Bernier}, {Dupuis}, {Earnshaw}, {Espinoza}, {Filippazzo}, {Fullerton}, {Gaudreau}, {Gawlik}, {Goudfrooij}, {Haley}, {Kammerer}, {Kendall}, {Lambros}, {Ignat}, {Maszkiewicz}, {McColgan}, {Morishita}, {Ouellette}, {Pacifici}, {Philippi}, {Radica}, {Ravindranath}, {Rowe}, {Roy}, {Roy}, {Saad}, {Sohn}, {Talens}, {Touahri}, {Thatte}, {Taylor}, {Vandal}, {Volk}, {Wander}, {Warner}, {Zheng}, {Zhou}, {Abraham}, {Beaulieu}, {Benneke}, {Ferrarese}, {Jayawardhana}, {Johnstone}, {Kaltenegger}, {Meyer}, {Pipher}, {Rameau}, {Rieke}, {Salhi}, \& {Sawicki}}]{doyonetal2023}
{Doyon}, R., {Willott}, C.~J., {Hutchings}, J.~B., {et~al.} 2023, \pasp, 135, 098001, \dodoi{10.1088/1538-3873/acd41b}

\bibitem[{{Dyrek} {et~al.}(2024){Dyrek}, {Min}, {Decin}, {Bouwman}, {Crouzet}, {Molli{\`e}re}, {Lagage}, {Konings}, {Tremblin}, {G{\"u}del}, {Pye}, {Waters}, {Henning}, {Vandenbussche}, {Ardevol Martinez}, {Argyriou}, {Ducrot}, {Heinke}, {van Looveren}, {Absil}, {Barrado}, {Baudoz}, {Boccaletti}, {Cossou}, {Coulais}, {Edwards}, {Gastaud}, {Glasse}, {Glauser}, {Greene}, {Kendrew}, {Krause}, {Lahuis}, {Mueller}, {Olofsson}, {Patapis}, {Rouan}, {Royer}, {Scheithauer}, {Waldmann}, {Whiteford}, {Colina}, {van Dishoeck}, {{\"O}stlin}, {Ray}, \& {Wright}}]{dyreketal2024}
{Dyrek}, A., {Min}, M., {Decin}, L., {et~al.} 2024, \nat, 625, 51, \dodoi{10.1038/s41586-023-06849-0}

\bibitem[{Espinoza(2022)}]{espinoza_nestor_2022_6960924}
Espinoza, N. 2022, TransitSpectroscopy, 0.3.11,  Zenodo, \dodoi{10.5281/zenodo.6960924}

\bibitem[{{Espinoza} \& {Jord{\'a}n}(2015)}]{EspinozaJordan2015}
{Espinoza}, N., \& {Jord{\'a}n}, A. 2015, \mnras, 450, 1879, \dodoi{10.1093/mnras/stv744}

\bibitem[{{Espinoza} \& {Jord{\'a}n}(2016)}]{EspinozaJordan2016}
---. 2016, \mnras, 457, 3573, \dodoi{10.1093/mnras/stw224}

\bibitem[{{Espinoza} {et~al.}(2023){Espinoza}, {{\'U}beda}, {Birkmann}, {Ferruit}, {Valenti}, {Sing}, {Rustamkulov}, {Regan}, {Kendrew}, {Sabbi}, {Schlawin}, {Beatty}, {Albert}, {Greene}, {Nikolov}, {Karakla}, {Keyes}, {Alves de Oliveira}, {B{\"o}ker}, {Pena-Guerrero}, {Giardino}, {Kumari}, {Manjavacas}, {Proffitt}, \& {Rawle}}]{Espinozaetal2023}
{Espinoza}, N., {{\'U}beda}, L., {Birkmann}, S.~M., {et~al.} 2023, \pasp, 135, 018002, \dodoi{10.1088/1538-3873/aca3d3}

\bibitem[{{Feinstein} {et~al.}(2023){Feinstein}, {Radica}, {Welbanks}, {Murray}, {Ohno}, {Coulombe}, {Espinoza}, {Bean}, {Teske}, {Benneke}, {Line}, {Rustamkulov}, {Saba}, {Tsiaras}, {Barstow}, {Fortney}, {Gao}, {Knutson}, {MacDonald}, {Mikal-Evans}, {Rackham}, {Taylor}, {Parmentier}, {Batalha}, {Berta-Thompson}, {Carter}, {Changeat}, {dos Santos}, {Gibson}, {Goyal}, {Kreidberg}, {L{\'o}pez-Morales}, {Lothringer}, {Miguel}, {Molaverdikhani}, {Moran}, {Morello}, {Mukherjee}, {Sing}, {Stevenson}, {Wakeford}, {Ahrer}, {Alam}, {Alderson}, {Allen}, {Batalha}, {Bell}, {Blecic}, {Brande}, {Caceres}, {Casewell}, {Chubb}, {Crossfield}, {Crouzet}, {Cubillos}, {Decin}, {D{\'e}sert}, {Harrington}, {Heng}, {Henning}, {Iro}, {Kempton}, {Kendrew}, {Kirk}, {Krick}, {Lagage}, {Lendl}, {Mancini}, {Mansfield}, {May}, {Mayne}, {Nikolov}, {Palle}, {Petit dit de la Roche}, {Piaulet}, {Powell}, {Redfield}, {Rogers}, {Roman}, {Roy}, {Nixon}, {Schlawin}, {Tan}, {Tremblin}, {Turner}, {Venot}, {Waalkes}, {Wheatley}, \&
  {Zhang}}]{feinsteinetal2023}
{Feinstein}, A.~D., {Radica}, M., {Welbanks}, L., {et~al.} 2023, \nat, 614, 670, \dodoi{10.1038/s41586-022-05674-1}

\bibitem[{{Feroz} {et~al.}(2009){Feroz}, {Hobson}, \& {Bridges}}]{ferozetal2009}
{Feroz}, F., {Hobson}, M.~P., \& {Bridges}, M. 2009, \mnras, 398, 1601, \dodoi{10.1111/j.1365-2966.2009.14548.x}

\bibitem[{{Feroz} {et~al.}(2013){Feroz}, {Hobson}, {Cameron}, \& {Pettitt}}]{ferozetal2013}
{Feroz}, F., {Hobson}, M.~P., {Cameron}, E., \& {Pettitt}, A.~N. 2013, ArXiv e-prints.
\newblock \doarXiv{1306.2144}

\bibitem[{{Foreman-Mackey} {et~al.}(2013){Foreman-Mackey}, {Hogg}, {Lang}, \& {Goodman}}]{Foreman-Mackey2013}
{Foreman-Mackey}, D., {Hogg}, D.~W., {Lang}, D., \& {Goodman}, J. 2013, \pasp, 125, 306, \dodoi{10.1086/670067}

\bibitem[{{Fortney}(2005)}]{Fortney2005}
{Fortney}, J.~J. 2005, \mnras, 364, 649, \dodoi{10.1111/j.1365-2966.2005.09587.x}

\bibitem[{{Fraine} {et~al.}(2014){Fraine}, {Deming}, {Benneke}, {Knutson}, {Jord{\'a}n}, {Espinoza}, {Madhusudhan}, {Wilkins}, \& {Todorov}}]{Fraine2014}
{Fraine}, J., {Deming}, D., {Benneke}, B., {et~al.} 2014, Natur, 513, 526, \dodoi{10.1038/nature13785}

\bibitem[{{GharibNezhad} {et~al.}(2013){GharibNezhad}, {Shayesteh}, \& {Bernath}}]{gharibnezhadetal2013}
{GharibNezhad}, E., {Shayesteh}, A., \& {Bernath}, P.~F. 2013, \mnras, 432, 2043, \dodoi{10.1093/mnras/stt510}

\bibitem[{{Goodman} \& {Weare}(2010)}]{GoodmanWeare2010}
{Goodman}, J., \& {Weare}, J. 2010, Communications in Applied Mathematics and Computational Science, 5, 65, \dodoi{10.2140/camcos.2010.5.65}

\bibitem[{{Gordon} \& {Trotta}(2007)}]{gordonandtrotta2007}
{Gordon}, C., \& {Trotta}, R. 2007, \mnras, 382, 1859, \dodoi{10.1111/j.1365-2966.2007.12707.x}

\bibitem[{{Gordon} {et~al.}(2017){Gordon}, {Rothman}, {Hill}, {Kochanov}, {Tan}, {Bernath}, {Birk}, {Boudon}, {Campargue}, {Chance}, {Drouin}, {Flaud}, {Gamache}, {Hodges}, {Jacquemart}, {Perevalov}, {Perrin}, {Shine}, {Smith}, {Tennyson}, {Toon}, {Tran}, {Tyuterev}, {Barbe}, {Cs{\'a}sz{\'a}r}, {Devi}, {Furtenbacher}, {Harrison}, {Hartmann}, {Jolly}, {Johnson}, {Karman}, {Kleiner}, {Kyuberis}, {Loos}, {Lyulin}, {Massie}, {Mikhailenko}, {Moazzen-Ahmadi}, {M{\"u}ller}, {Naumenko}, {Nikitin}, {Polyansky}, {Rey}, {Rotger}, {Sharpe}, {Sung}, {Starikova}, {Tashkun}, {Auwera}, {Wagner}, {Wilzewski}, {Wcis{\l}o}, {Yu}, \& {Zak}}]{gordonetal2017}
{Gordon}, I.~E., {Rothman}, L.~S., {Hill}, C., {et~al.} 2017, \jqsrt, 203, 3, \dodoi{10.1016/j.jqsrt.2017.06.038}

\bibitem[{{Gordon} {et~al.}(2022){Gordon}, {Rothman}, {Hargreaves}, {Hashemi}, {Karlovets}, {Skinner}, {Conway}, {Hill}, {Kochanov}, {Tan}, {Wcis{\l}o}, {Finenko}, {Nelson}, {Bernath}, {Birk}, {Boudon}, {Campargue}, {Chance}, {Coustenis}, {Drouin}, {Flaud}, {Gamache}, {Hodges}, {Jacquemart}, {Mlawer}, {Nikitin}, {Perevalov}, {Rotger}, {Tennyson}, {Toon}, {Tran}, {Tyuterev}, {Adkins}, {Baker}, {Barbe}, {Can{\`e}}, {Cs{\'a}sz{\'a}r}, {Dudaryonok}, {Egorov}, {Fleisher}, {Fleurbaey}, {Foltynowicz}, {Furtenbacher}, {Harrison}, {Hartmann}, {Horneman}, {Huang}, {Karman}, {Karns}, {Kassi}, {Kleiner}, {Kofman}, {Kwabia-Tchana}, {Lavrentieva}, {Lee}, {Long}, {Lukashevskaya}, {Lyulin}, {Makhnev}, {Matt}, {Massie}, {Melosso}, {Mikhailenko}, {Mondelain}, {M{\"u}ller}, {Naumenko}, {Perrin}, {Polyansky}, {Raddaoui}, {Raston}, {Reed}, {Rey}, {Richard}, {T{\'o}bi{\'a}s}, {Sadiek}, {Schwenke}, {Starikova}, {Sung}, {Tamassia}, {Tashkun}, {Vander Auwera}, {Vasilenko}, {Vigasin}, {Villanueva}, {Vispoel}, {Wagner}, {Yachmenev}, \&
  {Yurchenko}}]{gordonetal2022}
{Gordon}, I.~E., {Rothman}, L.~S., {Hargreaves}, R.~J., {et~al.} 2022, \jqsrt, 277, 107949, \dodoi{10.1016/j.jqsrt.2021.107949}

\bibitem[{{Gorman} {et~al.}(2019){Gorman}, {Yurchenko}, \& {Tennyson}}]{Gormanetal2019}
{Gorman}, M.~N., {Yurchenko}, S.~N., \& {Tennyson}, J. 2019, \mnras, 490, 1652, \dodoi{10.1093/mnras/stz2517}

\bibitem[{Grant \& Wakeford(2022)}]{david_grant_2022_ExoTiC-LD}
Grant, D., \& Wakeford, H.~R. 2022, Exo-TiC/ExoTiC-LD: ExoTiC-LD v3.0.0, v3.0.0,  Zenodo, \dodoi{10.5281/zenodo.7437681}

\bibitem[{{Greene} {et~al.}(2016){Greene}, {Line}, {Montero}, {Fortney}, {Lustig-Yaeger}, \& {Luther}}]{greeneetal2016}
{Greene}, T.~P., {Line}, M.~R., {Montero}, C., {et~al.} 2016, \apj, 817, 17, \dodoi{10.3847/0004-637X/817/1/17}

\bibitem[{{Gressier} {et~al.}(2025){Gressier}, {MacDonald}, {Espinoza}, {Wakeford}, {Lewis}, {Goyal}, {Louie}, {Radica}, {Batalha}, {Long}, {May}, {Mullens}, {Seager}, {Stevenson}, {Valenti}, {Alderson}, {Allen}, {Ca{\~n}as}, {Challener}, {Col{\'o}n}, {Glidden}, {Grant}, {Huang}, {Lin}, {Valentine}, {Mountain}, {Pueyo}, {Perrin}, \& {van der Marel}}]{Gressier_2025}
{Gressier}, A., {MacDonald}, R.~J., {Espinoza}, N., {et~al.} 2025, \aj, 169, 57, \dodoi{10.3847/1538-3881/ad97bf}

\bibitem[{{Guilluy} {et~al.}(2021){Guilluy}, {Gressier}, {Wright}, {Santerne}, {Jaziri}, {Edwards}, {Changeat}, {Modirrousta-Galian}, {Skaf}, {Al-Refaie}, {Baeyens}, {Bieger}, {Blain}, {Kiefer}, {Morvan}, {Mugnai}, {Pluriel}, {Poveda}, {Zingales}, {Whiteford}, {Yip}, {Charnay}, {Leconte}, {Drossart}, {Sozzetti}, {Marcq}, {Tsiaras}, {Venot}, {Waldmann}, \& {Beaulieu}}]{guilluyetal2021}
{Guilluy}, G., {Gressier}, A., {Wright}, S., {et~al.} 2021, \aj, 161, 19, \dodoi{10.3847/1538-3881/abc3c8}

\bibitem[{{Guilluy} {et~al.}(2023){Guilluy}, {Bourrier}, {Jaziri}, {Dethier}, {Mounzer}, {Giacobbe}, {Attia}, {Allart}, {Bonomo}, {Dos Santos}, {Rainer}, \& {Sozzetti}}]{guilluyetal2023}
{Guilluy}, G., {Bourrier}, V., {Jaziri}, Y., {et~al.} 2023, \aap, 676, A130, \dodoi{10.1051/0004-6361/202346419}

\bibitem[{{Guzm{\'a}n-Mesa} {et~al.}(2020){Guzm{\'a}n-Mesa}, {Kitzmann}, {Fisher}, {Burgasser}, {Hoeijmakers}, {M{\'a}rquez-Neila}, {Grimm}, {Mandell}, {Sznitman}, \& {Heng}}]{guzmanmesaetal2020}
{Guzm{\'a}n-Mesa}, A., {Kitzmann}, D., {Fisher}, C., {et~al.} 2020, \aj, 160, 15, \dodoi{10.3847/1538-3881/ab9176}

\bibitem[{{Hargreaves} {et~al.}(2019){Hargreaves}, {Gordon}, {Rothman}, {Tashkun}, {Perevalov}, {Lukashevskaya}, {Yurchenko}, {Tennyson}, \& {M{\"u}ller}}]{Hargreavesetal2019}
{Hargreaves}, R.~J., {Gordon}, I.~E., {Rothman}, L.~S., {et~al.} 2019, \jqsrt, 232, 35, \dodoi{10.1016/j.jqsrt.2019.04.040}

\bibitem[{{Helled} {et~al.}(2016){Helled}, {Lozovsky}, \& {Zucker}}]{helledetal2016}
{Helled}, R., {Lozovsky}, M., \& {Zucker}, S. 2016, \mnras, 455, L96, \dodoi{10.1093/mnrasl/slv158}

\bibitem[{{Hellier} {et~al.}(2019){Hellier}, {Anderson}, {Triaud}, {Bouchy}, {Burdanov}, {Collier Cameron}, {Delrez}, {Ehrenreich}, {Gillon}, {Jehin}, {Lendl}, {Linder}, {Nielsen}, {Maxted}, {Pepe}, {Pollacco}, {Queloz}, {S{\'e}gransan}, {Smalley}, {Spake}, {Temple}, {Udry}, {West}, \& {Wyttenbach}}]{Hellieretal2019}
{Hellier}, C., {Anderson}, D.~R., {Triaud}, A.~H.~M.~J., {et~al.} 2019, \mnras, 488, 3067, \dodoi{10.1093/mnras/stz1903}

\bibitem[{{Heng}(2017)}]{Heng2017}
{Heng}, K. 2017, {Exoplanetary Atmospheres: Theoretical Concepts and Foundations}

\bibitem[{{Hoeijmakers} {et~al.}(2019){Hoeijmakers}, {Ehrenreich}, {Kitzmann}, {Allart}, {Grimm}, {Seidel}, {Wyttenbach}, {Pino}, {Nielsen}, {Fisher}, {Rimmer}, {Bourrier}, {Cegla}, {Lavie}, {Lovis}, {Patzer}, {Stock}, {Pepe}, \& {Heng}}]{hoeijmakersetal2019}
{Hoeijmakers}, H.~J., {Ehrenreich}, D., {Kitzmann}, D., {et~al.} 2019, \aap, 627, A165, \dodoi{10.1051/0004-6361/201935089}

\bibitem[{{Horne}(1986)}]{Horne1986}
{Horne}, K. 1986, \pasp, 98, 609, \dodoi{10.1086/131801}

\bibitem[{{Howe} {et~al.}(2017){Howe}, {Burrows}, \& {Deming}}]{howeetal2017}
{Howe}, A.~R., {Burrows}, A., \& {Deming}, D. 2017, \apj, 835, 96, \dodoi{10.3847/1538-4357/835/1/96}

\bibitem[{{Jakobsen} {et~al.}(2022){Jakobsen}, {Ferruit}, {Alves de Oliveira}, {Arribas}, {Bagnasco}, {Barho}, {Beck}, {Birkmann}, {B{\"o}ker}, {Bunker}, {Charlot}, {de Jong}, {de Marchi}, {Ehrenwinkler}, {Falcolini}, {Fels}, {Franx}, {Franz}, {Funke}, {Giardino}, {Gnata}, {Holota}, {Honnen}, {Jensen}, {Jentsch}, {Johnson}, {Jollet}, {Karl}, {Kling}, {K{\"o}hler}, {Kolm}, {Kumari}, {Lander}, {Lemke}, {L{\'o}pez-Caniego}, {L{\"u}tzgendorf}, {Maiolino}, {Manjavacas}, {Marston}, {Maschmann}, {Maurer}, {Messerschmidt}, {Moseley}, {Mosner}, {Mott}, {Muzerolle}, {Pirzkal}, {Pittet}, {Plitzke}, {Posselt}, {Rapp}, {Rauscher}, {Rawle}, {Rix}, {R{\"o}del}, {Rumler}, {Sabbi}, {Salvignol}, {Schmid}, {Sirianni}, {Smith}, {Strada}, {te Plate}, {Valenti}, {Wettemann}, {Wiehe}, {Wiesmayer}, {Willott}, {Wright}, {Zeidler}, \& {Zincke}}]{jakobsenetal2022}
{Jakobsen}, P., {Ferruit}, P., {Alves de Oliveira}, C., {et~al.} 2022, \aap, 661, A80, \dodoi{10.1051/0004-6361/202142663}

\bibitem[{{Jenkins} {et~al.}(2020){Jenkins}, {D{\'\i}az}, {Kurtovic}, {Espinoza}, {Vines}, {Rojas}, {Brahm}, {Torres}, {Cort{\'e}s-Zuleta}, {Soto}, {Lopez}, {King}, {Wheatley}, {Winn}, {Ciardi}, {Ricker}, {Vanderspek}, {Latham}, {Seager}, {Jenkins}, {Beichman}, {Bieryla}, {Burke}, {Christiansen}, {Henze}, {Klaus}, {McCauliff}, {Mori}, {Narita}, {Nishiumi}, {Tamura}, {de Leon}, {Quinn}, {Villase{\~n}or}, {Vezie}, {Lissauer}, {Collins}, {Collins}, {Isopi}, {Mallia}, {Ercolino}, {Petrovich}, {Jord{\'a}n}, {Acton}, {Armstrong}, {Bayliss}, {Bouchy}, {Belardi}, {Bryant}, {Burleigh}, {Cabrera}, {Casewell}, {Chaushev}, {Cooke}, {Eigm{\"u}ller}, {Erikson}, {Foxell}, {G{\"a}nsicke}, {Gill}, {Gillen}, {G{\"u}nther}, {Goad}, {Hooton}, {Jackman}, {Louden}, {McCormac}, {Moyano}, {Nielsen}, {Pollacco}, {Queloz}, {Rauer}, {Raynard}, {Smith}, {Tilbrook}, {Titz-Weider}, {Turner}, {Udry}, {Walker}, {Watson}, {West}, {Palle}, {Ziegler}, {Law}, \& {Mann}}]{jenkinsetal2020}
{Jenkins}, J.~S., {D{\'\i}az}, M.~R., {Kurtovic}, N.~T., {et~al.} 2020, Nature Astronomy, 4, 1148, \dodoi{10.1038/s41550-020-1142-z}

\bibitem[{{Jensen} {et~al.}(2018){Jensen}, {Cauley}, {Redfield}, {Cochran}, \& {Endl}}]{jensenetal2018}
{Jensen}, A.~G., {Cauley}, P.~W., {Redfield}, S., {Cochran}, W.~D., \& {Endl}, M. 2018, \aj, 156, 154, \dodoi{10.3847/1538-3881/aadca7}

\bibitem[{{JWST Transiting Exoplanet Community Early Release Science Team} {et~al.}(2023){JWST Transiting Exoplanet Community Early Release Science Team}, {Ahrer}, {Alderson}, {Batalha}, {Batalha}, {Bean}, {Beatty}, {Bell}, {Benneke}, {Berta-Thompson}, {Carter}, {Crossfield}, {Espinoza}, {Feinstein}, {Fortney}, {Gibson}, {Goyal}, {Kempton}, {Kirk}, {Kreidberg}, {L{\'o}pez-Morales}, {Line}, {Lothringer}, {Moran}, {Mukherjee}, {Ohno}, {Parmentier}, {Piaulet}, {Rustamkulov}, {Schlawin}, {Sing}, {Stevenson}, {Wakeford}, {Allen}, {Birkmann}, {Brande}, {Crouzet}, {Cubillos}, {Damiano}, {D{\'e}sert}, {Gao}, {Harrington}, {Hu}, {Kendrew}, {Knutson}, {Lagage}, {Leconte}, {Lendl}, {MacDonald}, {May}, {Miguel}, {Molaverdikhani}, {Moses}, {Murray}, {Nehring}, {Nikolov}, {Petit dit de la Roche}, {Radica}, {Roy}, {Stassun}, {Taylor}, {Waalkes}, {Wachiraphan}, {Welbanks}, {Wheatley}, {Aggarwal}, {Alam}, {Banerjee}, {Barstow}, {Blecic}, {Casewell}, {Changeat}, {Chubb}, {Col{\'o}n}, {Coulombe}, {Daylan}, {de Val-Borro},
  {Decin}, {Dos Santos}, {Flagg}, {France}, {Fu}, {Garc{\'\i}a Mu{\~n}oz}, {Gizis}, {Glidden}, {Grant}, {Heng}, {Henning}, {Hong}, {Inglis}, {Iro}, {Kataria}, {Komacek}, {Krick}, {Lee}, {Lewis}, {Lillo-Box}, {Lustig-Yaeger}, {Mancini}, {Mandell}, {Mansfield}, {Marley}, {Mikal-Evans}, {Morello}, {Nixon}, {Ortiz Ceballos}, {Piette}, {Powell}, {Rackham}, {Ramos-Rosado}, {Rauscher}, {Redfield}, {Rogers}, {Roman}, {Roudier}, {Scarsdale}, {Shkolnik}, {Southworth}, {Spake}, {Steinrueck}, {Tan}, {Teske}, {Tremblin}, {Tsai}, {Tucker}, {Turner}, {Valenti}, {Venot}, {Waldmann}, {Wallack}, {Zhang}, \& {Zieba}}]{JWSTERSTEAM2023}
{JWST Transiting Exoplanet Community Early Release Science Team}, {Ahrer}, E.-M., {Alderson}, L., {et~al.} 2023, \nat, 614, 649, \dodoi{10.1038/s41586-022-05269-w}

\bibitem[{{Kempton} {et~al.}(2017){Kempton}, {Lupu}, {Owusu-Asare}, {Slough}, \& {Cale}}]{kemptonetal2017}
{Kempton}, E. M.~R., {Lupu}, R., {Owusu-Asare}, A., {Slough}, P., \& {Cale}, B. 2017, \pasp, 129, 044402, \dodoi{10.1088/1538-3873/aa61ef}

\bibitem[{{Kempton} {et~al.}(2014){Kempton}, {Perna}, \& {Heng}}]{Kemptonetal2014}
{Kempton}, E. M.~R., {Perna}, R., \& {Heng}, K. 2014, \apj, 795, 24, \dodoi{10.1088/0004-637X/795/1/24}

\bibitem[{{Kirk} {et~al.}(2024){Kirk}, {Stevenson}, {Fu}, {Lustig-Yaeger}, {Moran}, {Peacock}, {Alam}, {Batalha}, {Bennett}, {Gonzalez-Quiles}, {L{\'o}pez-Morales}, {Lothringer}, {MacDonald}, {May}, {Mayorga}, {Rustamkulov}, {Sing}, {Sotzen}, {Valenti}, \& {Wakeford}}]{Kirketal2024}
{Kirk}, J., {Stevenson}, K.~B., {Fu}, G., {et~al.} 2024, \aj, 167, 90, \dodoi{10.3847/1538-3881/ad19df}

\bibitem[{{Kitzmann} \& {Stock}(2018)}]{Kitzmannetal2018}
{Kitzmann}, D., \& {Stock}, J. 2018, {FastChem: An ultra-fast equilibrium chemistry}, Astrophysics Source Code Library, record ascl:1804.025

\bibitem[{{Koll}(2022)}]{Koll2022}
{Koll}, D. D.~B. 2022, \apj, 924, 134, \dodoi{10.3847/1538-4357/ac3b48}

\bibitem[{{Kopal}(1950)}]{Kopal1950}
{Kopal}, Z. 1950, Harvard College Observatory Circular, 454, 1

\bibitem[{Kramida {et~al.}(2013)Kramida, Ralchenko, \& Reader}]{kramida_et_al2013}
Kramida, A., Ralchenko, Y., \& Reader, J. 2013, Atomic Spectra Database.
\newblock \url{http://www.nist.gov/pml/data/asd.cfm}

\bibitem[{{Kreidberg}(2015)}]{Kreidbergetal2015}
{Kreidberg}, L. 2015, \pasp, 127, 1161, \dodoi{10.1086/683602}

\bibitem[{{Kreidberg} {et~al.}(2022){Kreidberg}, {Molli{\`e}re}, {Crossfield}, {Thorngren}, {Kawashima}, {Morley}, {Benneke}, {Mikal-Evans}, {Berardo}, {Kosiarek}, {Gorjian}, {Ciardi}, {Christiansen}, {Dragomir}, {Dressing}, {Fortney}, {Fulton}, {Greene}, {Hardegree-Ullman}, {Howard}, {Howell}, {Isaacson}, {Krick}, {Livingston}, {Lothringer}, {Morales}, {Petigura}, {Rodriguez}, {Schlieder}, \& {Weiss}}]{kreidbergetal2022}
{Kreidberg}, L., {Molli{\`e}re}, P., {Crossfield}, I. J.~M., {et~al.} 2022, \aj, 164, 124, \dodoi{10.3847/1538-3881/ac85be}

\bibitem[{{Lafarga} {et~al.}(2023){Lafarga}, {Brogi}, {Gandhi}, {Cegla}, {Seidel}, {Doyle}, {Allart}, {Buchschacher}, {Lendl}, {Lovis}, \& {Sosnowska}}]{lafargaetal2023}
{Lafarga}, M., {Brogi}, M., {Gandhi}, S., {et~al.} 2023, \mnras, 521, 1233, \dodoi{10.1093/mnras/stad480}

\bibitem[{{Lee} \& {Chiang}(2016)}]{leeandchiang2016}
{Lee}, E.~J., \& {Chiang}, E. 2016, \apj, 817, 90, \dodoi{10.3847/0004-637X/817/2/90}

\bibitem[{{Li} {et~al.}(2013){Li}, {Gordon}, {Le Roy}, {Hajigeorgiou}, {Coxon}, {Bernath}, \& {Rothman}}]{lietal2013}
{Li}, G., {Gordon}, I.~E., {Le Roy}, R.~J., {et~al.} 2013, \jqsrt, 121, 78, \dodoi{10.1016/j.jqsrt.2013.02.005}

\bibitem[{{Li} {et~al.}(2015){Li}, {Gordon}, {Rothman}, {Tan}, {Hu}, {Kassi}, {Campargue}, \& {Medvedev}}]{li_et_al2015}
{Li}, G., {Gordon}, I.~E., {Rothman}, L.~S., {et~al.} 2015, \apjs, 216, 15, \dodoi{10.1088/0067-0049/216/1/15}

\bibitem[{Louie {et~al.}(2025)Louie, Mullens, Alderson, Glidden, Lewis, Wakeford, Batalha, Colón, Gressier, Long, Radica, Espinoza, Goyal, MacDonald, May, Seager, Stevenson, Valenti, Allen, Cañas, Challener, Grant, Huang, Lin, Valentine, Clampin, Perrin, Pueyo, van~der Marel, \& Mountain}]{Louie_2025}
Louie, D.~R., Mullens, E., Alderson, L., {et~al.} 2025, AJ, 169, 86, \dodoi{10.3847/1538-3881/ad9688}

\bibitem[{{Lustig-Yaeger} {et~al.}(2023){Lustig-Yaeger}, {Fu}, {May}, {Ceballos}, {Moran}, {Peacock}, {Stevenson}, {Kirk}, {L{\'o}pez-Morales}, {MacDonald}, {Mayorga}, {Sing}, {Sotzen}, {Valenti}, {Redai}, {Alam}, {Batalha}, {Bennett}, {Gonzalez-Quiles}, {Kruse}, {Lothringer}, {Rustamkulov}, \& {Wakeford}}]{LustigYaegeretal2023}
{Lustig-Yaeger}, J., {Fu}, G., {May}, E.~M., {et~al.} 2023, Nature Astronomy, 7, 1317, \dodoi{10.1038/s41550-023-02064-z}

\bibitem[{{MacDonald}(2023)}]{macdonald2023}
{MacDonald}, R.~J. 2023, The Journal of Open Source Software, 8, 4873, \dodoi{10.21105/joss.04873}

\bibitem[{{MacDonald} {et~al.}(2020){MacDonald}, {Goyal}, \& {Lewis}}]{macdonaldetal2020}
{MacDonald}, R.~J., {Goyal}, J.~M., \& {Lewis}, N.~K. 2020, \apjl, 893, L43, \dodoi{10.3847/2041-8213/ab8238}

\bibitem[{{MacDonald} \& {Lewis}(2022)}]{macdonaldlewis2022}
{MacDonald}, R.~J., \& {Lewis}, N.~K. 2022, \apj, 929, 20, \dodoi{10.3847/1538-4357/ac47fe}

\bibitem[{{MacDonald} \& {Madhusudhan}(2017)}]{macdonaldandmadhusudhan2017}
{MacDonald}, R.~J., \& {Madhusudhan}, N. 2017, \mnras, 469, 1979, \dodoi{10.1093/mnras/stx804}

\bibitem[{{MacDonald} \& {Madhusudhan}(2019)}]{macdonaldandmadhusudhan2019}
---. 2019, \mnras, 486, 1292, \dodoi{10.1093/mnras/stz789}

\bibitem[{{Madhusudhan}(2019)}]{madhusudhan2019}
{Madhusudhan}, N. 2019, \araa, 57, 617, \dodoi{10.1146/annurev-astro-081817-051846}

\bibitem[{{Madhusudhan} {et~al.}(2016){Madhusudhan}, {Ag{\'u}ndez}, {Moses}, \& {Hu}}]{Madhusudhanetal2016}
{Madhusudhan}, N., {Ag{\'u}ndez}, M., {Moses}, J.~I., \& {Hu}, Y. 2016, \ssr, 205, 285, \dodoi{10.1007/s11214-016-0254-3}

\bibitem[{{Madhusudhan} {et~al.}(2017){Madhusudhan}, {Bitsch}, {Johansen}, \& {Eriksson}}]{madhusudhanetal2017}
{Madhusudhan}, N., {Bitsch}, B., {Johansen}, A., \& {Eriksson}, L. 2017, \mnras, 469, 4102, \dodoi{10.1093/mnras/stx1139}

\bibitem[{Magic {et~al.}(2015)Magic, Chiavassa, Collet, \& Asplund}]{magic2015stagger}
Magic, Z., Chiavassa, A., Collet, R., \& Asplund, M. 2015, Astronomy \& Astrophysics, 573, A90

\bibitem[{{Mant} {et~al.}(2018){Mant}, {Yachmenev}, {Tennyson}, \& {Yurchenko}}]{mantetal2018}
{Mant}, B.~P., {Yachmenev}, A., {Tennyson}, J., \& {Yurchenko}, S.~N. 2018, \mnras, 478, 3220, \dodoi{10.1093/mnras/sty1239}

\bibitem[{{Matsakos} \& {K{\"o}nigl}(2016)}]{matsakosandkonigl2016}
{Matsakos}, T., \& {K{\"o}nigl}, A. 2016, \apjl, 820, L8, \dodoi{10.3847/2041-8205/820/1/L8}

\bibitem[{{May} {et~al.}(2023){May}, {MacDonald}, {Bennett}, {Moran}, {Wakeford}, {Peacock}, {Lustig-Yaeger}, {Highland}, {Stevenson}, {Sing}, {Mayorga}, {Batalha}, {Kirk}, {L{\'o}pez-Morales}, {Valenti}, {Alam}, {Alderson}, {Fu}, {Gonzalez-Quiles}, {Lothringer}, {Rustamkulov}, \& {Sotzen}}]{Mayetal2023}
{May}, E.~M., {MacDonald}, R.~J., {Bennett}, K.~A., {et~al.} 2023, \apjl, 959, L9, \dodoi{10.3847/2041-8213/ad054f}

\bibitem[{{Mazeh} {et~al.}(2016){Mazeh}, {Holczer}, \& {Faigler}}]{mazehetal2016}
{Mazeh}, T., {Holczer}, T., \& {Faigler}, S. 2016, \aap, 589, A75, \dodoi{10.1051/0004-6361/201528065}

\bibitem[{{McKemmish} {et~al.}(2019){McKemmish}, {Masseron}, {Hoeijmakers}, {P{\'e}rez-Mesa}, {Grimm}, {Yurchenko}, \& {Tennyson}}]{Mckemmishetal2019}
{McKemmish}, L.~K., {Masseron}, T., {Hoeijmakers}, H.~J., {et~al.} 2019, \mnras, 488, 2836, \dodoi{10.1093/mnras/stz1818}

\bibitem[{{McKemmish} {et~al.}(2016){McKemmish}, {Yurchenko}, \& {Tennyson}}]{mckemmishetal2016}
{McKemmish}, L.~K., {Yurchenko}, S.~N., \& {Tennyson}, J. 2016, \mnras, 463, 771, \dodoi{10.1093/mnras/stw1969}

\bibitem[{{Moran} {et~al.}(2023){Moran}, {Stevenson}, {Sing}, {MacDonald}, {Kirk}, {Lustig-Yaeger}, {Peacock}, {Mayorga}, {Bennett}, {L{\'o}pez-Morales}, {May}, {Rustamkulov}, {Valenti}, {Adams Redai}, {Alam}, {Batalha}, {Fu}, {Gonzalez-Quiles}, {Highland}, {Kruse}, {Lothringer}, {Ortiz Ceballos}, {Sotzen}, \& {Wakeford}}]{Moranetal2023}
{Moran}, S.~E., {Stevenson}, K.~B., {Sing}, D.~K., {et~al.} 2023, \apjl, 948, L11, \dodoi{10.3847/2041-8213/accb9c}

\bibitem[{{Moses} {et~al.}(2011){Moses}, {Visscher}, {Fortney}, {Showman}, {Lewis}, {Griffith}, {Klippenstein}, {Shabram}, {Friedson}, {Marley}, \& {Freedman}}]{Mosesetal2011}
{Moses}, J.~I., {Visscher}, C., {Fortney}, J.~J., {et~al.} 2011, \apj, 737, 15, \dodoi{10.1088/0004-637X/737/1/15}

\bibitem[{{Mullens} {et~al.}(2024){Mullens}, {Lewis}, \& {MacDonald}}]{mullensetal2024}
{Mullens}, E., {Lewis}, N.~K., \& {MacDonald}, R.~J. 2024, \apj, 977, 105, \dodoi{10.3847/1538-4357/ad8575}

\bibitem[{{Nikolov} {et~al.}(2016){Nikolov}, {Sing}, {Gibson}, {Fortney}, {Evans}, {Barstow}, {Kataria}, \& {Wilson}}]{nikolovetal2016}
{Nikolov}, N., {Sing}, D.~K., {Gibson}, N.~P., {et~al.} 2016, \apj, 832, 191, \dodoi{10.3847/0004-637X/832/2/191}

\bibitem[{{{\"O}berg} {et~al.}(2011){{\"O}berg}, {Murray-Clay}, \& {Bergin}}]{obergetal2011}
{{\"O}berg}, K.~I., {Murray-Clay}, R., \& {Bergin}, E.~A. 2011, \apjl, 743, L16, \dodoi{10.1088/2041-8205/743/1/L16}

\bibitem[{{Owens} {et~al.}(2022){Owens}, {Dooley}, {McLaughlin}, {Tan}, {Zhang}, {Yurchenko}, \& {Tennyson}}]{Owensetal2022}
{Owens}, A., {Dooley}, S., {McLaughlin}, L., {et~al.} 2022, \mnras, 511, 5448, \dodoi{10.1093/mnras/stac371}

\bibitem[{{Owens} {et~al.}(2024){Owens}, {Yurchenko}, \& {Tennyson}}]{Owensetal2024}
{Owens}, A., {Yurchenko}, S.~N., \& {Tennyson}, J. 2024, \mnras, 530, 4004, \dodoi{10.1093/mnras/stae1110}

\bibitem[{{Polanski} {et~al.}(2022){Polanski}, {Crossfield}, {Howard}, {Isaacson}, \& {Rice}}]{polanskietal2022}
{Polanski}, A.~S., {Crossfield}, I. J.~M., {Howard}, A.~W., {Isaacson}, H., \& {Rice}, M. 2022, Research Notes of the American Astronomical Society, 6, 155, \dodoi{10.3847/2515-5172/ac8676}

\bibitem[{{Polyansky} {et~al.}(2018){Polyansky}, {Kyuberis}, {Zobov}, {Tennyson}, {Yurchenko}, \& {Lodi}}]{polyansky_et_al2018}
{Polyansky}, O.~L., {Kyuberis}, A.~A., {Zobov}, N.~F., {et~al.} 2018, \mnras, 480, 2597, \dodoi{10.1093/mnras/sty1877}

\bibitem[{{Qu} {et~al.}(2021){Qu}, {Yurchenko}, \& {Tennyson}}]{Quetal2021}
{Qu}, Q., {Yurchenko}, S.~N., \& {Tennyson}, J. 2021, \mnras, 504, 5768, \dodoi{10.1093/mnras/stab1154}

\bibitem[{{Rackham} {et~al.}(2018){Rackham}, {Apai}, \& {Giampapa}}]{rackhametal2018}
{Rackham}, B.~V., {Apai}, D., \& {Giampapa}, M.~S. 2018, \apj, 853, 122, \dodoi{10.3847/1538-4357/aaa08c}

\bibitem[{{Rackham} {et~al.}(2019){Rackham}, {Apai}, \& {Giampapa}}]{rackhametal2019}
---. 2019, \aj, 157, 96, \dodoi{10.3847/1538-3881/aaf892}

\bibitem[{{Radica}(2024)}]{Radica2024JOSS}
{Radica}, M. 2024, The Journal of Open Source Software, 9, 6898, \dodoi{10.21105/joss.06898}

\bibitem[{{Radica} {et~al.}(2025){Radica}, {Taylor}, {Wakeford}, {Lafreni{\`e}re}, {Allart}, {Cowan}, {Jenkins}, \& {Parmentier}}]{radicaetal2025}
{Radica}, M., {Taylor}, J., {Wakeford}, H.~R., {et~al.} 2025, \mnras, 538, 1853, \dodoi{10.1093/mnras/staf402}

\bibitem[{{Radica} {et~al.}(2023){Radica}, {Welbanks}, {Espinoza}, {Taylor}, {Coulombe}, {Feinstein}, {Goyal}, {Scarsdale}, {Albert}, {Baghel}, {Bean}, {Blecic}, {Lafreni{\`e}re}, {MacDonald}, {Zamyatina}, {Allart1}, {Artigau}, {Batalha}, {Cook}, {Cowan}, {Dang}, {Doyon}, {Fournier-Tondreau}, {Johnstone}, {Line}, {Moran}, {Mukherjee}, {Pelletier}, {Roy}, {Talens}, {Filippazzo}, {Pontoppidan}, \& {Volk}}]{radicaetal2023}
{Radica}, M., {Welbanks}, L., {Espinoza}, N., {et~al.} 2023, \mnras, 524, 835, \dodoi{10.1093/mnras/stad1762}

\bibitem[{{Radica} {et~al.}(2024){Radica}, {Coulombe}, {Taylor}, {Albert}, {Allart}, {Benneke}, {Cowan}, {Dang}, {Lafreni{\`e}re}, {Thorngren}, {Artigau}, {Doyon}, {Flagg}, {Johnstone}, {Pelletier}, \& {Roy}}]{radicaetal2024}
{Radica}, M., {Coulombe}, L.-P., {Taylor}, J., {et~al.} 2024, \apjl, 962, L20, \dodoi{10.3847/2041-8213/ad20e4}

\bibitem[{{Rauscher} {et~al.}(2014){Rauscher}, {Boehm}, {Cagiano}, {Delo}, {Foltz}, {Greenhouse}, {Hickey}, {Hill}, {Kan}, {Lindler}, {Mott}, {Waczynski}, \& {Wen}}]{rauscher2014}
{Rauscher}, B.~J., {Boehm}, N., {Cagiano}, S., {et~al.} 2014, \pasp, 126, 739, \dodoi{10.1086/677681}

\bibitem[{{Redfield} {et~al.}(2008){Redfield}, {Endl}, {Cochran}, \& {Koesterke}}]{redfieldetal2008}
{Redfield}, S., {Endl}, M., {Cochran}, W.~D., \& {Koesterke}, L. 2008, \apjl, 673, L87, \dodoi{10.1086/527475}

\bibitem[{{Rey} {et~al.}(2017){Rey}, {Nikitin}, \& {Tyuterev}}]{reyetal2017}
{Rey}, M., {Nikitin}, A.~V., \& {Tyuterev}, V.~G. 2017, \apj, 847, 105, \dodoi{10.3847/1538-4357/aa8909}

\bibitem[{{Rogers} {et~al.}(2011){Rogers}, {Bodenheimer}, {Lissauer}, \& {Seager}}]{rogersetal2011}
{Rogers}, L.~A., {Bodenheimer}, P., {Lissauer}, J.~J., \& {Seager}, S. 2011, \apj, 738, 59, \dodoi{10.1088/0004-637X/738/1/59}

\bibitem[{{Rothman} {et~al.}(2010){Rothman}, {Gordon}, {Barber}, {Dothe}, {Gamache}, {Goldman}, {Perevalov}, {Tashkun}, \& {Tennyson}}]{rothmanetal2010}
{Rothman}, L.~S., {Gordon}, I.~E., {Barber}, R.~J., {et~al.} 2010, \jqsrt, 111, 2139, \dodoi{10.1016/j.jqsrt.2010.05.001}

\bibitem[{{Rustamkulov} {et~al.}(2023){Rustamkulov}, {Sing}, {Mukherjee}, {May}, {Kirk}, {Schlawin}, {Line}, {Piaulet}, {Carter}, {Batalha}, {Goyal}, {L{\'o}pez-Morales}, {Lothringer}, {MacDonald}, {Moran}, {Stevenson}, {Wakeford}, {Espinoza}, {Bean}, {Batalha}, {Benneke}, {Berta-Thompson}, {Crossfield}, {Gao}, {Kreidberg}, {Powell}, {Cubillos}, {Gibson}, {Leconte}, {Molaverdikhani}, {Nikolov}, {Parmentier}, {Roy}, {Taylor}, {Turner}, {Wheatley}, {Aggarwal}, {Ahrer}, {Alam}, {Alderson}, {Allen}, {Banerjee}, {Barat}, {Barrado}, {Barstow}, {Bell}, {Blecic}, {Brande}, {Casewell}, {Changeat}, {Chubb}, {Crouzet}, {Daylan}, {Decin}, {D{\'e}sert}, {Mikal-Evans}, {Feinstein}, {Flagg}, {Fortney}, {Harrington}, {Heng}, {Hong}, {Hu}, {Iro}, {Kataria}, {Kempton}, {Krick}, {Lendl}, {Lillo-Box}, {Louca}, {Lustig-Yaeger}, {Mancini}, {Mansfield}, {Mayne}, {Miguel}, {Morello}, {Ohno}, {Palle}, {Petit dit de la Roche}, {Rackham}, {Radica}, {Ramos-Rosado}, {Redfield}, {Rogers}, {Shkolnik}, {Southworth}, {Teske}, {Tremblin},
  {Tucker}, {Venot}, {Waalkes}, {Welbanks}, {Zhang}, \& {Zieba}}]{Rustamkulovetal2023}
{Rustamkulov}, Z., {Sing}, D.~K., {Mukherjee}, S., {et~al.} 2023, \nat, 614, 659, \dodoi{10.1038/s41586-022-05677-y}

\bibitem[{{Ryabchikova} {et~al.}(2015){Ryabchikova}, {Piskunov}, {Kurucz}, {Stempels}, {Heiter}, {Pakhomov}, \& {Barklem}}]{ryabchikova_2015}
{Ryabchikova}, T., {Piskunov}, N., {Kurucz}, R.~L., {et~al.} 2015, \physscr, 90, 054005, \dodoi{10.1088/0031-8949/90/5/054005}

\bibitem[{{Sarkar} {et~al.}(2024){Sarkar}, {Madhusudhan}, {Constantinou}, \& {Holmberg}}]{Sarkaretal2024}
{Sarkar}, S., {Madhusudhan}, N., {Constantinou}, S., \& {Holmberg}, M. 2024, \mnras, 531, 2731, \dodoi{10.1093/mnras/stae1230}

\bibitem[{{Schlawin} {et~al.}(2020){Schlawin}, {Leisenring}, {Misselt}, {Greene}, {McElwain}, {Beatty}, \& {Rieke}}]{schlawin2020}
{Schlawin}, E., {Leisenring}, J., {Misselt}, K., {et~al.} 2020, \aj, 160, 231, \dodoi{10.3847/1538-3881/abb811}

\bibitem[{{Schlawin} {et~al.}(2024){Schlawin}, {Ohno}, {Bell}, {Murphy}, {Welbanks}, {Beatty}, {Greene}, {Fortney}, {Parmentier}, {Edelman}, {Gill}, {Anderson}, {Wheatley}, {Henry}, {Mehta}, {Kreidberg}, \& {Rieke}}]{schlawinetal2024}
{Schlawin}, E., {Ohno}, K., {Bell}, T.~J., {et~al.} 2024, \apjl, 974, L33, \dodoi{10.3847/2041-8213/ad7fef}

\bibitem[{{Seidel} {et~al.}(2019){Seidel}, {Ehrenreich}, {Wyttenbach}, {Allart}, {Lendl}, {Pino}, {Bourrier}, {Cegla}, {Lovis}, {Barrado}, {Bayliss}, {Astudillo-Defru}, {Deline}, {Fisher}, {Heng}, {Joseph}, {Lavie}, {Melo}, {Pepe}, {S{\'e}gransan}, \& {Udry}}]{seideletal2019}
{Seidel}, J.~V., {Ehrenreich}, D., {Wyttenbach}, A., {et~al.} 2019, \aap, 623, A166, \dodoi{10.1051/0004-6361/201834776}

\bibitem[{{Seidel} {et~al.}(2020){Seidel}, {Ehrenreich}, {Bourrier}, {Allart}, {Attia}, {Hoeijmakers}, {Lendl}, {Linder}, {Wyttenbach}, {Astudillo-Defru}, {Bayliss}, {Cegla}, {Heng}, {Lavie}, {Lovis}, {Melo}, {Pepe}, {dos Santos}, {S{\'e}gransan}, \& {Udry}}]{seideletal2020}
{Seidel}, J.~V., {Ehrenreich}, D., {Bourrier}, V., {et~al.} 2020, \aap, 641, L7, \dodoi{10.1051/0004-6361/202038497}

\bibitem[{{Seidel} {et~al.}(2022){Seidel}, {Cegla}, {Doyle}, {Lafarga}, {Brogi}, {Gandhi}, {Anderson}, {Allart}, {Buchschacher}, {Lovis}, \& {Sosnowska}}]{seideletal2022}
{Seidel}, J.~V., {Cegla}, H.~M., {Doyle}, L., {et~al.} 2022, \mnras, 513, L15, \dodoi{10.1093/mnrasl/slac027}

\bibitem[{{Serdyuchenko} {et~al.}(2014){Serdyuchenko}, {Gorshelev}, {Weber}, {Chehade}, \& {Burrows}}]{Serdyuchenkoetal2014}
{Serdyuchenko}, A., {Gorshelev}, V., {Weber}, M., {Chehade}, W., \& {Burrows}, J.~P. 2014, Atmospheric Measurement Techniques, 7, 625, \dodoi{10.5194/amt-7-625-201410.5194/amtd-6-6613-2013}

\bibitem[{{Sing} {et~al.}(2016){Sing}, {Fortney}, {Nikolov}, {Wakeford}, {Kataria}, {Evans}, {Aigrain}, {Ballester}, {Burrows}, {Deming}, {D{\'e}sert}, {Gibson}, {Henry}, {Huitson}, {Knutson}, {Lecavelier Des Etangs}, {Pont}, {Showman}, {Vidal-Madjar}, {Williamson}, \& {Wilson}}]{singetal2016}
{Sing}, D.~K., {Fortney}, J.~J., {Nikolov}, N., {et~al.} 2016, \nat, 529, 59, \dodoi{10.1038/nature16068}

\bibitem[{{Sing} {et~al.}(2024){Sing}, {Rustamkulov}, {Thorngren}, {Barstow}, {Tremblin}, {Alves de Oliveira}, {Beck}, {Birkmann}, {Challener}, {Crouzet}, {Espinoza}, {Ferruit}, {Giardino}, {Gressier}, {Lee}, {Lewis}, {Maiolino}, {Manjavacas}, {Rauscher}, {Sirianni}, \& {Valenti}}]{singetal2024}
{Sing}, D.~K., {Rustamkulov}, Z., {Thorngren}, D.~P., {et~al.} 2024, \nat, 630, 831, \dodoi{10.1038/s41586-024-07395-z}

\bibitem[{{Sousa-Silva} {et~al.}(2015){Sousa-Silva}, {Al-Refaie}, {Tennyson}, \& {Yurchenko}}]{Sousasilvaetal2015}
{Sousa-Silva}, C., {Al-Refaie}, A.~F., {Tennyson}, J., \& {Yurchenko}, S.~N. 2015, \mnras, 446, 2337, \dodoi{10.1093/mnras/stu2246}

\bibitem[{{Sousa-Silva} {et~al.}(2014){Sousa-Silva}, {Hesketh}, {Yurchenko}, {Hill}, \& {Tennyson}}]{sousa-silvaetal2014}
{Sousa-Silva}, C., {Hesketh}, N., {Yurchenko}, S.~N., {Hill}, C., \& {Tennyson}, J. 2014, \jqsrt, 142, 66, \dodoi{10.1016/j.jqsrt.2014.03.012}

\bibitem[{{Speagle}(2020)}]{Speagle2020}
{Speagle}, J.~S. 2020, \mnras, 493, 3132, \dodoi{10.1093/mnras/staa278}

\bibitem[{{Stock} {et~al.}(2022){Stock}, {Kitzmann}, \& {Patzer}}]{stocketal2022}
{Stock}, J.~W., {Kitzmann}, D., \& {Patzer}, A. B.~C. 2022, \mnras, 517, 4070, \dodoi{10.1093/mnras/stac2623}

\bibitem[{{Tashkun} \& {Perevalov}(2011)}]{tashkun+perevalov2011}
{Tashkun}, S.~A., \& {Perevalov}, V.~I. 2011, \jqsrt, 112, 1403, \dodoi{10.1016/j.jqsrt.2011.03.005}

\bibitem[{{Thao} {et~al.}(2024){Thao}, {Mann}, {Feinstein}, {Gao}, {Thorngren}, {Rotman}, {Welbanks}, {Brown}, {Duvvuri}, {France}, {Longo}, {Sandoval}, {Schneider}, {Wilson}, {Youngblood}, {Vanderburg}, {Barber}, {Wood}, {Batalha}, {Kraus}, {Murray}, {Newton}, {Rizzuto}, {Tofflemire}, {Tsai}, {Bean}, {Berta-Thompson}, {Evans-Soma}, {Froning}, {Kempton}, {Miguel}, \& {Pineda}}]{thaoetal2024}
{Thao}, P.~C., {Mann}, A.~W., {Feinstein}, A.~D., {et~al.} 2024, \aj, 168, 297, \dodoi{10.3847/1538-3881/ad81d7}

\bibitem[{{Thiabaud} {et~al.}(2015){Thiabaud}, {Marboeuf}, {Alibert}, {Leya}, \& {Mezger}}]{thiabaudetal2015}
{Thiabaud}, A., {Marboeuf}, U., {Alibert}, Y., {Leya}, I., \& {Mezger}, K. 2015, \aap, 574, A138, \dodoi{10.1051/0004-6361/201424868}

\bibitem[{{Tsai} {et~al.}(2017){Tsai}, {Lyons}, {Grosheintz}, {Rimmer}, {Kitzmann}, \& {Heng}}]{tsaietal2017}
{Tsai}, S.-M., {Lyons}, J.~R., {Grosheintz}, L., {et~al.} 2017, \apjs, 228, 20, \dodoi{10.3847/1538-4365/228/2/20}

\bibitem[{{Tsai} {et~al.}(2021){Tsai}, {Malik}, {Kitzmann}, {Lyons}, {Fateev}, {Lee}, \& {Heng}}]{tsaietal2021}
{Tsai}, S.-M., {Malik}, M., {Kitzmann}, D., {et~al.} 2021, \apj, 923, 264, \dodoi{10.3847/1538-4357/ac29bc}

\bibitem[{{Tsai} {et~al.}(2023){Tsai}, {Lee}, {Powell}, {Gao}, {Zhang}, {Moses}, {H{\'e}brard}, {Venot}, {Parmentier}, {Jordan}, \& et~al.}]{tsaietal2023}
{Tsai}, S.-M., {Lee}, E. K.~H., {Powell}, D., {et~al.} 2023, \nat, 617, 483, \dodoi{10.1038/s41586-023-05902-2}

\bibitem[{{Underwood} {et~al.}(2016){Underwood}, {Tennyson}, {Yurchenko}, {Huang}, {Schwenke}, {Lee}, {Clausen}, \& {Fateev}}]{underwood_et_al2016}
{Underwood}, D.~S., {Tennyson}, J., {Yurchenko}, S.~N., {et~al.} 2016, \mnras, 459, 3890, \dodoi{10.1093/mnras/stw849}

\bibitem[{{Wakeford} {et~al.}(2022){Wakeford}, {Alderson}, {Batalha}, {Grant}, {Lewis}, {Lopez-Morales}, {MacDonald}, {Marley}, {Moran}, \& {Ohno}}]{wakefordetal2022}
{Wakeford}, H., {Alderson}, L., {Batalha}, N., {et~al.} 2022, {Hubble Ultraviolet-optical Survey of Transiting Legacy Exoplanets (HUSTLE) treasury program}, HST Proposal. Cycle 30, ID. \#17183

\bibitem[{{Wakeford} {et~al.}(2017){Wakeford}, {Sing}, {Kataria}, {Deming}, {Nikolov}, {Lopez}, {Tremblin}, {Amundsen}, {Lewis}, {Mandell}, {Fortney}, {Knutson}, {Benneke}, \& {Evans}}]{wakefordetal2017}
{Wakeford}, H.~R., {Sing}, D.~K., {Kataria}, T., {et~al.} 2017, Science, 356, 628, \dodoi{10.1126/science.aah4668}

\bibitem[{{Welbanks} \& {Madhusudhan}(2019)}]{welbanksandmadhusudhan2019}
{Welbanks}, L., \& {Madhusudhan}, N. 2019, \aj, 157, 206, \dodoi{10.3847/1538-3881/ab14de}

\bibitem[{{Welbanks} {et~al.}(2019){Welbanks}, {Madhusudhan}, {Allard}, {Hubeny}, {Spiegelman}, \& {Leininger}}]{welbanksetal2019}
{Welbanks}, L., {Madhusudhan}, N., {Allard}, N.~F., {et~al.} 2019, \apjl, 887, L20, \dodoi{10.3847/2041-8213/ab5a89}

\bibitem[{{Welbanks} {et~al.}(2024){Welbanks}, {Bell}, {Beatty}, {Line}, {Ohno}, {Fortney}, {Schlawin}, {Greene}, {Rauscher}, {McGill}, {Murphy}, {Parmentier}, {Tang}, {Edelman}, {Mukherjee}, {Wiser}, {Lagage}, {Dyrek}, \& {Arnold}}]{welbanksetal2024}
{Welbanks}, L., {Bell}, T.~J., {Beatty}, T.~G., {et~al.} 2024, \nat, 630, 836, \dodoi{10.1038/s41586-024-07514-w}

\bibitem[{{Woitke} \& {Helling}(2021)}]{woitkeandhelling2021}
{Woitke}, P., \& {Helling}, C. 2021, {GGchem: Fast thermo-chemical equilibrium code}, Astrophysics Source Code Library, record ascl:2104.018

\bibitem[{{Woitke} {et~al.}(2018){Woitke}, {Helling}, {Hunter}, {Millard}, {Turner}, {Worters}, {Blecic}, \& {Stock}}]{woitkeetal2018}
{Woitke}, P., {Helling}, C., {Hunter}, G.~H., {et~al.} 2018, \aap, 614, A1, \dodoi{10.1051/0004-6361/201732193}

\bibitem[{{Wong} {et~al.}(2017){Wong}, {Yurchenko}, {Bernath}, {M{\"u}ller}, {McConkey}, \& {Tennyson}}]{wongetal2017}
{Wong}, A., {Yurchenko}, S.~N., {Bernath}, P., {et~al.} 2017, \mnras, 470, 882, \dodoi{10.1093/mnras/stx1211}

\bibitem[{{Wordsworth}(2015)}]{wordsworth2015}
{Wordsworth}, R. 2015, \apj, 806, 180, \dodoi{10.1088/0004-637X/806/2/180}

\bibitem[{{Wyttenbach} {et~al.}(2017){Wyttenbach}, {Lovis}, {Ehrenreich}, {Bourrier}, {Pino}, {Allart}, {Astudillo-Defru}, {Cegla}, {Heng}, {Lavie}, {Melo}, {Murgas}, {Santerne}, {S{\'e}gransan}, {Udry}, \& {Pepe}}]{wyttenbachetal2017}
{Wyttenbach}, A., {Lovis}, C., {Ehrenreich}, D., {et~al.} 2017, \aap, 602, A36, \dodoi{10.1051/0004-6361/201630063}

\bibitem[{{Yurchenko} {et~al.}(2011){Yurchenko}, {Barber}, \& {Tennyson}}]{yurchenkoetal2011}
{Yurchenko}, S.~N., {Barber}, R.~J., \& {Tennyson}, J. 2011, \mnras, 413, 1828, \dodoi{10.1111/j.1365-2966.2011.18261.x}

\bibitem[{{Yurchenko} {et~al.}(2018{\natexlab{a}}){Yurchenko}, {Bond}, {Gorman}, {Lodi}, {McKemmish}, {Nunn}, {Shah}, \& {Tennyson}}]{yurchenkoetal2018a}
{Yurchenko}, S.~N., {Bond}, W., {Gorman}, M.~N., {et~al.} 2018{\natexlab{a}}, \mnras, 478, 270, \dodoi{10.1093/mnras/sty939}

\bibitem[{{Yurchenko} {et~al.}(2020){Yurchenko}, {Mellor}, {Freedman}, \& {Tennyson}}]{yurchenko_et_al2020}
{Yurchenko}, S.~N., {Mellor}, T.~M., {Freedman}, R.~S., \& {Tennyson}, J. 2020, \mnras, 496, 5282, \dodoi{10.1093/mnras/staa1874}

\bibitem[{{Yurchenko} {et~al.}(2024){Yurchenko}, {Owens}, {Kefala}, \& {Tennyson}}]{yurchenko_et_al2024}
{Yurchenko}, S.~N., {Owens}, A., {Kefala}, K., \& {Tennyson}, J. 2024, \mnras, 528, 3719, \dodoi{10.1093/mnras/stae148}

\bibitem[{{Yurchenko} {et~al.}(2018{\natexlab{b}}){Yurchenko}, {Sinden}, {Lodi}, {Hill}, {Gorman}, \& {Tennyson}}]{Yurchenkoetal2018b}
{Yurchenko}, S.~N., {Sinden}, F., {Lodi}, L., {et~al.} 2018{\natexlab{b}}, \mnras, 473, 5324, \dodoi{10.1093/mnras/stx2738}

\bibitem[{{Yurchenko} {et~al.}(2018{\natexlab{c}}){Yurchenko}, {Szab{\'o}}, {Pyatenko}, \& {Tennyson}}]{yurchenkoetal2018c}
{Yurchenko}, S.~N., {Szab{\'o}}, I., {Pyatenko}, E., \& {Tennyson}, J. 2018{\natexlab{c}}, \mnras, 480, 3397, \dodoi{10.1093/mnras/sty2050}

\bibitem[{{Yurchenko} {et~al.}(2022){Yurchenko}, {Tennyson}, {Syme}, {Adam}, {Clark}, {Cooper}, {Dobney}, {Donnelly}, {Gorman}, {Lynas-Gray}, {Meltzer}, {Owens}, {Qu}, {Semenov}, {Somogyi}, {Upadhyay}, {Wright}, \& {Zapata Trujillo}}]{Yurchenkoetal2022}
{Yurchenko}, S.~N., {Tennyson}, J., {Syme}, A.-M., {et~al.} 2022, \mnras, 510, 903, \dodoi{10.1093/mnras/stab3267}

\bibitem[{{Zhang} {et~al.}(2019){Zhang}, {Chachan}, {Kempton}, \& {Knutson}}]{zhangetal2019}
{Zhang}, M., {Chachan}, Y., {Kempton}, E. M.~R., \& {Knutson}, H.~A. 2019, \pasp, 131, 034501, \dodoi{10.1088/1538-3873/aaf5ad}

\bibitem[{{Zhang} {et~al.}(2020){Zhang}, {Chachan}, {Kempton}, {Knutson}, \& {Chang}}]{zhangetal2020}
{Zhang}, M., {Chachan}, Y., {Kempton}, E. M.~R., {Knutson}, H.~A., \& {Chang}, W.~H. 2020, \apj, 899, 27, \dodoi{10.3847/1538-4357/aba1e6}

\end{thebibliography}
\bibliographystyle{aasjournal}

\end{document}